# Present knowledge of the Cabibbo-Kobayashi-Maskawa matrix

M. BARGIOTTI(1), A. BERTIN(1), M. BRUSCHI(1), M. CAPPONI(1), S. DE CASTRO(1), R. DONÀ(1), P. FACCIOLI(1), D. GALLI(1), B. GIACOBBE(1), U. MARCONI(1), I. MASSA(1), M. PICCININI(1), M. POLI(2), N. SEMPRINI CESARI(1), R. SPIGHI(1), V. VAGNONI(1), S. VECCHI(1), M. VILLA(1), A. VITALE(1) AND A. ZOCCOLI(1)

(1) *Dipartimento di Fisica dell'Università di Bologna – Bologna, Italy*
*Istituto Nazionale di Fisica Nucleare Sezione di Bologna – Bologna, Italy*

(2) *Dipartimento di Energetica 'Sergio Stecco' dell'Università di Firenze – Firenze, Italy*
*Istituto Nazionale di Fisica Nucleare Sezione di Bologna – Bologna, Italy*

## Abstract

A complete review of the Cabibbo-Kobayashi-Maskawa (CKM) matrix elements and of the experimental methods for their determination is presented. A critical analysis of the relevant experimental results, and in particular of the most recent ones, allows to improve the accuracies of all the matrix elements. A $\chi^2$ minimization with the three-family unitarity constraint on the CKM matrix is performed to test the current interpretation of the CP violating phenomena inside the Standard Model. A complete and unambiguous solution satisfying all the imposed constraints is found. As a by-product of the fit, the precision on the values of the matrix elements is further increased and it is possible to obtain estimates for the important CP violation observables $\sin 2\beta$, $\sin 2\alpha$ and $\gamma$. Finally, an independent estimation of the CKM elements based on a Bayesian approach is performed. This complementary method constitutes a check of the results obtained, providing also the probability functions of the CKM elements and of the related quantities.

# 1 Introduction

One of the most interesting and mysterious aspects of nature is its behaviour under Charge-Parity (CP) transformation. For more than 35 years after the discovery of CP violation in the neutral kaon sector, no conclusive evidence of violation of this symmetry has been found in other phenomena, but a huge experimental effort is being made in this direction. The main reason for the interest in this field is related to the question whether the Standard Model is able to account for the observed magnitude of these phenomena. CP violation is introduced, in a general way, into the charged-current weak interactions between quarks existing in at least three different families. In the Standard Model, it originates from an unremovable complex phase in the Cabibbo-Kobayashi-Maskawa[1,2] (CKM) matrix, which describes the flavour mixing between the quark states. With this formulation, it is possible to foresee many different CP violating phenomena, which will be a matter of study in the current (as well as in the next) generation of *B* factories. In order to have reliable estimations for these phenomena, as well as to test this sector of the Standard Model, a precise knowledge of all the matrix elements is necessary.
The values of the CKM matrix elements are not fixed by theory and must be evaluated on the basis of the available experimental information. This can be performed by studying specific classes of processes, such as, for example, the semileptonic decays of mesons or baryons, and by exploiting the relations between the experimental determinations of decay rates or branching ratios and the relevant matrix elements. The theoretical uncertainties in the hadronic system description usually limit the final precision obtained in the determination of the CKM matrix. Notwithstanding this difficulty, which in some cases makes the related experimental information even unserviceable, all the CKM matrix elements of the first two rows can be safely determined.
The requirement that the CKM matrix be a unitary matrix imposes strong constraints between its elements. This characteristic allows one to test the goodness of the values found and of the assumptions made in the Standard Model in order to account for the observed CP violation.

The main purposes of this work are a survey of the CKM matrix elements as determined one by one from the experimental data and the refinement of their values by means of a unitarity-constrained minimum-$\chi^2$ fit. The main results of this study already appeared in Refs. 3 and 4. The impact of new measurements foreseen in the next years is also emphasized.

In the following paragraphs, a short introduction to the CKM matrix and the unitarity triangle is presented. In Section 2, the direct measurements of the CKM matrix elements and other experimental constraints are discussed taking into account the most recent experimental, theoretical and phenomenological achievements. In Section 3, we present the methodology used to obtain the maximum precision for the CKM parameters from the results presented in Section 2, through the requirement that the CKM matrix be a $3\times3$ unitary matrix. Finally, the determination of the imaginary part of the matrix and the best estimation of the CP violating observables are presented.

## 1.1 The CKM matrix

In the Standard Model, the quark mass eigenstates (physical states) do not take part as pure states in weak interactions. The unitary transformation connecting the two bases of mass and weak eigenstates is represented by the Cabibbo-Kobayashi-Maskawa (CKM) matrix. By convention, the charge +2/3 quarks (*u, c* and *t*) are chosen to be pure states, and flavour mixing is described in terms of a 3×3 matrix operating on the *d, s* and *b* quark states:

$$\begin{pmatrix} d' \\ s' \\ b' \end{pmatrix} = V_{CKM} \begin{pmatrix} d \\ s \\ b \end{pmatrix} = \begin{pmatrix} V_{ud} & V_{us} & V_{ub} \\ V_{cd} & V_{cs} & V_{cb} \\ V_{td} & V_{ts} & V_{tb} \end{pmatrix} \begin{pmatrix} d \\ s \\ b \end{pmatrix} \tag{1.1-1}$$

Thus *d', s'* and *b'*, instead of *d, s* and *b*, are partners of *u, c* and *t* respectively within the weak isospin doublets.

All the properties of weak quark interaction are codified inside the CKM matrix, which makes it possible to extend the Cabibbo[1] model, preserving the weak coupling universality while explaining the existing priority scale among the transitions occurring inside one quark family and those connecting two neighbouring families, or the first with the third one. Its structure incorporates the GIM[5] mechanism, which suppresses the flavour-changing neutral-current (FCNC) processes. Finally, the imaginary part of the CKM matrix is the source of all the CP-violating phenomena which the Standard Model is able to account for.

CP symmetry, which is equivalent, according to the CPT theorem, to time reversal invariance T, would be conserved if the matrix were real. On the other hand, to account for CP violation $V_{CKM}$ must be complex independently of the phase convention of the fermionic fields, which it acts on. The Lagrangian density term which represents the ($W^{\pm}$-mediated) charge current processes involving quarks depends on $V_{CKM}$ according to the expression

$$\mathcal{L}_{W^{\pm}} \propto G_F \begin{pmatrix} \bar{u} & \bar{c} & \bar{t} \end{pmatrix} \gamma^{\mu} \left(1-\gamma^5\right) V_{CKM} \begin{pmatrix} d \\ s \\ b \end{pmatrix} W_{\mu} + h.c. \tag{1.1-2}$$

Let us consider initially the most general case. As an $N \times N$ unitary matrix, $V_{CKM}$ depends on $N^2$ real parameters, $2N - 1$ of which can be reabsorbed by the quark fields which multiply the right and left sides of the matrix, by means of a global redefinition of the arbitrary phases. Of the remaining $(N - 1)^2$ free parameters, $N(N - 1)/2$ are the Euler angles, common to the real (orthogonal) matrix and to the complex (unitary) one, the others are unremovable complex phases; they are therefore physically meaningful and can be measured as signals of CP violation. The $N = 2$ matrix, which contains only one parameter (the Cabibbo angle), is real; consequently, it cannot give rise to CP violation. This is precisely the reason why Kobayashi and Maskawa[3] conjectured the existence of three quark doublets in 1973. The authors presented the hypothesis as one of several possible explanations, which were able to account for CP non-conservation with a minimal extension of the two-generation model. But the idea began to be held in high consideration when Perl and collab.[6] discovered the third lepton ($\tau$) in 1975, following up the original intuition and preceding experience of Zichichi and collab.[7], and after the discovery of the fifth quark (*b*) at Fermilab[8] in 1977.

The CKM matrix ($N = 3$) can be parametrized in terms of three Euler angles and one phase; the latter is entirely responsible for the CP violation in the Standard Model. Several possible parametrizations differ both in the choice of the Euler angles and in the positioning of the phases; the one proposed by Chau and Keung[9] combining notations already used by Maiani[10] and Wolfenstein[11] is adopted by the PDG[12] as the 'canonical' parametrization:

$$V_{CKM} = \begin{pmatrix} c_{12}c_{13} & s_{12}c_{13} & s_{13}e^{-i\delta_{13}} \\ -s_{12}c_{23} - c_{12}s_{23}s_{13}e^{i\delta_{13}} & c_{12}c_{23} - s_{12}s_{23}s_{13}e^{i\delta_{13}} & s_{23}c_{13} \\ s_{12}s_{23} - c_{12}c_{23}s_{13}e^{i\delta_{13}} & -c_{12}s_{23} - s_{12}c_{23}s_{13}e^{i\delta_{13}} & c_{23}c_{13} \end{pmatrix} \qquad (1.1\text{-}3)$$

where $c_{ij} = \cos\vartheta_{ij}$, $s_{ij} = \sin\vartheta_{ij}$, $\vartheta_{ij}$ is the mixing angle between the *i*th and the *j*th generation ($\vartheta_{12}$ is the Cabibbo angle), and $\delta_{13}$ is the phase angle.

Given the experimentally observed hierarchy among couplings

$$|V_{ub}| < |V_{cb}| < |V_{us}|, |V_{cd}| < 1, \qquad (1.1\text{-}4)$$

the matrix elements can be written in terms of powers of the sine of the Cabibbo angle

$$\lambda = s_{12} \cong |V_{us}| \cong |V_{cd}| \cong 0.22. \qquad (1.1\text{-}5)$$

If one defines the parameters $A$, $\rho$, $\eta$ according to the relations

$$s_{23} = A\lambda^2, \qquad s_{13}e^{-i\delta_{13}} = A\lambda^3(\rho - i\eta) \qquad (1.1\text{-}6)$$

and neglects, for example, terms of order $\mathcal{O}(\lambda^6)$, the following expression

$$V_{CKM} = \begin{pmatrix} 1 - \frac{\lambda^2}{2} - \frac{\lambda^4}{8} & \lambda & A\lambda^3(\rho - i\eta) \\ -\lambda\left[1 - A^2\lambda^4\left(\frac{1}{2} - \rho\right) + iA^2\lambda^4\eta\right] & 1 - \frac{\lambda^2}{2} - \frac{\lambda^4}{8}(1 + 4A^2) & A\lambda^2 \\ A\lambda^3\left[(1-\rho) + \lambda^2\frac{\rho}{2} - i\eta\left(1 - \frac{\lambda^2}{2}\right)\right] & -A\lambda^2\left[1 - \lambda^2\left(\frac{1}{2} - \rho\right) + i\eta\lambda^2\right] & 1 - \frac{A^2\lambda^4}{2} \end{pmatrix}$$

(1.1-7)

is obtained. This parametrization, formulated by Wolfenstein[11], has the advantage of making the characters which differentiate the CKM matrix from a common unitary matrix particularly evident: it is almost diagonal (the diagonal elements are close to unity) and its elements decrease in magnitude with increasing distance from the diagonal, according to a nearly symmetrical pattern ($|V_{ji}| \cong |V_{ij}|$ at the first non-zero order in $\lambda$). The second part of Eq. (1.1-6) yields the following relation (for $\rho \neq 0$)

$$\tan\delta_{13} = \frac{\eta}{\rho} \qquad (1.1\text{-}8)$$

between the phase and the parameters $\eta$ and $\rho$: therefore $\eta \neq 0$ is the CP symmetry-breaking condition in the Standard Model, as expressed in Wolfenstein's parametrization.

## 1.2 The unitarity triangle

Among the orthonormality relations of the row-vectors and those of the column-vectors of the CKM matrix (9 independent relations in all),

$$\left|V_{id}\right|^2+\left|V_{is}\right|^2+\left|V_{ib}\right|^2=1 \qquad i=u,c,t \tag{1.2-1}$$

$$V_{ud}V_{us}^*+V_{cd}V_{cs}^*+V_{td}V_{ts}^*=0 \tag{1.2-2}$$

$$V_{ud}V_{ub}^*+V_{cd}V_{cb}^*+V_{td}V_{tb}^*=0 \tag{1.2-3}$$

$$\vdots$$

(the first provides a way of testing the unitarity condition, with reference to direct measurements of the elements of the same row), Eq. (1.2-3) is of particular interest, as it defines on the complex plane a triangle whose sides have dimensions of the same order in $\lambda$ ($\mathcal{O}(\lambda^3)$) and thus subtend angles having comparable amplitudes. The length of one of the sides can be normalised to the real value 1: dividing Eq. (1.2-3) by $V_{cd}V_{cb}^*$, one obtains the so-called *unitarity triangle*, defined by the relation

$$1+\frac{V_{td}V_{tb}^*}{V_{cd}V_{cb}^*}+\frac{V_{ud}V_{ub}^*}{V_{cd}V_{cb}^*}=0\,, \tag{1.2-4}$$

and represented in Figure 1, where the angles are indicated, according to a widespread convention, by the letters $\alpha$, $\beta$, $\gamma$.

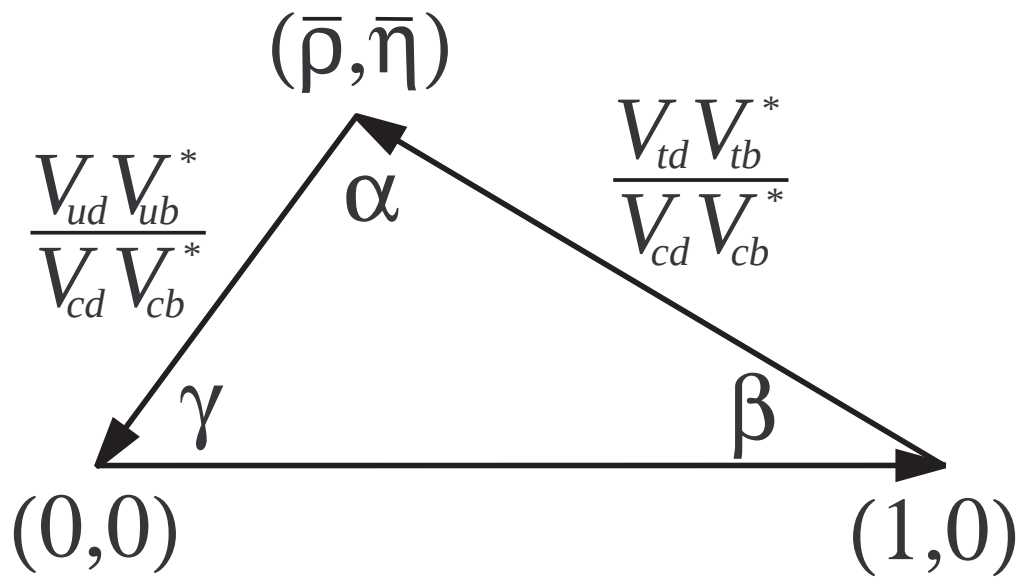

**Figure 1.** – *The unitarity triangle*

The vertex of the triangle is the complex vector

$$1+\frac{V_{td}V_{tb}^*}{V_{cd}V_{cb}^*}=(\rho+i\eta)\left(1-\frac{\lambda^2}{2}\right)+\mathcal{O}(\lambda^4). \tag{1.2-5}$$

The correction of order $\lambda^4$ is equal to $\frac{1}{8}\lambda^4 f(\rho,\eta,A)\cong 3\cdot 10^{-4}$; thus

$$\bar\rho=\rho\left(1-\frac{\lambda^2}{2}\right) \quad \text{and} \quad \bar\eta=\eta\left(1-\frac{\lambda^2}{2}\right) \tag{1.2-6}$$

are, almost exactly, the coordinates of the vertex in the complex plane. The angle $\gamma$ is equal to the phase $\delta_{13}$ of the canonical parametrization:

$$\gamma=\arctan\frac{\bar\eta}{\bar\rho}=\arctan\frac{\eta}{\rho}=\delta_{13}\,. \tag{1.2-7}$$

$\alpha$, $\beta$ and $\gamma$ are in direct relation to the CP asymmetries which are peculiar to $B$ meson decays. Forthcoming experiments will measure sin2$\alpha$, sin2$\beta$ and sin$^2\gamma$ or sin2$\gamma$. The first two observables can be expressed as functions of $\bar{\rho}$ and $\bar{\eta}$ in the following way:

$$\sin 2\alpha = \frac{2\bar{\eta}\left(\bar{\eta}^2 + \bar{\rho}^2 - \bar{\rho}\right)}{\left(\bar{\eta}^2 + \bar{\rho}^2 - \bar{\rho}\right)^2 + \bar{\eta}^2}; \qquad \sin 2\beta = \frac{2\bar{\eta}\left(1-\bar{\rho}\right)}{\left(1-\bar{\rho}\right)^2 + \bar{\eta}^2}. \tag{1.2-8}$$

The CP asymmetry which should be measured in the decays $B_d^0, \bar{B}_d^0 \to J/\psi\, K_S^0$ is defined as

$$a_{J/\psi K_S^0}^{CP} = \frac{N\left(\bar{B}_d^0 \to J/\psi\, K_S^0\right) - N\left(B_d^0 \to J/\psi\, K_S^0\right)}{N\left(\bar{B}_d^0 \to J/\psi\, K_S^0\right) + N\left(B_d^0 \to J/\psi\, K_S^0\right)} \tag{1.2-9}$$

($N$ is the number of events in a given proper time interval) and is proportional to sin2$\beta$. A very preliminary measurement of this asymmetry has been performed recently by CDF[13], but given the lack of precision it cannot be interpreted as a definitive proof of CP violation in the $b$ quark sector. New, precise measurements of this quantity are expected in the forthcoming years from dedicated experiments like BaBar, Belle and HERA-B, which should start their physics data taking since the beginning of the year 2000. Moreover, the upgraded version of the CDF detector should provide new results from the beginning of 2001. Finally, the LHCb experiment should be operative from year 2006. These experiments will be active at different machines ($e^+e^-$ colliders, $pN$ fixed target, $p\bar{p}$ and $pp$ colliders) and with different experimental techniques. This will lead to independent determinations of the angles of the unitarity triangle. The attainable precision will also depend on the technique used and on the actual value of these quantities.

# 2 Experimental determination of the CKM matrix elements

As it will become explicit in the following Subsections, the possibility of carrying out independent measurements of the CKM matrix elements and other observables connected with weak decay amplitudes is due to the great precision with which the Fermi constant $G_F$ was measured.

The measurement of the muon lifetime, by which $G_F$ is determined, has been the subject of nearly half a century of experimental inquiries[14], which began in the early Forties. The first accurate measurement was carried out at CERN[15] in 1962 and, for the first time, an inconsistency with the coupling constant of the nuclear transition $^{14}\mathrm{O} \to {}^{14}\mathrm{N} + e^+ + \nu_e$ clearly emerged. One year later this evidence led Cabibbo[1] to a new formulation of the weak-coupling universality principle. The present level of precision was achieved in 1984 by the Saclay-CERN-Bologna[16] and TRIUMPH[17] groups. The SCB group also provided one of the most meaningful tests of CPT invariance through a determination of the ratio $\tau_{\mu^+} / \tau_{\mu^-}$. From the world average of these measurements, the following value of $G_F$ is obtained[12] taking into account the radiative corrections:

$$G_F = (1.16639 \pm 0.00001) \cdot 10^{-5} GeV^{-2} \quad (^*). \tag{2-1}$$

Sections 2.1 to 2.6 describe the direct measurements of the CKM matrix elements of the first two rows. In most cases, the main source of uncertainty in the values obtained is the theoretical interpretation of the experimental results. In fact, detailed calculations of hadronic terms are generally needed to relate the measured quantities to the CKM elements. Moreover, different theoretical approaches do not always give univocal predictions. For example, in the recent investigations of rare *B* decays the data analysis is strongly dependent on the model used. The problems of theoretical interpretation are not restricted to the latest results: the extraction of $|V_{us}|$ from the hyperon decay measurements, for example, still cannot rely on a solid agreement between independent theoretical calculations. As a further example, the lack of consistency between the very precise $|V_{ud}|$ value obtained from the analysis of nuclear decays and the unitarity of the CKM matrix has led to the formulation of a number of (not yet fully established) hypotheses which can explain the discrepancy with the expected value by predicting additional nuclear structure effects. The discussion of these aspects is essential in obtaining reliable estimations of the CKM matrix elements.

From a different point of view, the main interest in the measurement of the observables which are related to the elements of the third row (*B* meson mixing, CP violation parameters) is the chance of revealing the existence of new physics, since the hypotheses involved are subject to a wide range of possible modifications in the extensions of the Standard Model. The information available on the elements of the third row will be collectively discussed in section 2.7, since all the measurements give combinations rather than single CKM elements.

---

[*] Units $\hbar = c = 1$ are used here and throughout the text.

## 2.1 $|V_{ud}|$

Super-allowed nuclear transitions, neutron decay and pion $\beta$ decay are the experimental sources which provide basic information for the determination of $|V_{ud}|$. Several difficulties are encountered in the theoretical description of these processes when tracing back from the hadron to the quark level ($u \leftrightarrow d$) transition. The result obtained from the study of nuclear decays is sufficient to make $|V_{ud}|$ the CKM matrix element known with the greatest precision; nevertheless, additional nuclear corrections, which have been proposed in order to account for a 2 σ incompatibility with the value expected assuming unitarity, may still have to be applied. The deduction of $|V_{ud}|$ from neutron decay has to deal with the historical inconsistency between the independent measurements of the axial coupling constant as well as of the neutron mean-life. Pion $\beta$ decay provides the cleanest way of extracting $|V_{ud}|$, but a suitable level of precision has still to be attained through a refined measurement of the decay rate.

### 2.1.1 Super-allowed nuclear decays

The most precise and the largest amount of experimental information on $|V_{ud}|$ comes from the analysis of *super-allowed nuclear β decays,* in which the spin-parity $J^P = 0^+$ and isospin of the nuclei are preserved. Due to the energies released in these processes, the so-called *allowed approximation* assumes that the emitted leptons have zero relative angular momentum. Under this hypothesis, the $0^+ \to 0^+$ nuclear decays can be simply described considering only the *vector* part of the weak interaction (Fermi transitions). Moreover, the nucleus involved in a transition occurring within an isospin multiplet is labelled with the same quantum numbers in the final state as in the initial one; then, assuming an exact isospin symmetry and that the other nucleons behave as spectators, the initial and final nuclear states are completely superimposed, so that the nuclear matrix element can be assumed independent of any detail of the nuclear structure and is given by a simple isospin coefficient. In this way the experimental values of *ft* – the product of the phase space factor *f,* depending on the atomic number and the end-point energy, and the half life *t* of the specific process – are expected to be nearly the same in all the super-allowed transitions.

For the super-allowed transitions occurring within isospin $T$= 1 multiplets, *ft* is defined by

$$ft = \frac{K}{G_V^2 |M_{fi}|^2} = \frac{K}{2G_F^2 |V_{ud}|^2} \tag{2.1-1}$$

where $|M_{fi}|^2 = 2$ is the matrix element, $G_V$ the vector coupling constant, which depends on the $|V_{ud}|$ matrix element, $G_V = G_F |V_{ud}|$, and the constant $K$ is

$$K = 2\pi^3 \ln 2 / m_e^5 = (8120.271 \pm 0.012) \cdot 10^{-10} GeV^{-4} s\,. \tag{2.1-2}$$

A more accurate description of these processes, which includes the small (of the order of 1%) and calculable contribution of electromagnetic effects and radiative corrections, leads to the following expression:

$$\mathcal{F}t \equiv ft\left(1+\delta_R\right)\left(1-\delta_C\right) = \frac{K}{2G_F^2\left|V_{ud}\right|^2\left(1+\Delta_R\right)} \tag{2.1-3}$$

where $\delta_R$ and $\Delta_R$ are the nucleus-dependent and nucleus-independent radiative corrections and $\delta_C$ is the charge-dependent correction to $\left|M_{fi}\right|^2$ due to isospin symmetry breaking.

The nucleus independence of the $\mathcal{F}t$ value and its direct correspondence to the $\left|V_{ud}\right|$ matrix element are two important theoretical aspects which make the study of super-allowed nuclear transitions particularly useful for the determination of $\left|V_{ud}\right|$. From an experimental point of view, the study of these reactions is simpler than that of neutron or pion $\beta$ decays and more precise measurements can be obtained. The *ft* value of a super-allowed transition ${}^{A}X_{0^+} \to {}^{A}Y^*_{0^+} + e^+\nu_e$ is determined by measuring the total half-life, the branching fraction (*R*) and the end-point energy of the specific process, which can be clearly identified by measuring the relative yield and the energy of the monochromatic $\gamma$ rays emitted in the $Y^*$ nuclear de-excitations.

The averaged experimental results of nine super-allowed decays, including the recent data for ${}^{10}C$ (particularly relevant as it is close to the *Z*= 0 limit where the effects of the nuclear structure vanish) have been collected and updated by Towner and Hardy[18] (see Table I). The radiative and nuclear structure corrections have been evaluated taking into account the results of several independent calculations[19]. As expected, the corrected *ft*-values are in fact mutually compatible. Their weighted average

$$\overline{\mathcal{F}t} = 3072.3 \pm 0.9 \pm 1.1 = 3072.3 \pm 2.0\ s \tag{2.1-4}$$

(where the second error is due to the systematic difference between two independent computations of $\delta_C$ and has been added linearly to the statistical error) leads, with

$$\Delta_R = (2.40 \pm 0.08) \cdot 10^{-2}, \tag{2.1-5}$$

to the following measurement of $\left|V_{ud}\right|$:

$$\left|V_{ud}\right| = 0.9740 \pm 0.00014_{\ exp} \pm 0.00048_{\ th} = 0.9740 \pm 0.0005 \tag{2.1-6}$$

(the second error represents the uncertainty on the radiative corrections).

In spite of its precision, the value of $\left|V_{ud}\right|$ obtained from nuclear $\beta$ decays has been often criticized due to the inconsistency it presents with the unitarity of the 3×3 CKM matrix. Combining the result 2.1-6 with the best values of $\left|V_{us}\right|$ (2.2-16) and $\left|V_{ub}\right|$ (2.4-4), in fact, a deviation of 1.9 $\sigma$ from unity is found:

$$\left|V_{ud}\right|^2 + \left|V_{us}\right|^2 + \left|V_{ub}\right|^2 = 0.9971 \pm 0.0015\,. \tag{2.1-7}$$

This result may indicate that the value 2.1-6 is underestimated. Various attempts have been made to explain this inconsistency by including additional corrective factors.

On pure phenomenological grounds, Wilkinson[20] admitted a residual dependence of $\mathcal{F}t$ on *Z* and tried to extract it from the experimental data. After extrapolating $\mathcal{F}t$ to $Z = 0$ the value $\left|V_{ud}\right| = 0.9746 \pm 0.0006$ was obtained.

Saito and Thomas[21] proposed a physical model of the infinite nuclear matter, in which the influence of nuclear binding, mediated by meson exchange, is studied in terms of the quark degrees of freedom. A non-negligible charge-symmetry violating effect was found to involve the quarks inside bound nucleons, resulting in an additional correction to the *ft* value, which should raise $|V_{ud}|$ by 0.06 to 0.08%,

$$|V_{ud}| = 0.9747 \pm 0.0006, \tag{2.1-8}$$

lowering the discrepancy with the unitarity condition to 1 σ.
Since none of the model proposed have been clearly established yet, they are not applied directly neither in Ref. 18 nor by the PDG[12]. However, we follow the PDG in enlarging the uncertainty in the determination of $|V_{ud}|$:

$$|V_{ud}|_{nuclear} = 0.974 \pm 0.001. \tag{2.1-9}$$

| | $Z'$ | $Q_{EC}$ (*keV*) | $t_{1/2}$ (*ms*) | $R$ (%) | $P_{EC}$ (%) | *ft* (*s*) | $\mathcal{F}t$ (*s*) |
|---|---|---|---|---|---|---|---|
| $^{10}C$ | 5 | 1907.77(9) | 19290(12) | 1.4645(19) | 0.296 | 3038.7(45) | 3072.9(48) |
| $^{14}O$ | 7 | 2830.51(22) | 70603(18) | 99.336(10) | 0.087 | 3038.1(18) | 3069.7(26) |
| $^{26m}Al$ | 12 | 4232.42(35) | 6344.9(19) | ≥ 99.97 | 0.083 | 3035.8(17) | 3070.0(21) |
| $^{34}Cl$ | 16 | 5491.71(22) | 1525.76(88) | ≥ 99.988 | 0.078 | 3048.4(19) | 3070.1(24) |
| $^{38m}K$ | 18 | 6044.34(12) | 923.95(64) | ≥ 99.998 | 0.082 | 3049.5(21) | 3071.1(27) |
| $^{42}Sc$ | 20 | 6425.58(28) | 680.72(26) | 99.9941(14) | 0.095 | 3045.1(14) | 3077.3(23) |
| $^{46}V$ | 22 | 7050.63(69) | 422.51(11) | 99.9848(13) | 0.096 | 3044.6(18) | 3074.4(27) |
| $^{50}Mn$ | 24 | 7632.39(28) | 283.25(14) | 99.942(3) | 0.100 | 3043.7(16) | 3073.8(27) |
| $^{54}Co$ | 26 | 8242.56(28) | 193.270(63) | 99.9955(6) | 0.104 | 3045.8(11) | 3072.2(27) |
| | | | | | | *Average* $\overline{\mathcal{F}t} = 3072.3(9)$ | |
| | | | | | | $\chi^2/8 = 1.10$ | |

**Table I.** – *Experimental results, computed electron-capture probabilities ($P_{EC}$) and ft-values of nine super-allowed $\beta^+$ decays (see Ref. 18). The values of the total half-lives ($t_{1/2}$), the branching ratios (R) and the electron-capture probabilities computed for $0^+ \to 0^+$ transitions have been used to determine the partial half-life $t = t_{1/2}(1+P_{EC})/R$. f is the value of the phase-space integral corresponding to the atomic number $Z'$ of the residual nucleus and the Q-value of each reaction ($Q_{EC}$ is the Q-value of the electron-capture process, $Q_{EC} \cong Q_{\beta^+} + 2m_e$). The last column lists the ft-values after correction for nucleus-dependent effects.*

### *2.1.2 Neutron decay*

The free neutron mean life ($\tau_n$) and the ratio $g_A/g_V$ between the coupling constants of axial and vector currents (both take part in the decay of neutrons) are in this case the experimental parameters required for the determination of $|V_{ud}|$[18]:

$$\left|V_{ud}\right|^2 = \frac{K/\ln 2}{G_F^2(1+\Delta_R)\left(1+3(g_A/g_V)^2\right)f(1+\delta_R)\tau_n} \qquad (2.1\text{-}10)$$

where $K$ is the typical constant of $\beta$ decays (Eq. 2.1-2), $\Delta_R$ the nucleus-independent radiative correction (2.1-5) and $f(1+\delta_R)$ the phase-space function (which depends on the Q-value) after radiative correction, $f(1+\delta_R)$=1.71489 ± 0.00002[22]. The experimental data concerning neutron $\beta$ decay are shown in Table II.

The results of several measurements, based on two different experimental techniques, are used for the determination of the neutron mean life.

i. The method used in the first experiments consists in counting the $\beta$ decays which occur in a definite portion of a continuous or pulsed neutron beam. The precision of the result depends mainly on the accuracy with which it is possible to determine the spatial dimensions of the observed part of the beam and its density of neutrons. From these quantities the initial population, $N_0$ , can be computed; the mean life is derived from the relation $(N_0 - N_t)/N_0 = e^{-t/\tau}$, where $N_t$ is the number of the neutrons which have decayed in the transit time *t*.
ii. The recent development of techniques for the production and the accumulation of ultra-cold neutrons (having kinetic energy lower than $3\cdot 10^{-7} eV$) has considerably reduced the sources of systematic errors in the measurement of the mean life. Very slow neutrons can be confined within material walls or magnetic traps. By counting the number of neutrons inside the cell as a function of time, $\tau_\mathrm{n}$ can be determined as the only parameter of the exponential decay law. The efficiency of the detector (which does not affect the measured ratio between the populations at different times), the volume of the cell and the initial density do not need to be known and it is not necessary to detect the decay products.

The ratio $g_A/g_V$ between the axial and vector coupling constants can be determined by measuring the parameters describing the correlations between neutron spin and lepton momenta. The decay probability per unit time has the following expression:

$$P(\mathbf{p}_e,\mathbf{p}_{\bar\nu},\boldsymbol{\sigma}) \propto 1 + a\frac{\mathbf{p}_e\cdot\mathbf{p}_{\bar\nu}}{E_e E_{\bar\nu}} + A\frac{\boldsymbol{\sigma}\cdot\mathbf{p}_e}{E_e} + B\frac{\boldsymbol{\sigma}\cdot\mathbf{p}_{\bar\nu}}{E_{\bar\nu}} + D\frac{\boldsymbol{\sigma}\cdot(\mathbf{p}_e\times\mathbf{p}_{\bar\nu})}{E_e E_{\bar\nu}} \qquad (2.1\text{-}11)$$

where $E_e$, $E_{\bar\nu}$, $\mathbf{p}_e$ and $\mathbf{p}_{\bar\nu}$ are the energies and the momenta of the electron and the antineutrino, $\boldsymbol{\sigma}$ the neutron spin).

i. The axial part of weak interaction is responsible for an anisotropic production of the electrons. The number of electrons emitted in the same direction as the nuclear spin ($N^\uparrow$) is smaller than the number of electrons going in the opposite direction ($N^\downarrow$). Consequently, when the neutrons are polarized the term which correlates neutron spin and electron momentum is non-zero and its coefficient

$$A \propto \frac{N^\uparrow - N^\downarrow}{N^\uparrow + N^\downarrow} \qquad (2.1\text{-}12)$$

is measurable; then $\lambda = g_A/g_V$ can be determined from the relation

$$A = -2\frac{\lambda(\lambda+1)}{1+3\lambda^2}. \qquad (2.1\text{-}13)$$

| Experimental techniques | $\tau_n\,(s)$ | References |
|---|---|---|
| *β decay count of beamed neutrons* | $918 \pm 14$ | 1972 Christensen *et al.*[23] |
| | $891 \pm 9$ | 1988 Spivak[24] |
| | $876 \pm 21$ | 1988 Last et *al.*[25] |
| | $878 \pm 30$ | 1989 Kossakowski *et al.*[26] |
| | $889.2 \pm 4.8$ | 1996 Byrne *et al.*[27] |
| *Accumulation of ultra-cold neutrons* | $903 \pm 13$ | 1986 Kosvintsev *et al.*[28] |
| | $877 \pm 10$ | 1989 Paul *et al.*[29] |
| | $887.6 \pm 3.0$ | 1989 Mampe *et al.*[30] |
| | $888.4 \pm 3.3$ | 1992 Nesvizhevskii *et al.*[31] |
| | $882.6 \pm 2.7$ | 1993 Mampe *et al.*[32] |
| Average | $886.7 \pm 1.9$ ($\chi^2/7 = 1.53,\ s = 1.24$) | |

| Measured quantities | $g_A/g_V$ | References |
|---|---|---|
| *Decay asymmetry of free polarized neutrons* | $-1.258 \pm 0.015$ | 1975 Krohn *et al.*[33] |
| | $-1.261 \pm 0.012$ | 1979 Erozolimskii *et al.*[34] |
| | $-1.262 \pm 0.005$ | 1986 Bopp *et al.*[35] |
| | $-1.2594 \pm 0.0038$ | 1997 Yerozolimsky *et al.*[36] |
| | $-1.266 \pm 0.004$ | 1997 Liaud *et al.*[37] |
| | $-1.274 \pm 0.003$ | 1997 Abele *et al.*[38] |
| *Electron-neutrino correlation coefficient (from the analysis of the proton energy spectrum)* | $-1.259 \pm 0.017$ | 1978 Stratowa *et al.*[39] |
| *Rate of muon capture in hydrogen* | $-1.24 \pm 0.04$ | 1984 Bertin *et al.*[40] |
| Average | $-1.2665 \pm 0.0025$ ($\chi^2/6 = 1.88,\ s = 1.37$) | |

**Table II.** – *Neutron β decay: measurement of the parameters necessary for the determination of $|V_{ud}|$ (statistical and systematic errors have been added in quadrature). The results of more than twenty years of experimental investigation are listed. The mutual determinations of $\tau_n$ and $g_A/g_V$ (obtained fixing the value of $\cos\vartheta_C$ or $|V_{ud}|$) have obviously not been taken into account. All the data have been used in the calculation of the averages; however, given the wide range of the error values, the compatibility test ($\chi^2$) has been performed without taking into account the least precise measurements. The degrees of freedom of the measurements whose uncertainties exceed the (arbitrarily defined) threshold value of ten times the unscaled error in the average have not been counted. The $g_A/g_V$ value independently obtained from the measured rate of nuclear muon capture by a proton is also quoted for comparison.*

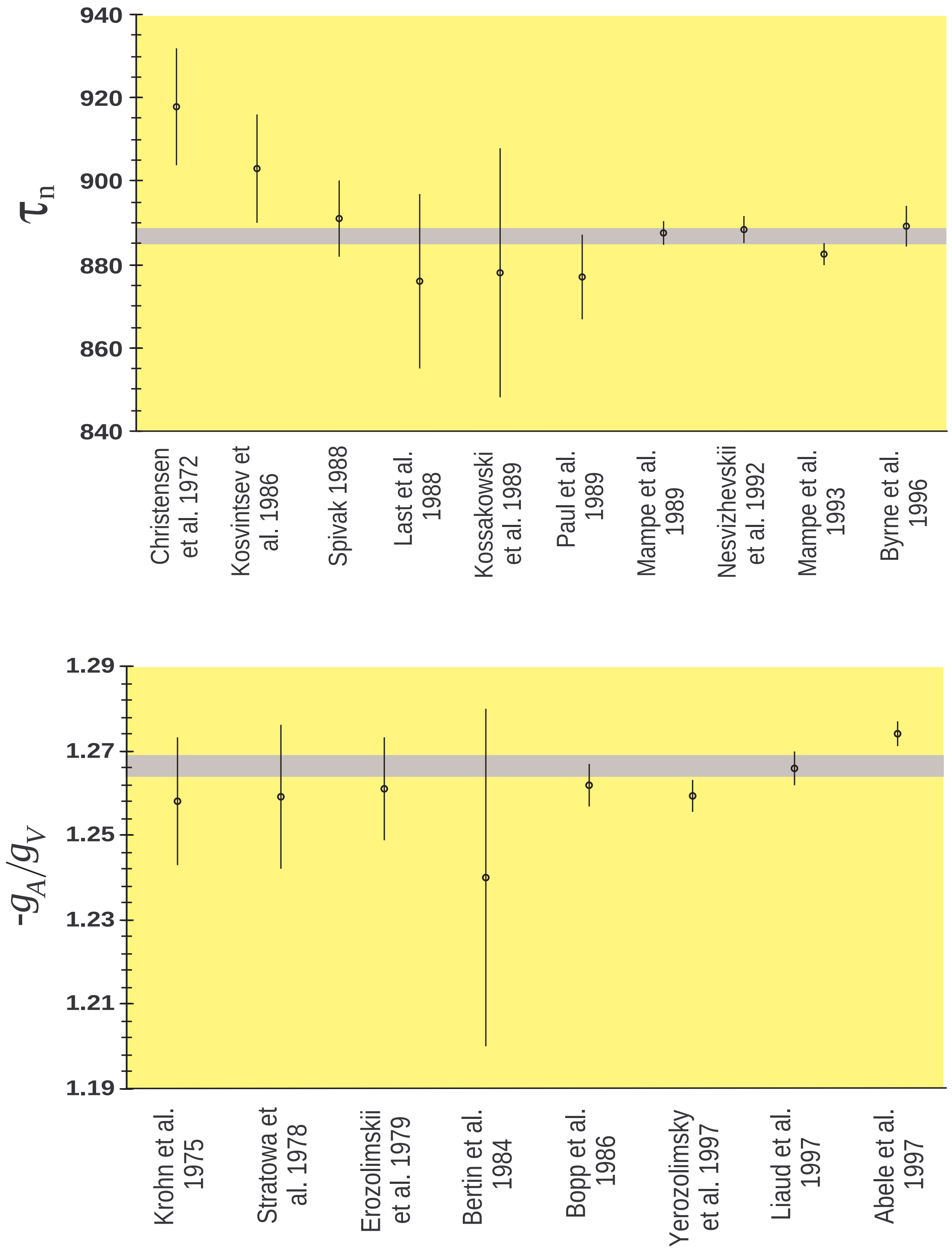

**Figure 2.** – *Neutron mean life and* $g_A/g_V$ *measurements used in the determination of* $|V_{ud}|$. *The dark bands represent the average values (with rescaled errors, see Table II).*

ii. In the experiments using non-polarized neutrons the coefficient

$$a = -\frac{\lambda^2 - 1}{1 + 3\lambda^2} \qquad (2.1\text{-}14)$$

of the electron-neutrino angular correlation is measured, since it is the only term of the decay probability which does not depend on neutron spin and therefore does not vanish on the average. The parameter $a$ is extracted by comparing the measured energy distribution of the outgoing protons to the spectrum shape calculated as a function of $a$.

iii. A determination of $g_A/g_V$ was deduced by the experimental data regarding muon capture in hydrogen[40] — a process which, according to the hypothesis of lepton universality, is analogous to the reversed neutron $\beta$ decay.

To compute the average values of $\tau_n$ and $g_A/g_V$, scale factors have been applied to the errors because of the disagreement between independent determinations. The dispersion of the measurements is shown in Figure 2.
Substituting the best values of $\tau_n$ and $g_A/g_V$ in Eq. (2.1-10), the result

$$\left|V_{ud}\right|_{neutron} = 0.9755 \pm 0.0016_{g_A/g_V} \pm 0.0010_{\tau_n} \pm 0.0004_{\Delta_R} = 0.9755 \pm 0.0019 \qquad (2.1\text{-}15)$$

is obtained. The experimental uncertainty, which is the most substantial contribution in the total error, exceeds the one of the measurement (2.1-6) deduced from nuclear decays (which is mainly theoretical in origin) by more than one order of magnitude. This determination is compatible with the nuclear result (2.1-9) and with unitarity (1-2-1):

$$\left|V_{ud}\right|^2 + \left|V_{us}\right|^2 + \left|V_{ub}\right|^2 = 1.0000 \pm 0.0039\,. \qquad (2.1\text{-}16)$$

It should however be noticed that single measurements of $g_A/g_V$ would lead to different determinations of $\left|V_{ud}\right|$ which in some cases are incompatible with each other, with the nuclear result and with unitarity. For example, when the measurement by Abele *at al.*[38] is used alone (not averaged), it leads to the value $\left|V_{ud}\right| = 0.9712 \pm 0.0020$, which is even lower than the result given by nuclear decays (2.1-6) and deviate from unitarity by 2.1 standard deviations ( $\sum_{i=d,s,b}\left|V_{ui}\right|^2 = 0.9916 \pm 0.0040$). On the contrary, if the average of the remaining seven measurements is considered ($g_A/g_V = -1.2621 \pm 0.0023$ with $\chi^2/6 = 0.26$), the resulting value $\left|V_{ud}\right| = 0.9788 \pm 0.0016$ actually exceeds unitarity by 2σ: $\sum_{i=d,s,b}\left|V_{ui}\right|^2 = 1.0065 \pm 0.0033$. Most likely, therefore, the determination 2.1-15 is the result of a number of accidental compensations between opposite systematic effects.

### *2.1.3 Pion β decay*

With respect to the already described methods for determining $\left|V_{ud}\right|$, the advantages of studying pion $\beta$ decay are twofold: it is a pure vector transition and, obviously, it does not depend on effects due to nuclear structure. Unfortunately, the small ($\approx 10^{-8}$) branching ratio of the channel $\pi^+ \to \pi^0 e^+ \nu_e$ has not yet been measured precisely enough to provide a competitive determination of $|V_{ud}|$. In the relation[18]

$$|V_{ud}|^2 = \frac{(K/\ln 2)\mathcal{Br}(\pi^+ \to \pi^0 e^+ \nu_e)}{2G_F^2(1+\Delta_R) f_1 f_2 f (1+\delta_R)\tau_\pi} \tag{2.1-17}$$

$K$, $G_F$ and $\Delta_R$ are given by Eqs. (2.1-2), (2-1) and (2.1-5) respectively, $\delta_R = (1.05 \pm 0.15)\%$, $f = [(m_{\pi^\pm} - m_{\pi^0})/m_e]^5/30$ where[12] $m_{\pi^\pm} - m_{\pi^0} = (4.5936 \pm 0.0005)$ $MeV$ and the corrections $f_1$ and $f_2$, calculated as functions of $(m_{\pi^\pm} - m_{\pi^0})/m_{\pi^\pm}$, are[41] $f_1 = 0.941039$ and $f_2 = 0.951439$. Using the values of the branching ratio of the reaction and the pion mean life,

$$\mathcal{Br}(\pi^+ \to \pi^0 e^+ \nu_e) = (1.025 \pm 0.034)\cdot 10^{-8} \quad [41, 12] \tag{2.1-18}$$

$$\tau_\pi = (2.6033 \pm 0.0005)^{-8} s \quad [42,12], \tag{2.1-19}$$

from Eq. (2.1-17) the result

$$|V_{ud}|_{pion} = 0.9670 \pm 0.0160_{\mathcal{Br}} \pm 0.0009 = 0.967 \pm 0.016, \tag{2.1-20}$$

is obtained, which is consistent with the other measurements of $|V_{ud}|$ and with unitarity:

$$\sum_{i=d,s,b} |V_{ui}|^2 = 0.984 \pm 0.031. \tag{2.1-21}$$

The experimental uncertainty should be reduced by a factor of 10 before it becomes comparable to the theoretical error. A new, more precise measurement of the decay rate, therefore, would be essential. It would make it possible to get around the controversial aspects of the theoretical description of nuclear decays, and provide an unambiguous determination of $|V_{ud}|$ without requiring the knowledge of additional experimental parameters such as the axial coupling constant in the case of neutron decay.

### *2.1.4 Summary of $|V_{ud}|$ determinations*

The best value of $|V_{ud}|$ is calculated as the weighted average of the independent determinations 2.1-9, 2.1-15 and 2.1-21:

$$|V_{ud}|_{nuclear} = 0.9740 \pm 0.0010. \tag{2.1-22}$$

$$|V_{ud}|_{neutron} = 0.9755 \pm 0.0019. \tag{2.1-23}$$

$$|V_{ud}|_{pion} = 0.967 \pm 0.016. \tag{2.1-24}$$

$$\textit{Average} \quad |V_{ud}| = 0.9743 \pm 0.0008. \tag{2.1-25}$$

## 2.2 $|V_{us}|$

A precise determination of $|V_{us}|$ can be obtained from the analysis of kaon and hyperon semileptonic decays. In this Section, we will describe the results and point out the main theoretical uncertainties connected with the two methods.

### 2.2.1 Kaon semileptonic decays

The $K^+ \to \pi^0 \ell^+ \nu_\ell$ and $K_L^0 \to \pi^- \ell^+ \nu_\ell$ processes, called respectively $K_{\ell 3}^+$ and $K_{\ell 3}^0$, are pure vector $\bar{s} \to \bar{u} \ell^+ \nu_\ell$ transitions. From the observed decay rates $\Gamma$ the value of $|V_{us}|$ can be extracted according to the relation[43]

$$\Gamma = \frac{G_F^2 |V_{us}|^2}{192\pi^3} m_K^5 C^2 |f_1(0)|^2 I (1+\delta)(1+\Delta). \; (^*) \qquad (2.2\text{-}1)$$

Here $G_F$ is the Fermi constant, $I$ the phase-space integral, $m_K$ the kaon mass, $C$ a normalization constant ($C^2$=1/2 and $C^2$=1 for the $K_{\ell 3}^+$ and $K_{\ell 3}^0$ transitions respectively), $\delta$ (channel-dependent) and $\Delta$ the radiative corrections[44, 45]

$$\Delta = (2.12 \pm 0.08)\% \qquad \delta = \begin{cases} -2.0\% \text{ for } K_{e3}^+ \\ \;\;0.5\% \text{ for } K_{e3}^0 \end{cases}. \qquad (2.2\text{-}2)$$

The form factor $f_1(t)$ appearing in Eq. (2.2-1) and the analogous $f_2(t)$ account for the dependence of the matrix element of the vector current on the square momentum transfer $t = (p_K - p_\pi)^2$:

$$\langle K(p_K) | \bar{u}\gamma^\mu s | \pi(p_\pi) \rangle = C\{ (p_K^\mu + p_\pi^\mu) f_1(t) + (p_K^\mu - p_\pi^\mu) f_2(t) \}. \qquad (2.2\text{-}3)$$

To compute the integral $I$, the $t$-dependence of both form factors has to be known:

$$I = \frac{1}{m_K^8} \int_{m_\ell^2}^{(m_K - m_\pi)^2} dt\, \lambda^{3/2} \left(1 + \frac{m_\ell^2}{2t}\right)\left(1 - \frac{m_\ell^2}{t}\right)^2 \left\{ \bar{f}_1(t)^2 + \frac{3m_\ell^2 (m_K^2 - m_\pi^2)^2}{(2t + m_\ell^2)\cdot \lambda} \bar{f}(t)^2 \right\} \qquad (2.2\text{-}4)$$

where

$$\lambda \equiv (m_K^2 + m_\pi^2 - t)^2 - 4 m_K^2 m_\pi^2, \qquad f(t) \equiv f_1(t) + t\, f_2(t)/(m_K^2 - m_\pi^2),$$
$$\bar{f}_1(t) \equiv f_1(t)/f_1(0), \qquad \bar{f}(t) \equiv f(t)/f(0)$$

and $m_K$, $m_\pi$, $m_\ell$ ($\ell = e, \mu$) are the masses of the particles involved in the process. One can determine the reduced form factor $\bar{f}_1(t)$ from the experimental data, which can be fitted with an $\bar{f}_1(t)$ depending linearly on $t$. On the contrary, the linear form does not give a good description of the data in the case of $\bar{f}(t)$. However, since $\bar{f}(t)$ is multiplied by the square lepton mass in the expression of $I$ (2.2-4), its contribution can be made negligible by considering only the $K_{e3}$ decay yields ($K^+ \to \pi^0 e^+ \nu_e$ and $K_L^0 \to \pi^- e^+ \nu_e$). In this way, the integrals can be computed as[43]

$$I_{K_{e3}^+} = 0.1605 \pm 0.0009, \qquad I_{K_{e3}^0} = 0.1561 \pm 0.0008. \qquad (2.2\text{-}5)$$

---

$^*$ This formula is valid if the small (3‰) CP violating effect which differentiates the rates of $K_L^0 \to \pi^- e^+ \nu_e$ and $K_L^0 \to \pi^+ e^- \bar{\nu}_e$ decays is neglected. This is certainly an allowed approximation given the current precision (≈1%) of the experimental data and the presence of larger theoretical uncertainties.

Substituting these values and the latest average determination of the rates for the $K_{e3}^+$ and $K_{e3}^0$ processes[12]

$$\Gamma_{K_{e3}^+} = (2.560 \pm 0.033) 10^{-15} MeV \qquad \Gamma_{K_{e3}^0} = (4.937 \pm 0.053) 10^{-15} MeV \qquad (2.2\text{-}6)$$

into Eq. (2.2-1), one obtains

$$f_1^{K^+\pi^0}(0)|V_{us}| = 0.2181 \pm 0.0015 \pm 0.0001 \qquad f_1^{K^0\pi^-}(0)|V_{us}| = 0.2101 \pm 0.0013 \pm 0.0001 \qquad (2.2\text{-}7)$$

where the theoretical uncertainty of the radiative corrections (a relative error of the same order as that on $\Delta$ has been assumed for $\delta$) is quoted separately to show that it is negligible. The two results can be combined after taking into account the effects of isospin symmetry breaking: with[43]

$$f_1^{K^+\pi^0}(0) / f_1^{K^0\pi^-}(0) = 1.022 \pm 0.002 \qquad (2.2\text{-}8)$$

the weighted average

$$f_1^{K^0\pi^-}(0)|V_{us}| = 0.2114 \pm 0.0016 \qquad (2.2\text{-}9)$$

is obtained (*).

Two calculations were performed to determine the form factor of the $K_{e3}^+$ decay at zero momentum transfer, giving the following results:

$$f_1^{K^0\pi^-}(0) = 0.961 \pm 0.008^{43}, \qquad (2.2\text{-}10)$$

$$f_1^{K^0\pi^-}(0) = 0.963 \pm 0.004^{46}. \qquad (2.2\text{-}11)$$

The reliability of the error estimate in Eq. (2.2-10) was questioned by the authors of Ref. 47 due to the uncertainties which are inherent in all chiral perturbation theory computations. However, a perfectly consistent result was provided by the second, independent calculation (2.2-11), performed in the framework of a relativistic constituent quark model which proved to be successful in the description of the electro-weak properties of light mesons.

Using the result 2.2-10, from Eq. (2.2-9) one obtains

$$|V_{us}| = 0.2200 \pm 0.0017_{exp} \pm 0.0018_{th} = 0.2200 \pm 0.0025. \qquad (2.2\text{-}12)$$

---

* The two determinations of $f_1^{K^0\pi^-}(0)|V_{us}|$, though consistent at the 11% confidence level ($\chi^2/1 = 2.6$), have been averaged, as suggested by the PDG[12], after multiplying the errors by the scale-factor $s = \sqrt{\chi^2/1} = 1.6$. This procedure is followed also in those cases where other authors prefer not to average the results at all. Incidentally, better agreeing results would be obtained for $f_1^{K^0\pi^-}(0)|V_{us}|$ if the channel-dependent radiative corrections in Eq. 2.2-2 would *not* be applied.

## 2.2.2 Hyperon semileptonic decays

The deduction of $|V_{us}|$ from hyperon semileptonic decay rates requires a more complex theoretical analysis. While the kaon decays are described by the vector interaction alone and thus the SU(3) symmetry-breaking occurs at the second order only (Ademollo-Gatto's theorem), an accurate study of hyperon decays requires the additional computation of the more substantial (first-order) correction for the axial form factor. Donoghue, Holstein and Klimt[48] (DHK) computed SU(3)-breaking corrections up to the second order for the vector part and to the first order for the axial one, and applied them to the WA2 data (SPS) combined with the previous world averages (ten distinct hyperon decay channels in all)[49]. Finally, they quoted the result

$$|V_{us}| = 0.220 \pm 0.001_{exp} \pm 0.003_{th}, \qquad (2.2\text{-}13)$$

where the error is evidently dominated by the theoretical uncertainty. The PDG[12] do not use the hyperon decay result to determine the best value of $|V_{us}|$. Flores-Mendieta *et al.*[50] tested the reliability of four distinct theoretical models which provide second order SU(3)-breaking corrections for hyperon semileptonic transitions and pointed out a strong model-dependence of the results. In particular, using the DHK model they found a value of $|V_{us}|$ noticeably higher than the one originally determined by the authors themselves (2.2-13), and much less consistent with the kaon decay data. This disagreement may be explained as the result of a different treatment of the experimental data. For example, in Ref. 50 only the three most precisely measured decay channels instead of the original ten were taken into account. This reduction in the experimental basis increases considerably the relative statistical weight of the $\Xi^- \to \Lambda e^- \bar{\nu}_e$ data, which is the one showing by far the worst agreement with the DHK model, as already found by the authors themselves. Moreover, the weighted average of two disagreeing results

$$\Gamma(\Xi^- \to \Lambda e^- \bar{\nu}_e) = (1.83 \pm 0.79)\cdot 10^6 s^{-1} \ [49]$$
$$\Gamma(\Xi^- \to \Lambda e^- \bar{\nu}_e) = (3.44 \pm 0.19)\cdot 10^6 s^{-1} \ [51] \qquad (2.2\text{-}14)$$
$$\text{average} \quad \Gamma(\Xi^- \to \Lambda e^- \bar{\nu}_e) = (3.36 \pm 0.19)\cdot 10^6 s^{-1} \quad \text{with } \chi^2 = 4$$

(the first of which is consistent with the DHK fitted value of $2.65\cdot 10^6\ s^{-1}$) is used for the $\Xi^- \to \Lambda e^- \bar{\nu}_e$ decay width in both the analyses, but only DHK apply a scale factor of 2 to the errors to make the two measurements compatible. The authors of the second analysis[50] quote the value obtained using the chiral perturbation model of Ref. 52 as the best determination of $|V_{us}|$ from hyperon decays:

$$|V_{us}| = 0.2176 \pm 0.0026. \qquad (2.2\text{-}15)$$

However, they also show that values ranging from 0.21 to 0.25 can be obtained by applying different models.

Since apparently unresolved theoretical uncertainties still affect the description of hyperon decays, we prefer not to use these results. Therefore, we assume the determination obtained from kaon decays

$$|V_{us}| = 0.2200 \pm 0.0025 \qquad (2.2\text{-}16)$$

as the best value, noting however that it is in good agreement with the results of two different analyses of hyperon semileptonic decays.

## 2.3 $|V_{cd}|$ and $|V_{cs}|$

A measurement of the matrix elements $|V_{cd}|$ and $|V_{cs}|$ can be obtained from the study of the deep inelastic scattering of neutrinos on nucleons. For the determination of $|V_{cs}|$ the experimental data concerning the $D \rightarrow \overline{K}\, e^+ \nu_e$ semileptonic transitions and the recent measurements of the $W^\pm$ hadronic decays are also used.

### 2.3.1 $|V_{cd}|$

$|V_{cd}|$ and $|V_{cs}|$ can be extracted by comparing the cross sections of the charm production induced by neutrino and anti-neutrino scattering on nucleons.

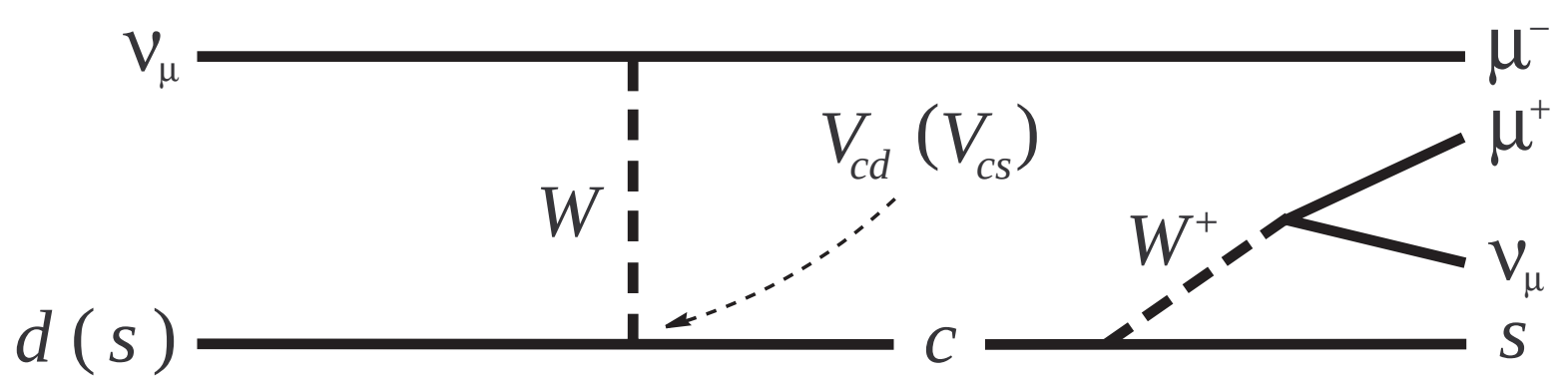

**Figure 3.** – *Neutrino-induced dimuon production.*

The dimuon production from neutrino-nucleon scattering can be described as a two-stage process (see Figure 3). The charge-current interaction of the neutrino with a *d* or *s* quark yields the first muon and changes the quark flavour into charm. The second muon, having the opposite charge, is the result of the semileptonic decay of the *c* quark:

$$\begin{aligned} &\nu_\mu + d\,(s) \rightarrow c + \mu^- \quad \left(c \rightarrow s + \mu^+ + \nu_\mu\right) \\ &\overline{\nu}_\mu + \overline{d}\,(\overline{s}) \rightarrow \overline{c} + \mu^+ \quad \left(\overline{c} \rightarrow \overline{s} + \mu^- + \overline{\nu}_\mu\right) \end{aligned} \qquad (2.3\text{-}1)$$

The cross-sections of the dimuon processes were measured by the collaborations CDHS[53], CCFR[54] and CHARM II[55]. The cross-section terms containing the dependence on $|V_{cd}|$ and $|V_{cs}|$ were extracted by fitting the neutrino and anti-neutrino data simultaneously:

$$\text{CDHS} \qquad |V_{cd}|^2 B_c = (4.1 \pm 0.7) \cdot 10^{-3} \qquad (2.3\text{-}2)$$

$$\text{CCFR} \qquad |V_{cd}|^2 B_c = \left(5.34^{+0.38}_{-0.39\,stat}\;{}^{+0.37}_{-0.55\,syst}\right) \cdot 10^{-3} \qquad (2.3\text{-}3)$$

$$\text{CHARM II} \qquad |V_{cd}|^2 B_c = \left(4.42^{+0.35}_{-0.34\,stat} \pm 0.34_{syst}\right) \cdot 10^{-3} \qquad (2.3\text{-}4)$$

where $B_c$ is the weighted average of the semileptonic branching ratios of the charmed hadrons produced by the neutrino-nucleon interaction. The analysis performed by CHDS made use of leading-order QCD calculations, while the CCFR and CHARM II results were obtained in the context of the more accurate next-to-leading-order formalism.

Combining Eqs. (2.3-2), (2.3-3) and (2.3-4) (*), the following weighted average is obtained:

$$|V_{cd}|^2 B_c = (4.63 \pm 0.34)\cdot 10^{-3} . \qquad (2.3\text{-}5)$$

The value of $B_c$ was calculated[56] using the independently measured rates of charmed hadron production from neutrinos[57] and the semileptonic branching ratios[58] of charmed hadrons:

$$B_c = 0.0919 \pm 0.0094. \qquad (2.3\text{-}6)$$

Thus from Eq. (2.3-5) the value

$$|V_{cd}| = 0.224 \pm 0.014 \qquad (2.3\text{-}7)$$

is obtained.

### 2.3.2 $|V_{cs}|$ from neutrino charm production

From dimuon cross-section measurements, the above-mentioned CDHS, CCFR and CHARM II experiments obtained also the following results:

CDHS $$\left|\frac{V_{cs}}{V_{cd}}\right|^2 \kappa = 9.3 \pm 1.6 \qquad (2.3\text{-}8)$$

CCFR $$\frac{\kappa}{\kappa+2}|V_{cs}|^2 B_c = \left(2.00 \pm 0.10_{stat}\,{}^{+0.09}_{-0.15}{}_{syst}\right)\cdot 10^{-2} \qquad (2.3\text{-}9)$$

CHARM II $$\frac{1}{\kappa+2}\left(1+\left|\frac{V_{cs}}{V_{cd}}\right|^2 \kappa\right) = 3.58\,{}^{+0.49}_{-0.41}{}_{stat} \pm 0.44_{syst} \qquad (2.3\text{-}10)$$

Here the parameter $\kappa$, which quantifies the relative size of the strange quark sea, is defined as the ratio of the integrated distribution function of strange quarks to that of $\bar{u}$ and $\bar{d}$ quarks:

$$\kappa = \int_0^1 [x\,s(x) + x\,\bar{s}(x)]dx \Big/ \int_0^1 [x\bar{u}(x) + x\bar{d}(x)]dx \qquad (2.3\text{-}11)$$

where $s(x)$, $\bar{s}(x)$, $\bar{u}(x)$ and $\bar{d}(x)$ are the quark density distributions in momentum space and $x$ is the fraction of nucleon momentum carried by the quark. All three analyses assumed the equality $\bar{s}(x) = s(x)$. Although this assumption can be contradicted[59,60], CCFR verified that the removal of the $\bar{s}(x) = s(x)$ constraint does not lead to a significant modification of the results.

A new analysis by CCFR[61], which has determined the value of $\kappa$ with an independent measurement of the inclusive neutrino cross-section, makes it possible to extract a measurement of $|V_{cs}|$ from these data. The result,

$$\kappa = 0.453 \pm 0.106^{+0.028}_{-0.096} , \qquad (2.3\text{-}12)$$

indicates a marked SU(3)-flavour-symmetry violation (an SU(3)-symmetric sea would have $\kappa = 1$). In order to compare the CDHS, CCFR and CHARM II results for $|V_{cs}|$, we

* For the estimate of $|V_{cd}|$ (and of $|V_{cs}|$ in the next Subsection), we symmetrize the errors in the CCFR result, taking the average of the positive and negative errors (statistical and systematic added in quadrature) while keeping the same central value.

multiply Eqs. (2.3-9) and (2.3-10) by $\kappa + 2 = 2.453 \pm 0.126$, Eq. (2.3-8) and the determination of $\left|V_{cs}/V_{cd}\right|^2 \kappa$ resulting from Eq. (2.3-10) by Eq. (2.3-5), obtaining:

| | | |
|---|---|---|
| CDHS | $\kappa\left\|V_{cs}\right\|^2 B_c = (4.31 \pm 0.81)\cdot 10^{-2}$ | (2.3-13) |
| CCFR | $\kappa\left\|V_{cs}\right\|^2 B_c = (4.91 \pm 0.46)\cdot 10^{-2}$ | (2.3-14) |
| CHARM II | $\kappa\left\|V_{cs}\right\|^2 B_c = (3.60 \pm 0.79)\cdot 10^{-2}$ | (2.3-15) |
| *Average* | $\kappa\left\|V_{cs}\right\|^2 B_c = (4.53 \pm 0.37)\cdot 10^{-2}$ | (2.3-16) |

From Eqs. (2.3-6), (2.3-12) and (2.3-16) $\left|V_{cs}\right|$ is determined as follows:

$$\left|V_{cs}\right| = 1.04 \pm 0.16. \qquad (2.3\text{-}17)$$

## *2.3.3 |$V_{cs}$| from semileptonic decays of D mesons*

$\left|V_{cs}\right|$ can also be obtained by comparing the measured decay amplitude of the semileptonic processes $D_{e3}$ ($D \to \overline{K}\, e^+ \nu_e$) to the corresponding expression in the theory of weak interaction. The method is very similar to the one used for the derivation of $\left|V_{us}\right|$ from $K_{e3}$ decays. The decay rate is described by the relation

$$\Gamma\left(D \to \overline{K} e^+ \nu_e\right) = \frac{G_F^2 \left|V_{cs}\right|^2}{192\pi^3} m_D^5 \left|f_1(0)\right|^2 I\ (1+\Delta) \qquad (2.3\text{-}18)$$

where the meanings of the single quantities are the same as in Eq. (2.2-1). The phase-space density *I* is obtained by calculation of the integral 2.2-4 with $m_K$ and $m_\pi$ replaced respectively by $m_D$ and $m_K$; in this case, the observed square-momentum spectrum can be fitted[62] using the single-pole parametrization of the reduced form factor, $\bar{f}_1(t) = f_1(t)/f_1(0) = m_{pole}^2 / \left(m_{pole}^2 - t\right)$, with $m_{pole} \cong m_{D_s^*} \cong 2.1 GeV$. Assuming this form exactly and neglecting the term proportional to $m_e^2\, \bar{f}(t)^2$ in the integral, we get $I = 0.195$. Although isospin symmetry requires equal transition rates for $D^0 \to K^- e^+ \nu_e$ and $D^+ \to \overline{K}^0 e^+ \nu_e$ decays, the measured values differ considerably outside their errors. In averaging the available experimental results for the $D_{e3}$ branching ratios and *D* lifetimes[12], we rescale the error by the factor $s = \sqrt{\chi^2} = 2.6$, obtaining:

$$\Gamma\left(D \to \overline{K} e^+ \nu_e\right) = (8.3 \pm 1.0)\cdot 10^{10}\, s^{-1}. \qquad (2.3\text{-}19)$$

After substituting into Eq. (2.3-18), we get

$$\left|f_1(0)\right|^2 \left|V_{cs}\right|^2 = 0.53 \pm 0.06. \qquad (2.3\text{-}20)$$

Most of the theoretical evaluations for the form factor at zero-recoil (see Ref. 62 for a review) give results included between 0.6 and 0.8. Assuming $f_1(0) = 0.7 \pm 0.1$, we obtain

$$\left|V_{cs}\right| = 1.04 \pm 0.16, \qquad (2.3\text{-}21)$$

in full agreement with Eq. (2.3-17).

## 2.3.4 $|V_{cs}|$ from $W^\pm$ boson decays

New measurements have been obtained from the recent study of $W^\pm$ boson decays at LEP2. Their precision is already higher than that of the previous independent results; moreover, the errors are dominated by the experimental uncertainty and are bound to be further reduced. DELPHI[63], ALEPH[64], L3[65] and OPAL[66] extracted the value of $|V_{cs}|$ using two independent procedures. The ratio

$$R_c^W = \frac{\Gamma\left(W^+ \to c\bar{q}\right)_{q=d,s,b}}{\Gamma\left(W^+ \to hadrons\right)} = \frac{|V_{cd}|^2 + |V_{cs}|^2 + |V_{cb}|^2}{|V_{ud}|^2 + |V_{us}|^2 + |V_{ub}|^2 + |V_{cd}|^2 + |V_{cs}|^2 + |V_{cb}|^2} \qquad (2.3\text{-}22)$$

(in the second equality, corrections of order $m^2_{q=u,d,c,s,b}/m_W^2$ are neglected; the transitions with a top quark in the final state are energetically forbidden) was determined by tagging the flavours of the hadronic jets which arise from the fragmentation of the primary quarks coming from $W^\pm$ decays. The results obtained,

$$R_c^W = \begin{cases} 0.46^{+0.18}_{-0.14\,stat} \pm 0.07_{syst} & \text{DELPHI} \\ 0.51 \pm 0.05_{stat} \pm 0.03_{syst} & \text{ALEPH} \\ 0.50 \pm 0.11_{stat} \pm 0.04_{syst} & \text{L3} \\ 0.47 \pm 0.04_{stat} \pm 0.06_{syst} & \text{OPAL} \end{cases} \qquad (2.3\text{-}23)$$

are consistent with the value ½ expected assuming the unitarity of the CKM matrix. The ratio of hadronic $W^\pm$ decays to leptonic $W^\pm$ decays,

$$\frac{\mathcal{Br}\left(W^+ \to hadrons\right)}{\mathcal{Br}\left(W^+ \to leptons\right)} = \frac{\mathcal{Br}\left(W^+ \to hadrons\right)}{1-\mathcal{Br}\left(W^+ \to hadrons\right)} = \left(|V_{ud}|^2 + |V_{us}|^2 + |V_{ub}|^2 + |V_{cd}|^2 + |V_{cs}|^2 + |V_{cb}|^2\right)\cdot\left(1+\frac{\alpha_S(m_W)}{\pi}\right), \qquad (2.3\text{-}24)$$

where $\alpha_S$ is the strong coupling constant and once again the quark phase-space corrections have been neglected, was also measured. The results

$$\mathcal{Br}\left(W^+ \to hadrons\right) = \begin{cases} \left(66.0^{+3.6}_{-3.7\,stat} \pm 0.9_{syst}\right)\% & \text{DELPHI} \\ \left(66.89 \pm 0.67_{stat} \pm 0.31_{syst}\right)\% & \text{ALEPH} \\ \left(68.69 \pm 0.68_{stat} \pm 0.39_{syst}\right)\% & \text{L3} \\ \left(68.34 \pm 0.61_{stat} \pm 0.31_{syst}\right)\% & \text{OPAL} \end{cases} \qquad (2.3\text{-}25)$$

correspond (using $\alpha_S(m_W) = 0.120 \pm 0.005$ [12]) to the determinations

$$\sum_{\substack{i=u,c \\ j=d,s,b}} |V_{ij}|^2 = \begin{cases} 1.87 \pm 0.31_{stat+syst} & \text{DELPHI} \\ 1.946 \pm 0.065_{stat+syst} & \text{ALEPH} \\ 2.113 \pm 0.077_{stat+syst} & \text{L3} \\ 2.079 \pm 0.066_{stat+syst} & \text{OPAL} \end{cases} \qquad (2.3\text{-}26)$$

which have to be compared with the Standard Model expectation of 2.

From the measurements of the observables 2.3-22 and 2.3-24 each experiment obtained two independent determinations of $|V_{cs}|$ (see Table III). The values compiled by the PDG[67,12] were assigned to the other CKM matrix elements. However, their uncertainties only contribute to a small fraction of the $|V_{cs}|$ systematic errors (for example, ±0.003 in DELPHI's measurement, to be compared with ±0.05), so that the correlation of these results with the other measurements of the CKM matrix elements is negligible.

| | $\vert V_{cs}\vert$ from $\mathcal{Br}(W \to c\bar{q})$ | $\vert V_{cs}\vert$ from $\mathcal{Br}(W \to hadrons)$ |
|---|---|---|
| DELPHI[63] | $0.94^{+0.32}_{-0.26}\,{}_{stat} \pm 0.13_{syst}$ | $0.90 \pm 0.17_{stat} \pm 0.04_{syst}$ |
| ALEPH[64] | $1.00 \pm 0.11_{stat} \pm 0.07_{syst}$ | $0.947 \pm 0.031_{stat} \pm 0.015_{syst}$ |
| L3[65] | $0.98 \pm 0.22_{stat} \pm 0.08_{syst}$ | $1.032 \pm 0.033_{stat} \pm 0.018_{syst}$ |
| OPAL[66] | $0.91 \pm 0.07_{stat} \pm 0.11_{syst}$ | $1.015 \pm 0.029_{stat} \pm 0.015_{syst}$ |

**Table III.** – *Measurements of $\vert V_{cs}\vert$ obtained using W decays at LEP2. In most cases, the results are still preliminary.*

The average determinations (with statistical and systematic errors added in quadrature) are

$$\text{DELPHI} \qquad \vert V_{cs}\vert = 0.91^{+0.16}_{-0.15\,stat+syst}\,, \tag{2.3-27}$$

$$\text{ALEPH} \qquad \vert V_{cs}\vert = 0.950 \pm 0.033_{stat+syst}\,, \tag{2.3-28}$$

$$\text{L3} \qquad \vert V_{cs}\vert = 1.031 \pm 0.037_{stat+syst}\,, \tag{2.3-29}$$

$$\text{OPAL} \qquad \vert V_{cs}\vert = 1.009 \pm 0.032_{stat+syst}\,. \tag{2.3-30}$$

Finally, we average them allowing for a common systematic error of $\pm 0.015$ and obtain

$$\vert V_{cs}\vert = 0.993 \pm 0.025 \qquad (\chi^2 = 1.4), \tag{2.3-31}$$

where the scale factor s= 1.2 has been applied to (the uncorrelated part of) the errors.

### *2.3.5 Determination of the best values of $\vert V_{cd}\vert$ and $\vert V_{cs}\vert$*

The best values of $\vert V_{cd}\vert$ and $\vert V_{cs}\vert$ are obtained as the result of an overall fit to the measurements mentioned in the preceding Subsections. 8 constraints (Eqs. 2.3-2,3,4,6,8,9,10,12) and 2 direct determinations (Eqs. 2.3-21,31) are included in the fit (*). The maximum likelihood estimates of the 4 parameters of the fit, $\vert V_{cd}\vert$, $\vert V_{cs}\vert$, $B_c$ and $\kappa$, are

$$\vert V_{cd}\vert = 0.225^{+0.013}_{-0.011} \tag{2.3-32}$$

$$\vert V_{cs}\vert = 0.996 \pm 0.024 \tag{2.3-33}$$

$$B_c = 0.0935^{+0.0086}_{-0.0080} \qquad \kappa = 0.489^{+0.067}_{-0.059} \tag{2.3-34}$$

with $\chi^2/_{d.o.f.} = \chi^2/6 = 0.7$ and a 10.5% correlation between $\vert V_{cd}\vert$ and $\vert V_{cs}\vert$.

These results can be used to test the unitarity of the CKM matrix with reference to the elements of the second row: using Eqs. (2.3-32), (2.3-33) and (2.5-10) ($\vert V_{cb}\vert$) we obtain

$$\vert V_{cd}\vert^2 + \vert V_{cs}\vert^2 + \vert V_{cb}\vert^2 = 1.044 \pm 0.048\,. \tag{2.3-35}$$

* The following method is used to take the asymmetric errors into account. Each error is treated as a function of the parameters, which is constant and equal to the positive or negative error when the current value of the unknown term of the constraint falls outside the positive or the negative margin; in the intermediate region, the function is matched continuously by a straight line.

## 2.4 $|V_{ub}|$

CP violation can be included in the framework of the Standard Model only if all elements of the CKM quark-mixing matrix have non-zero values. This fact justified the strong experimental and theoretical effort devoted to the determination of $|V_{ub}|$, which is expected, on the basis of the unitarity condition, to be very small.
The observation of an excess of events in the lepton end-point spectra, interpreted as the result of charmless semileptonic decays of $B$ mesons, was made by CLEO[68] at CESR and ARGUS[69] at DESY in 1990 and gave the first experimental evidence[70] of a non-zero $|V_{ub}|$. The subsequent experiments were faced with the hard task of a precise determination of its value. The main difficulty in measuring signals from $b \to u\ell^- \bar{\nu}_\ell$ processes arises from the large background induced by the stronger $b \to c\ell^- \bar{\nu}_\ell$ transitions. The contribution of the charmed $B$ decays can be suppressed by looking for exclusive decays were charmless final states can be directly observed by means of invariant-mass peaks. Such measurements, though easier from an experimental point of view, involve the transition between the heavy $B$ meson and a light daughter hadron, which is extremely challenging to describe precisely, and lead to model-dependent results. This approach, successfully exploited by the CLEO collaboration in the measurement of the $B^0 \to \pi^- \ell^+ \nu_\ell$ and $B^0 \to \rho^- \ell^+ \nu_\ell$ branching ratios, led to the following average determination of $|V_{ub}|$ (see Table IV):

$$|V_{ub}|_{CLEO\,exclusive} = \left(3.25 \pm 0.14^{+0.21}_{-0.29} \pm 0.55\right) \cdot 10^{-3}, \qquad (2.4\text{-}1)$$

where the errors are respectively statistical, systematic and theoretical.

Exclusive measurements, feasible with the $B$ production at the $\Upsilon(4S)$, are not effective at LEP, where the lepton momenta in the $b$-hadron rest frame cannot be reconstructed with sufficient accuracy, the efficiency of the $B$ decay vertex reconstruction depends on the charge multiplicities of the final state and the fragmentation of the $b$ quarks produces different particles which dilute the signal. These disadvantages are compensated for by the back-to-back topology of the $b$ quark production from $Z$ decays, which avoids the mixing of the $B$ decay products. This characteristic is fully exploited by a novel technique based on the study of the shape of the invariant mass of the hadronic system recoiling against the lepton in $b \to u\ell^- \bar{\nu}_\ell$ transitions. The analyses, up to now performed by the ALEPH and L3 collaborations, have allowed the determination of the $b \to X_u \ell \bar{\nu}_\ell$ inclusive branching ratios quoted in Table IV. The basic remark is that a smaller model-dependence is expected in predicting the shape of this invariant-mass distribution, while the recent progress in the theoretical calculations has succeeded in containing the uncertainties at the level of few percent[71]. The weighted average of the inclusive branching ratios obtained by the LEP experiments leads to

$$\mathcal{Br}\left(b \to X_u \ell \bar{\nu}_\ell\right) = \left(2.0 \pm 0.8\right) \cdot 10^{-3} \qquad \chi^2 = 0.99 \qquad (2.4\text{-}2)$$

(the correlations between systematic errors have been taken into account following the indications found in Ref. 72), from which, using the average $b$ quark lifetime[12] $\tau_b = (1.564 \pm 0.014) ps$, the value of $|V_{ub}|$ can be extracted in the context of the Heavy Quark Theory[71]:

$$|V_{ub}|_{LEP\ inclusive} = (4.3 \pm 0.9_{exp} \pm 0.3_{th}) \cdot 10^{-3} = (4.3 \pm 0.9) \cdot 10^{-3}. \qquad (2.4\text{-}3)$$

This determination can be averaged with the CLEO mean value coming from exclusive branching-ratio measurements, giving the following reference value of $|V_{ub}|$:

$$|V_{ub}| = (3.6 \pm 0.5) \cdot 10^{-3} \qquad \chi^2 = 0.92. \qquad (2.4\text{-}4)$$

| Experiment | Decays measured | $\mathcal{B}r$ | $\lvert V_{ub}\rvert$ |
|---|---|---|---|
| CLEO[73] | $B^0 \to \pi^- \ell^+ \nu_\ell$<br>$B^0 \to \rho^- \ell^+ \nu_\ell$ | $(1.8 \pm 0.4 \pm 0.3 \pm 0.2) \cdot 10^{-4}$<br>$(2.5 \pm 0.4^{+0.5}_{-0.7} \pm 0.5) \cdot 10^{-4}$ | $(3.3 \pm 0.2^{+0.3}_{-0.4} \pm 0.7) \cdot 10^{-3}$ |
| CLEO[74] | $B^0 \to \rho^- \ell^+ \nu_\ell$ | $(2.69 \pm 0.41^{+0.35}_{-0.40} \pm 0.50) \cdot 10^{-4}$ | $(3.23 \pm 0.24^{+0.23}_{-0.26} \pm 0.58) \cdot 10^{-3}$ |
| L3[75] | inclusive $b \to X_u \ell \bar{\nu}_\ell$ | $(3.3 \pm 1.0 \pm 1.7) \cdot 10^{-3}$ | $(6.0^{+0.8}_{-1.0}\,^{+1.4}_{-1.9} \pm 0.2) \cdot 10^{-3}$ |
| ALEPH[76] | inclusive $b \to X_u \ell \bar{\nu}_\ell$ | $(1.73 \pm 0.55 \pm 0.55) \cdot 10^{-3}$ | $\lvert V_{ub}\rvert^2 = (18.68 \pm 5.94 \pm 5.94 \pm 1.45) \cdot 10^{-6}$ |

**Table IV.** – *Summary of the experimental determinations of $|V_{ub}|$. The branching ratios measurements have statistical and systematic errors quoted separately. Where a third error is given, it accounts for the model-dependence of the results.*

## 2.5 $|V_{cb}|$

The charmed semileptonic decays of $B$ mesons constitute the experimental basis for the direct determination of the matrix element $|V_{cb}|$. In this Section, two different methods, which make use of exclusive and inclusive measurements, are described.

### 2.5.1 *Exclusive decays*

A number of research groups have carried out measurements of $|V_{cb}|$ studying the $B \to D^* \ell \bar{\nu}_\ell$ and $B \to D \ell \bar{\nu}_\ell$ exclusive decays in the kinematic configuration in which the $D$ meson is produced at rest (*zero recoil*) and the energy of the lepton-neutrino system is at its maximum.

In the context of the *Heavy Quark Effective Theory* (HQET)[77], which approximates the masses of the heavy quarks involved in the process ($b$ and $c$ in this case) as infinite, the hadronic form factors appearing in the expressions of the differential semileptonic decay rates can be defined using one universal function $\mathcal{F}(w)$ (Isgur-Wise function), which is independent of the initial and final heavy mesons. This form factor describes the superposition of the light quark wave functions. The maximum overlap is reached in the limit of zero recoil, for which the normalization $\mathcal{F}(w=1)=\mathcal{F}(q^2=q^2_{max})=1$ is adopted; here $w=v_B\cdot v_D=\left(m_B^2+m_D^2-q^2\right)/\left(2m_Bm_D\right)$, and $q^2=\left(p_B-p_D\right)^2$ is the square momentum transferred to the leptonic system, with $q^2_{max}=\left(m_B-m_D\right)^2$. However, the analytic form of the function cannot be predicted within HQET. Therefore, to extract the value of $\left|V_{cb}\right|$ the differential decay rate

$$\frac{d\Gamma}{dw}=\frac{G_F^2}{48\pi^3}f(w,m_B,m_D)\mathcal{F}^2(w)\left|V_{cb}\right|^2 \qquad (2.5\text{-}1)$$

($f$ is a known function measuring the density of states) is measured in the limit of zero recoil, where the matrix element is the only unknown factor. Due to the reduced observable statistics at the point of zero recoil ($w$= 1), the data are collected in the range $w$> 1; the value of $\mathcal{F}(1)\cdot\left|V_{cb}\right|$ is then obtained by extrapolating to the limit $w\rightarrow 1^+$. For this purpose the function $\mathcal{F}(w)$, *a priori* unknown, is parametrized as a Taylor expansion around the point $w=1$:

$$\mathcal{F}(w)=\mathcal{F}(1)[1-\hat{\rho}^2(w-1)+\hat{c}(w-1)^2+\ldots] \qquad (2.5\text{-}2)$$

(the first derivative must be negative, since the condition $w=1$ corresponds to the maximum superposition of the wave functions and thus to the maximum of $\mathcal{F}(w)$). The value of the intercept $\mathcal{F}(1)\cdot\left|V_{cb}\right|$ is then extracted from a fit to the differential decay data. Although not all the experiments are sensitive to the quadratic term ($\mathcal{F}(w)$ is often assumed to depend linearly on $w$), basic QCD considerations favour a positive value of the curvature $\hat{c}$, having a definite correlation with the slope $\hat{\rho}^2$ ($\hat{c}\cong 0.66\hat{\rho}^2-0.11$)[78]. The form factor $\mathcal{F}(1)$, which equals 1 at the leading order in the HQ expansion, is then determined by calculating perturbatively the effect of the finiteness of the heavy quark masses. While the second order $1/m_Q$ ($m_Q=m_b$ or $m_c$) power corrections have been computed in the case of $B\rightarrow D^*\ell\bar{\nu}_\ell$ decays, for which the $\mathcal{O}(1/m_Q)$ term is absent[79], at present the $B\rightarrow D\ell\bar{\nu}_\ell$ decays can be described with less theoretical accuracy, since only the first order of the expansion is known. Moreover, both the overall branching ratio and the rate near the point of zero-recoil are smaller for the $B\rightarrow D\ell\bar{\nu}_\ell$ mode, thus limiting the precision of the measurements.

From the experimental analysis of the $B\rightarrow D^*\ell\bar{\nu}_\ell$ exclusive decays, CLEO and the LEP experiments obtained accurate determinations of the parameter $\mathcal{F}(1)\cdot\left|V_{cb}\right|$. The data are displayed in Table V. OPAL and DELPHI derived their results using a quadratic parametrization of the Isgur-Wise function, while a linear dependence was assumed by CLEO and ALEPH. We do not report the results of other measurements carried out by ARGUS[80] ($B\rightarrow D^*\ell\bar{\nu}_\ell$), ALEPH[84] and CLEO[81] ($B\rightarrow D\ell\bar{\nu}_\ell$), which provide less precise (though consistent) information.

Since the LEP measurements, though presumably highly correlated, are scarcely consistent even within their *total* errors, we compute their weighted average separately (in this way we avoid diluting the reduced $\chi^2$ with the contribution of the independent measurement by CLEO) and rescale the error by the factor $s=\sqrt{\chi^2/2}=\sqrt{2.0}$, obtaining

$$\mathcal{F}_{D^*}(1)\cdot\left|V_{cb}\right|_{LEP}=(35\pm 2)\cdot 10^{-3}. \qquad (2.5\text{-}3)$$

Further averaging with the CLEO measurement leads to the result given in Table V.

| | | $\mathcal{F}_{D^*}(1)\cdot\|V_{cb}\|\cdot 10^3$ |
|---|---|---|
| $\overline{B}^0 \to D^{*+}\ell^-\overline{\nu}_\ell$, $B^- \to D^{*0}\ell^-\overline{\nu}_\ell$ | CLEO II[82] | $35.1 \pm 1.9 \pm 1.9$ |
| $\overline{B}^0 \to D^{*+}\ell^-\overline{\nu}_\ell$ | OPAL[83] | $32.8 \pm 1.9 \pm 2.2$ |
| | ALEPH[84] | $31.9 \pm 1.8 \pm 1.9$ |
| | DELPHI[85] | $37.95 \pm 1.34 \pm 1.59$ |
| *Average* | | **$35.0 \pm 1.6$** |

**Table V.** – $\mathcal{F}_{D^*}(1)\cdot|V_{cb}|$ *measurements. The errors are statistical and systematic respectively.*

Detailed calculations of the QCD and finite-mass corrections which have to be applied to the form factor normalization give the result[78]

$$\mathcal{F}_{D^*}(1) = 0.91 \pm 0.03. \qquad (2.5\text{-}4)$$

Then we can extract $|V_{cb}|$ from the best value of $\mathcal{F}_{D^*}(1)\cdot|V_{cb}|$:

$$|V_{cb}|_{exclusive} = (38.5 \pm 1.8_{exp} \pm 1.3_{th})\cdot 10^{-3} = (38.5 \pm 2.2)\cdot 10^{-3}. \qquad (2.5\text{-}5)$$

## 2.5.2 *Inclusive decays*

The measurements of the inclusive branching ratio of $B$ semileptonic decays to charmed final states and of the $B$ meson mean life provide another independent determination of $|V_{cb}|$. The relation between the matrix element and the experimental observables is calculated in the framework of the Heavy Quark Theory[71]:

$$|V_{cb}| = 41.1\cdot 10^{-3}\sqrt{\frac{\Gamma(B \to X_c\ell\overline{\nu}_\ell)}{6.77\cdot 10^{-2}\,ps^{-1}}}\cdot(1 \pm 0.06). \qquad (2.5\text{-}6)$$

The measurements on the $\Upsilon(4S)$ and those at the $Z$ give quite different results (*):

$$\Gamma(B^0/B^\pm \to X_c\ell\overline{\nu}_\ell)_{\Upsilon(4S)} = (6.39 \pm 0.18)\cdot 10^{-2}\,ps^{-1} \quad |V_{cb}|_{\Upsilon(4S)} = (39.91 \pm 0.55)\cdot 10^{-3} \qquad (2.5\text{-}7)$$

$$\Gamma(b \to X_c\ell\overline{\nu}_\ell)_Z = (6.90 \pm 0.17)\cdot 10^{-2}\,ps^{-1} \quad |V_{cb}|_Z = (41.48 \pm 0.50)\cdot 10^{-3} \qquad (2.5\text{-}8)$$

(only the experimental errors are quoted in the $|V_{cb}|$ values). Their average ($\chi^2 = 4.5$, error rescaled by 2.1) leads to

$$\left|V_{cb}\right|_{inclusive} = (40.8 \pm 0.8_{\,exp} \pm 2.4_{\,th}) \cdot 10^{-3} = (40.8 \pm 2.5) \cdot 10^{-3}. \qquad (2.5\text{-}9)$$

---

* The world-average total semileptonic branching-ratios have been corrected by the $b \to u\ell\overline{\nu}_\ell$ contribution (Eq. 2.4-2). The average of the $B^0$ and $B^\pm$ mean lives[12] has been used to calculate $\Gamma(B \to X_c\ell\overline{\nu}_\ell)_{\Upsilon(4S)}$.

Finally, we obtain our best value of $\left|V_{cb}\right|$ averaging the results obtained from the inclusive (Eq. 2.5-5) and exclusive (2.5-9) measurements:

$$\left|V_{cb}\right| = (39.5 \pm 1.7) \cdot 10^{-3}. \qquad (2.5\text{-}10)$$

## 2.6 $|V_{ub}|/|V_{cb}|$

The first observations of $b \to u\ell^-\bar{\nu}_\ell$ transitions, carried out by the CLEO and, subsequently, the ARGUS collaboration, were based on the gain in sensitivity obtained by restricting the data selection to the final part of the phase-space, where the lepton momenta are beyond the end-point of the $b \to c\ell^-\bar{\nu}_\ell$ processes. An excess of leptons with momenta in the range $2.3\,GeV < p_\ell < 2.6\,GeV$ is interpreted as evidence of charmless *B* decays. The drawback of this approach is represented by the limited portion of the observed phase-space, which requires an extrapolation to the low momentum region in order to recover the full decay rate, leading to a model-dependent measurement. This technique led CLEO, ARGUS, and CLEO II to the direct measurement of the ratio $\left|V_{ub}\right|/\left|V_{cb}\right|$ (Table VI). Recently, the DELPHI collaboration, by means of a technique based on the study of the shape of the invariant-mass distribution of the hadronic system recoiling against the lepton, has produced a new measurement of the $\left|V_{ub}\right|/\left|V_{cb}\right|$ ratio, which takes into account the whole lepton spectrum (Table VI).

We get a further estimation of the $\left|V_{ub}\right|/\left|V_{cb}\right|$ ratio using our reference values for $\left|V_{ub}\right|$ and $\left|V_{b}\right|$ derived in Sections 2.4 and 2.5 (Eqs. 2.4-4 and 2.5-10):

$$\left|V_{ub}\right|/\left|V_{b}\right| = 0.091 \pm 0.013. \qquad (2.6\text{-}1)$$

By taking the weighted average of this value with the DELPHI determination we obtain the following evaluation of the $\left|V_{ub}\right|/\left|V_{b}\right|$ ratio:

$$\left|V_{ub}\right|/\left|V_{b}\right| = 0.093 \pm 0.011. \qquad (2.6\text{-}2)$$

We note that the CLEO and ARGUS $\left|V_{ub}\right|/\left|V_{b}\right|$ measurements, quoted in the first part of Table VI, are spread out by the considerable differences of the theoretical models available at that time. On the contrary, the theoretical uncertainties connected to our estimation (2.6-2) of the $\left|V_{ub}\right|/\left|V_{b}\right|$ ratio are at the level of few percent[71]. Therefore, we can use this value to reject, by means of a $\chi^2$ test, the models which give conflicting predictions. The $\chi^2$ values indicating the compatibility with the reference value 2.6-2,

$$\chi^2(\text{KS})=9.4, \quad \chi^2(\text{WSB})=6.9, \quad \chi^2(\text{ACCMM})=2.4, \quad \chi^2(\text{ISGW})=24,$$

(with 3 degrees of freedom) induce to reject the KS and IGSW models at the 5% level and to recover the WSB and ACCMM predictions. In particular, the ACCMM model, which is also able to reproduce a good agreement between the three measurements (the $\chi^2$ of the average is 2.5, see Table VI), leads to the following average determination:

$$\left|V_{ub}\right|/\left|V_{b}\right| = 0.088 \pm 0.009, \qquad (2.6\text{-}3)$$

where the error has been rescaled by the factor $\sqrt{2.5}$.

This result is finally averaged with the DELPHI measurement, giving our $|V_{ub}|/|V_{cb}|$ reference value:

$$|V_{ub}|/|V_{cb}| = 0.090 \pm 0.008. \qquad (2.6\text{-}4)$$

| Model ► / ▼ Experiment | KS[86] | WSB[87] | ACCMM[88] | ISGW[89] |
|---|---|---|---|---|
| CLEO[68] 2.2< $p_\ell$< 2.4 *GeV* 2.4< $p_\ell$< 2.6 *GeV* | $0.095 \pm 0.011$ | $0.114 \pm 0.018$ | $0.089 \pm 0.011$ | $0.148 \pm 0.020$ |
| ARGUS[70] 2.3< $p_\ell$< 2.6 *GeV* | $0.110 \pm 0.012$ | $0.130 \pm 0.015$ | $0.110 \pm 0.012$ | $0.200 \pm 0.023$ |
| CLEO II[90] 2.3< $p_\ell$< 2.4 *GeV* 2.4< $p_\ell$< 2.6 *GeV* | $0.057 \pm 0.006$ | $0.075 \pm 0.007$ | $0.078 \pm 0.008$ | $0.104 \pm 0.010$ |
| Average ($\chi^2$/2) | $0.073 \pm 0.005$ (10.3) | $0.088 \pm 0.006$ (6.7) | $0.088 \pm 0.006$ (2.5) | $0.124 \pm 0.008$ (8.2) |

| | |
|---|---|
| DELPHI[91] | $0.100 \pm 0.011_{stat} \pm 0.018_{syst} \pm 0.009_{model}$ |

**Table VI.** – *Summary of the experimental determinations of* $|V_{ub}|/|V_{cb}|$. *The upper part compares the* $|V_{ub}|/|V_{cb}|$ *measurements obtained by CLEO, ARGUS and CLEO II using four distinct theoretical models. The lepton momentum (* $p_\ell$ *) ranges observed by each experiment is also indicated. The CLEO and CLEO II results have been renormalized to the present value of the inclusive branching ratio of the background processes,* $\mathcal{Br}(B \to X_c \ell \bar{\nu}_\ell) = \mathcal{Br}(B \to X \ell \bar{\nu}_\ell) - \mathcal{Br}(B \to X_u \ell \bar{\nu}_\ell) = (10.25 \pm 0.22)\%$ *(obtained using Eq. (2.4-2) and the value of* $\mathcal{Br}(B \to X \ell \bar{\nu}_\ell)$ *quoted in Ref. 12).*

## 2.7 Elements of the third row

The $\Delta m_{B_d}$ and $\Delta m_{B_s}$ meson-antimeson oscillation frequencies, the parameter $\varepsilon_K$ measuring the mass-matrix CP violation in the neutral-kaon system and the inclusive rate of the $b \to s\gamma$ rare decays constitute the main experimental constraints on the values of $V_{td}$, $V_{ts}$ and $V_{tb}$.

All the measurements involving elements of the third row implicitly contain the hypothesis of a three-generation CKM matrix. For example, CDF[92] measured the ratio $\mathcal{Br}(t \to Wb) / \sum_{q=d,s,b} \mathcal{Br}(t \to Wq)$, which can be translated into a direct measurement of $|V_{tb}|$,

$$|V_{tb}| = \frac{|V_{tb}|}{\sqrt{|V_{td}|^2 + |V_{ts}|^2 + |V_{tb}|^2}} = \sqrt{\frac{\mathcal{Br}(t \to Wb)}{\sum_{q=d,s,b} \mathcal{Br}(t \to Wq)}} 0.96^{+0.16}_{-0.12}, \qquad (2.7\text{-}1)$$

only if the assumption $|V_{td}|^2 + |V_{ts}|^2 + |V_{tb}|^2 = 1$ is made. It has to be stressed that, if this condition is relaxed, no determination is possible for $|V_{tb}|$ using such an experimental procedure (based on $t\bar{t}$ production) and only very loose bounds can be set ($|V_{tb}| > 0.045$ at 95% C.L.)[92]. On the other hand, a measurement of the single top production cross section, which is directly proportional to $|V_{tb}|^2$ without any underlying hypothesis, will be possible in Runs 2 and 3 of CDF and D0[93].

The mixing and CP violating phenomena are directly related to the existence of new physics. If the $d$, $s$ and $b$ quarks were allowed to couple with the members of a fourth quark family (or with another unknown species of heavy fermions) or a different kind of highly massive gauge bosons could replace the exchanged $W^{\pm}$, new, significant contributions would have to be expected in the description of these processes. A common property of the weak transitions of which these observables are characteristic parameters is the ability to change the flavour of the quarks involved without changing their charge (*effective FCNC processes*). The corresponding amplitudes can be obtained, within the Standard Model, by iterating the basic, charged-current weak couplings. In this way, the $\Delta S = 2$ and/or $\Delta B = 2$ transitions on which the mixing phenomena of neutral mesons are based, are traced back to the elementary $\Delta S = 1$ and $\Delta B = 1$ processes, which are represented by single CKM matrix elements. The description of mixing by means of the so-called ‘box’ diagrams (see Figure 4), in which two virtual $W^{\pm}$ bosons are exchanged between two quark lines, cannot lead to a quantification of the transition amplitudes without the additional assumption that only three quark families exist. The hypothesis of the 3×3 unitarity of the CKM matrix, which has not been used for the deduction of the elements of the first two rows, plays an essential role in the measurement of the elements of the third row, even if it not imposed as a direct constraint.

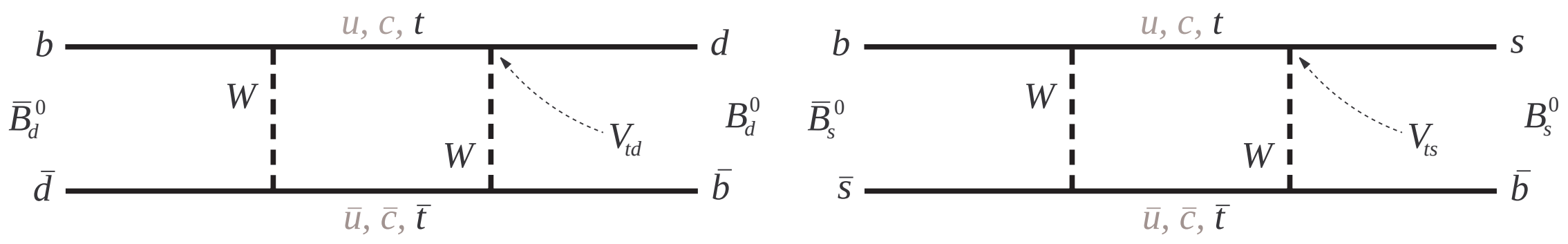

**Figure 4.** – *Examples of box diagrams describing* $B_d^0 - \overline{B}_d^0$ *and* $B_s^0 - \overline{B}_s^0$ *mixing.*

The effective Hamiltonian for the $\Delta B = 2$ transitions ($B_d^0 - \overline{B}_d^0$ mixing), for example, is proportional (neglecting corrective terms) to

$$\sum_{i,j} V_{id} V_{ib}^* V_{jd} V_{jb}^* S(x_i, x_j)\,, \qquad (2.7\text{-}2)$$

where $S(x_i, x_j)$ are the Inami-Lim[94] functions (apart from the sign, irrelevant for the final formulae), which depend on quark masses ($x_i = m_i^2 / m_W^2$),

$$S(x_i, x_j) = x_i x_j \left[ \left( \frac{1}{4} + \frac{3}{2(1-x_i)} - \frac{3}{4(1-x_i)^2} \right) \frac{\log x_i}{x_i - x_j} + \left( x_i \leftrightarrow x_j \right) - \frac{3}{4} \frac{1}{(1-x_i)(1-x_j)} \right] \quad i \neq j \qquad (2.7\text{-}3)$$

$$S(x_i, x_i) \equiv S(x_i) = x_i \left( \frac{1}{4} + \frac{9}{4(1-x_i)} - \frac{3}{2(1-x_i)^2} \right) - \frac{3}{2} \left( \frac{x_i}{1-x_i} \right)^3 \log x_i \qquad (2.7\text{-}4)$$

and the indexes in the summation assume in turn all the values corresponding to the *up*-type quarks which take part in the intermediate virtual state; therefore, in the three-family model, *i, j = u, c, t.* The unitarity condition is then used again to simplify the expression 2.7-2. The approximate equality between orders of magnitude

$$\left| V_{ud} V_{ub}^* \right| \approx \left| V_{cd} V_{cb}^* \right| \approx \left| V_{td} V_{tb}^* \right| = \mathcal{O}\left(\lambda^3\right) \qquad (2.7\text{-}5)$$

applies in Wolfenstein's parametrization; the relative weight of the summation terms is thus determined by the functions $S(x_i, x_j)$, the magnitudes of which fulfil the following relation (the values of the masses and all other parameters occurring in the expressions quoted in this paragraph will be given in Table XV):

$$S(x_t, x_t) : S(x_t, x_c) : S(x_c, x_c) \cong 10^4 : 10^1 : 10^0\,, \qquad (2.7\text{-}6)$$

while all other combinations are negligible in comparison with $S(x_c, x_c)$. Therefore, the box diagram is dominated by the top quark contribution and, when all other terms are neglected, the $B_d^0$ mass difference is proportional to $\left| V_{td} V_{tb}^* \right|^2 S(x_t)$. Replacing the *d* quark by an *s* quark does not alter the proportions between the members of Eq. (2.7-6), so that the same approximation applies to $B_s^0 - \overline{B}_s^0$ mixing:

$$\Delta m_{B_d} = \frac{G_F^2}{6\pi^2} m_W^2 m_{B_d} f_{B_d}^2 B_{B_d} \eta_B S(x_t) \left| V_{td} V_{tb}^* \right|^2 \qquad (2.7\text{-}7)$$

$$\Delta m_{B_s} = \frac{G_F^2}{6\pi^2} m_W^2 m_{B_s} f_{B_s}^2 B_{B_s} \eta_B S(x_t) \left| V_{ts} V_{tb}^* \right|^2 \qquad (2.7\text{-}8)$$

Here $\eta_B S(x_t)$ is the loop function (2.7-4) corrected for the effects of perturbative QCD; its value does not depend on the convention chosen for the top quark mass, provided the same convention is used to compute $\eta_B$; the result[95, 96] $\eta_B = 0.55 \pm 0.01$ is valid within the $\overline{MS}$ (Minimal Subtraction) renormalization scheme and, for reasons of consistency, the mass $m_t$ has to be computed by rescaling the pole mass[7]: $\overline{m}_t(m_{t\,pole}) = (166 \pm 5) GeV$.

$f_{B_d}$ and $f_{B_s}$ are the $B$ meson decay constants; the products $f_{B_d}^2 B_{B_d}$ and $f_{B_s}^2 B_{B_s}$ ($B_{B_d}$ and $B_{B_s}$ are called 'bag' factors) parametrize the matrix elements between the initial and the final hadronic state. Their values are the main theoretical uncertainty in the relations 2.7-7 and 2.7-8. In order to determine these parameters, most theoretical analyses make use of QCD computations performed on a discretized space-time (*lattice-QCD*). (A comparison between some recent results will be shown in Table IX and in Table X).

Even though no direct experimental determinations of $f_{B_d}$ and $f_{B_s}$ exist at present, a number of experiments have measured the rates of the processes $D_s^+ \to \mu^+ \nu_\mu$ and $D_s^+ \to \tau^+ \nu_\tau$, leading to an estimate of the decay constant $f_{D_s}$. An indirect measurement of $f_{B_d}$ and $f_{B_s}$ can then be obtained by extrapolating from $D$ to $B$ sector[97] by means of the theoretical determinations of $f_{B_d} / f_{D_s}$ and $f_{B_s} / f_{D_s}$, which are quite accurate. Moreover, the measurement of $f_{D_s}$ can be used to check the results of lattice simulations, which generally determine $f_{D_s}$ and the $B$ meson decay constants simultaneously.

## *2.7.1 Experimental determination of* $f_{D_s}$

According to the Standard Model, $D_s^+ \to \ell^+ \nu_\ell$ semileptonic decays occur by pair annihilation of the constituent quark ($c$) and antiquark ($\bar{s}$) into a virtual $W$; the decay constant $f_{D_s}$ parametrizes the matrix element between the quark-antiquark wave function and vacuum. The strength of the coupling is the same as in a $c \to s$ charge-current process and the decay amplitude is proportional to $V_{cs}$:

$$\mathcal{Br}(D_s^+ \to \ell^+ \nu_\ell) = \tau_{D_s} \frac{G_F^2 |V_{cs}|^2}{8\pi} f_{D_s}^2 m_{D_s} m_\ell^2 \left(1 - \frac{m_\ell^2}{m_{D_s}^2}\right)^2 . \qquad (2.7\text{-}9)$$

The dependence on the square lepton mass causes the suppression of the decays having an electron (or positron) in the final state. Table VII shows the measured values of the relevant branching ratios.

E653 and CLEO respectively determined the ratios

$$\mathcal{Br}(D_s^+ \to \mu^+ \nu_\mu) / \mathcal{Br}(D_s^+ \to \varphi \mu^+ \nu_\mu) = 0.16 \pm 0.06 \pm 0.03 \qquad (2.7\text{-}10)$$

and $$\mathcal{Br}(D_s^+ \to \mu^+ \nu_\mu) / \mathcal{Br}(D_s^+ \to \varphi \pi^+) = 0.173 \pm 0.023 \pm 0.035 \qquad (2.7\text{-}11)$$

from which the values listed in the Table have been obtained using[12]

$$\mathcal{Br}(D_s^+ \to \varphi \mu^+ \nu_\mu) = (2.0 \pm 0.5)\% , \qquad (2.7\text{-}12)$$

$$\mathcal{Br}(D_s^+ \to \varphi \pi^+) = (3.6 \pm 0.9)\% . \qquad (2.7\text{-}13)$$

The measurements of the channel $D_s^+ \to \tau^+ \nu_\tau$,

L3: $$\mathcal{Br}(D_s^+ \to \tau^+ \nu_\tau) = (7.4 \pm 2.8 \pm 2.4)\% \qquad (2.7\text{-}14)$$

DELPHI: $$\mathcal{Br}(D_s^+ \to \tau^+ \nu_\tau) = (8.5 \pm 4.2 \pm 2.6)\% , \qquad (2.7\text{-}15)$$

have been converted according to the relation

$$\mathcal{Br}\left(D_s^+ \to \mu^+ \nu_\mu\right) \cong \frac{\mathcal{Br}\left(D_s^+ \to \tau^+ \nu_\tau\right)}{9.743} \qquad (2.7\text{-}16)$$

which follows from Eq. (2.7-9) (see Table VIII for the mass values). The same fixed ratio between the two branching ratios was assumed by BES and ALEPH, which determined the value of $\mathcal{Br}\left(D_s^+ \to \mu^+ \nu_\mu\right)$ by fitting the data provided by the observation of both leptonic channels.

| Experiment / Decays | | $\mathcal{Br}\left(D_s^+ \to \mu^+ \nu_\mu\right)$ (%) |
|---|---|---|
| WA75[98] | $D_s^+ \to \mu^+ \nu_\mu$ | $0.40\,^{+0.18}_{-0.14}\,^{+0.19}_{-0.18}$ |
| BES[99] | $\begin{cases} D_s^+ \to \mu^+ \nu_\mu \\ D_s^+ \to \tau^+ \nu_\tau \end{cases}$ (**) | $1.5\,^{+1.3}_{-0.6}\,^{+0.3}_{-0.2}$ |
| E653[100] | $\begin{cases} D_s^+ \to \mu^+ \nu_\mu \\ D_s^+ \to \varphi\mu^+ \nu_\mu \end{cases}$ (*) | $0.32 \pm 0.12 \pm 0.10$ |
| L3[101] | $D_s^+ \to \tau^+ \nu_\tau$ (**) | $0.76 \pm 0.29 \pm 0.25$ |
| DELPHI[102] | $D_s^+ \to \tau^+ \nu_\tau$ (**) | $0.87 \pm 0.43 \pm 0.27$ |
| CLEO[103] | $\begin{cases} D_s^+ \to \mu^+ \nu_\mu \\ D_s^+ \to \varphi\pi^+ \end{cases}$ (*) | $0.62 \pm 0.08 \pm 0.20$ |
| ALEPH[104] | $\begin{cases} D_s^+ \to \mu^+ \nu_\mu \\ D_s^+ \to \tau^+ \nu_\tau \end{cases}$ (**) | $0.64 \pm 0.08 \pm 0.26$ |
| *Weighted average* | | $0.512 \pm 0.098$<br>$\chi^2/6 = 0.93$<br>$f_{D_s} = (254 \pm 25) MeV$ |

**Table VII.** – *Branching ratio measurements for the* $D_s^+ \to \mu^+ \nu_\mu$ *decay. The errors quoted are statistical and systematic respectively. See the footnote on page 24 for the description of how the asymmetric errors are included in the average.* (*) *E653 and CLEO measured the ratios between the branching fractions of the decays indicated (see text).* (**) *Lepton universality has been assumed in converting the* $D_s^+ \to \tau^+ \nu_\tau$ *result into a* $\mathcal{Br}\left(D_s^+ \to \mu^+ \nu_\mu\right)$ *measurement.*

From the average branching ratio, using the data given in Table VIII the result

$$f_{D_s} = (254 \pm 25) MeV \qquad (2.7\text{-}17)$$

is obtained. The error is entirely due to the experimental determination of the decay rate. The correlation between this result and the measurement of $|V_{cs}|$ is completely negligible. The value of $|V_{cs}|$ used for the deduction of Eq. (2.7-17) is the output value of a unitarity-constrained fit performed by the PDG[12] using all the directly measured CKM matrix elements (it is perfectly consistent with the result we will obtain in Sect. 3 after imposing the unitarity constraint); in this case, given its precision, it can be regarded as a constant (*). The result is in good agreement with a value ( $f_{D_s} = (241 \pm 36) MeV$ ) obtained in Ref. 105 as an application of the Isgur-Wise Theory to the measured rates of the decays $\overline{B} \to D\overline{D}_s$ and $\overline{B} \to D^*\overline{D}_s$ .

| | |
|---|---|
| $m_\mu$ | $(105.658389 \pm 0.000034) MeV$ |
| $m_\tau$ | $(1777.05^{+0.29}_{-0.26}) MeV$ |
| $m_{D_s^\pm}$ | $(1968.5 \pm 0.6) MeV$ |
| $\tau_{D_s^\pm}$ | $(0.467 \pm 0.017) \cdot 10^{-12} s$ |
| $\lvert V_{cs} \rvert$ | $0.9745 \pm 0.0008 \quad 90\% C.L.$ |

**Table VIII.** – *List of the parameters used to extract* $f_{D_s}$ *from the measurements of* $\mathcal{Br}(D_s^+ \to \mu^+ \nu_\mu)$ *and* $\mathcal{Br}(D_s^+ \to \tau^+ \nu_\tau)$ *(see Ref. 12).*

### *2.7.2 Lattice QCD results*

Some recent lattice computations of heavy meson decay constants and bag factors are shown in Table IX and X. The large systematic errors are due to several types of approximations. For example, the results have been obtained on a finite space-time lattice and therefore are dependent on the lattice spacing. Moreover, the bag factors have been computed in a renormalization scheme which is peculiar to the lattice gauge theory, and have to be converted using a continuum renormalization scheme (such as the $\overline{MS}$ scheme), within which the experimental data are analysed. The extreme SU(3) symmetry, often assumed for the light quarks, and the use of perturbative techniques are two other sources of uncertainty which the systematic errors usually account for. On the other hand, the evaluation of the consequences of the so-called *quenched approximation* is still at a preliminary stage. In this approximation, which is common to all the computations, the contribution of the sea quarks in closed loops is neglected, leading to a substantial reduction in computing time. However, only the MILC collaboration[106] has performed an estimation of the resulting systematic errors; moreover, its investigation of quenching effects was limited to a very simplified case[107], in which only two quark flavours were considered and the extrapolations to the physical masses and to the continuum were omitted. While the theoretical panorama is still evolving, at present a truly reliable estimate of the uncertainties on the currently available results cannot be given.

* The correlation between two measurements can obviously be eliminated if the margin of error corresponding to the 100% confidence level is assumed for one of them. The error in the determination of $f_{D_s}$ does not increase if that for $\lvert V_{cs} \rvert$ (already corresponding to a 90% confidence level) is rescaled by a factor of 25.

| $f_{D_s}\,(MeV)$ | $f_{B_d}/f_{D_s}$ | $f_{B_d}\,(MeV)$ | $f_{B_s}/f_{D_s}$ | $f_{B_s}\,(MeV)$ | $f_{B_s}/f_{B_d}$ | Ref. |
|---|---|---|---|---|---|---|
| $224 \pm 2 \pm 16 \pm 11$ | $(0.77 \pm 0.09)$ | $173 \pm 4 \pm 9 \pm 9$ | $(0.89 \pm 0.10)$ | $199 \pm 3 \pm 10 \pm 10$ | | [108] |
| | | $162 \pm 7 \pm 5 \pm 5 \pm 11 \pm 6^{+31}_{-8}$ | | $190 \pm 5 \pm 5 \pm 5 \pm 13 \pm 6^{+39\;+4}_{-9\;-0}$ | $1.18 \pm 0.03 \pm 0.05^{+0.02}_{-0}$ | [109] |
| $221 \pm 9$ | $(0.73 \pm 0.08)$ | $161 \pm 16$ | $(0.86 \pm 0.06)$ | $190 \pm 12$ | $1.18 \pm 0.08$ | [110] |
| $213^{+14}_{-11} \pm 11$ | $\frac{f_{B_d}}{f_D}\frac{f_D}{f_{D_s}} = 0.76^{+0.04}_{-0.05}$ | $164^{+14}_{-11} \pm 8$ | $\frac{1}{f_{D_s}/f_{B_s}} = 0.88^{+0.02}_{-0.03}$ | $185^{+13}_{-8} \pm 9$ | $1.13^{+0.05}_{-0.04}$ | [111] |
| | | | | $201 \pm 6 \pm 15 \pm 7$ | | [112] |
| $210 \pm 9^{+25\;+17}_{-9\;-1}$ | $0.75 \pm 0.03^{+0.04\;+0.08}_{-0.02\;-0}$ | $157 \pm 11^{+25\;+23}_{-9\;-0}$ | $0.85 \pm 0.03^{+0.05\;+0.05}_{-0.03\;-0}$ | $171 \pm 10^{+34\;+27}_{-9\;-2}$ | $1.11 \pm 0.02^{+0.04}_{-0.03} \pm 0.03$ | [106] |
| | | | | | $1.16 \pm 0.03$ | [113] |
| $237 \pm 16$ | $(0.76 \pm 0.14)$ | $180 \pm 32$ | $(0.87 \pm 0.17)$ | | $1.14 \pm 0.08$ | [114] |
| | | | | | $1.17 \pm 0.03$ | [115] |
| | | $147 \pm 11^{+8}_{-12} \pm 9 \pm 6$ | | $175 \pm 8^{+7}_{-10} \pm 11 \pm 7^{+7}_{-0}$ | $1.20 \pm 0.04^{+0.04}_{-0}$ | [116] |
| $231 \pm 12^{+6}_{-0}$ | $0.78 \pm 0.04^{+0.11}_{-0}$ | $179 \pm 18^{+26}_{-9}$ | $0.88 \pm 0.03^{+0.12}_{-0}$ | $204 \pm 16^{+28}_{-0}$ | $1.14 \pm 0.03^{+0}_{-0.01}$ | [117] |

**Table IX.** – *Recent lattice QCD results for the heavy-meson decay constants. The errors reflect many different causes of systematic uncertainties which are inherent in lattice QCD calculations. However, their evaluation is incomplete in almost all cases. In particular, the effects of the quenched approximation, in which only the valence quark contribution is taken into account, has not been estimated except for a partial evaluation by the authors of Ref. 106 (see the last error quoted). Moreover, some results are still preliminary. Where direct determinations are not available, the values of* $f_{B_d}/f_{D_s}$ *and* $f_{B_s}/f_{D_s}$ *obtained by combining the results listed in the same row are shown (in brackets) for comparison.*

The independent predictions for the ratios $f_{B_d}/f_{D_s}$, $f_{B_s}/f_{D_s}$ and $f_{B_s}/f_{B_d}$ are fully compatible. This evidence can be made even more significant if the values calculated from $f_{D_s}$, $f_{B_d}$ and $f_{B_s}$ are indicated where direct determinations are not available. Apparently, the systematic errors tend to cancel out in the ratios. Assuming that a similar compensatory effect occurs between the quenching errors, the following estimates will be used for the purposes of the present discussion:

$$f_{B_d}/f_{D_s} = 0.76 \pm 0.04, \qquad (2.7\text{-}18)$$

$$f_{B_s}/f_{D_s} = 0.87 \pm 0.04. \qquad (2.7\text{-}19)$$

They correspond to the central value $f_{B_s}/f_{B_d}$ =1.145 (ratio of 2.7-19 to 2.7-18), which all the direct determinations are consistent with. Almost all predictions for $f_{B_s}/f_{B_d}$ are included in the range described by the estimate

$$f_{B_s}/f_{B_d} = 1.15 \pm 0.04. \qquad (2.7\text{-}20)$$

The comparison of the $f_{D_s}$ predictions from all models with the experimental value (2.7-17) shows a systematic theoretical underestimation: this goes in the direction to confirm the evaluation of the quenching errors made by MILC. Using the experimental value of $f_{D_s}$ and Eqs. (2.7-18,19) (which are valid under the assumption that the systematic errors in the ratios $f_{B_d}/f_{D_s}$, $f_{B_s}/f_{D_s}$ are small) the following indirect measurements of the $B$ meson decay constants are obtained:

$$f_{B_d} = (193 \pm 22)MeV, \qquad (2.7\text{-}21)$$

$$f_{B_s} = (221 \pm 24)MeV. \qquad (2.7\text{-}22)$$

The differences between the central values of these measurements and the average quenched results $f_{B_d} = 165\,MeV$, $f_{B_s} = 191\,MeV$ are nearly equal to the magnitudes of the quenching errors computed by MILC ($+23\,MeV$ and $+27\,MeV$ respectively).

On the other and, the direct calculations based on QCD sum rules favour slightly lower values of the decay constants: for example, the average $f_{B_d} = (160 \pm 30)MeV$ is quoted in a review[118] of the latest theoretical developments in this field. However, only the theoretical determinations of the *ratios* between decay constants (Eqs. 2.7-18,19,20), which are expected to be less model-dependent, will be used in the present analysis.

Since no information can be obtained from experiments about the bag factors $B_{B_d}$ and $B_{B_s}$, a value of $B_{B_d}$ with which all values in Table X are compatible is assumed:

$$B_{B_d} = 1.30 \pm 0.15; \qquad (2.7\text{-}23)$$

The error estimate is conservative enough to cover a probable discrepancy which may be due to quenching effects (roughly 4%[119]).

No calculation has highlighted any difference between $B_{B_s}$ and $B_{B_d}$ due to SU(3)-breaking effects: they will be assumed to be nearly equal:

$$\frac{B_{B_s}}{B_{B_d}} = 1.00 \pm 0.01 \qquad (2.7\text{-}24)$$

that is, using (2.7-23),

$$B_{B_s} = B_{B_d} = B_B = 1.30 \pm 0.15. \qquad (2.7\text{-}25)$$

| $B_{B_d}$ | $B_{B_s}/B_{B_d}$ | Ref. |
|---|---|---|
| | $1.01 \pm 0.01$ | [115] |
| $1.29 \pm 0.08 \pm 0.05 \pm 0.03$ | | [120] |
| $1.40 \pm 0.06^{+0.04}_{-0.26}$ | $0.99 \pm 0.01 \pm 0.01$ | [121] |
| $1.17 \pm 0.09 \pm 0.04 \pm 0.03$ | $1.00 \pm 0.02$ | [113] |
| $1.46 \pm 0.19$ | $1$ | [122] |
| | $0.99 \pm 0.03$ | [110] |

**Table X.** – *Lattice-QCD evaluations for the bag parameters entering into the expressions of the* $B_d^0$ *and* $B_s^0$ *mass differences.*

From Eqs. (2.7-17), (2.7-18), (2.7-19) and (2.7-25) the following estimates of the hadronic factors occurring in Eqs. (2.7-7) and (2.7-8) are obtained:

$$f_{B_d}\sqrt{B_{B_d}} = \left(220 \pm 22_{exp} \pm 17_{th}\right) MeV = \left(220 \pm 28\right) MeV \qquad (2.7\text{-}26)$$

$$f_{B_s}\sqrt{B_{B_s}} = \left(252 \pm 25_{exp} \pm 19_{th}\right) MeV = \left(252 \pm 31\right) MeV \qquad (2.7\text{-}27)$$

where the first error is due to the experimental uncertainty in $f_{D_s}$ .

## *2.7.3 Measurements of $\Delta m_{B_d}$ and $\Delta m_{B_s}$*

The world average value of $\Delta m_{B_d}$ ,

$$\Delta m_{B_d} = \left(0.473 \pm 0.016\right) ps^{-1}, \qquad (2.7\text{-}28)$$

was computed by the LEP B Oscillations Working Group[123] using the results of the measurements carried out by CLEO and ARGUS (at the $\Upsilon(4S)$), CDF, SLD, and the LEP experiments . An accurate procedure was followed in order to take all common systematic uncertainties into account. The same method was applied to the $B_s^0$ oscillation data collected by ALEPH, DELPHI, OPAL, CDF and SLD. The fast oscillations $B_s^0 - \overline{B}_s^0$ have not been resolved yet, but a lower limit on $\Delta m_{B_s}$ ,

$$\Delta m_{B_s} > 14.3\, ps^{-1} \quad 95\%\, C.L., \qquad (2.7\text{-}29)$$

has been set using the amplitude method, of which a detailed theoretical account was given by Moser and Roussarie[124]. The time evolution of an initially pure $B_s^0$ state is described by the functions

$$P(B_s^0) = \frac{1}{2\tau_{B_s}} e^{-t/\tau_{B_s}} \left(1 + \cos \Delta m_{B_s} t\right)$$
$$P(\overline{B}_s^0) = \frac{1}{2\tau_{B_s}} e^{-t/\tau_{B_s}} \left(1 - \cos \Delta m_{B_s} t\right) \qquad (2.7\text{-}30)$$

They are the probability density functions representing the likelihood that at the time $t$ the meson is found in the $B_s^0$ and $\overline{B}_s^0$ states respectively; the amplitude $\mathcal{A}$ is defined as the coefficient of the oscillating term:

$$P(B_s^0,\overline{B}_s^0)=\frac{1}{2\tau_{B_s}}e^{-t/\tau_{B_s}}\left(1\pm\mathcal{A}\cos\Delta m_{B_s}t\right). \qquad (2.7\text{-}31)$$

The experiments measure $\mathcal{A}$ at each fixed value of $\Delta m_{B_s}$; the expected physical values are $\mathcal{A}=1$ if $\Delta m_{B_s}$ is the actual oscillation frequency, otherwise $\mathcal{A}=0$. The amplitude spectrum obtained from the combined results is shown in Figure 5. The lower limit $\Delta m_{B_s}=14.3\,ps^{-1}$ is the value at which $\mathcal{A}$ is incompatible with 1 at a 95% confidence level:

$$\mathcal{A}(\Delta m_{B_s})+1.645\,\sigma_{\mathcal{A}}(\Delta m_{B_s})=1. \qquad (2.7\text{-}32)$$

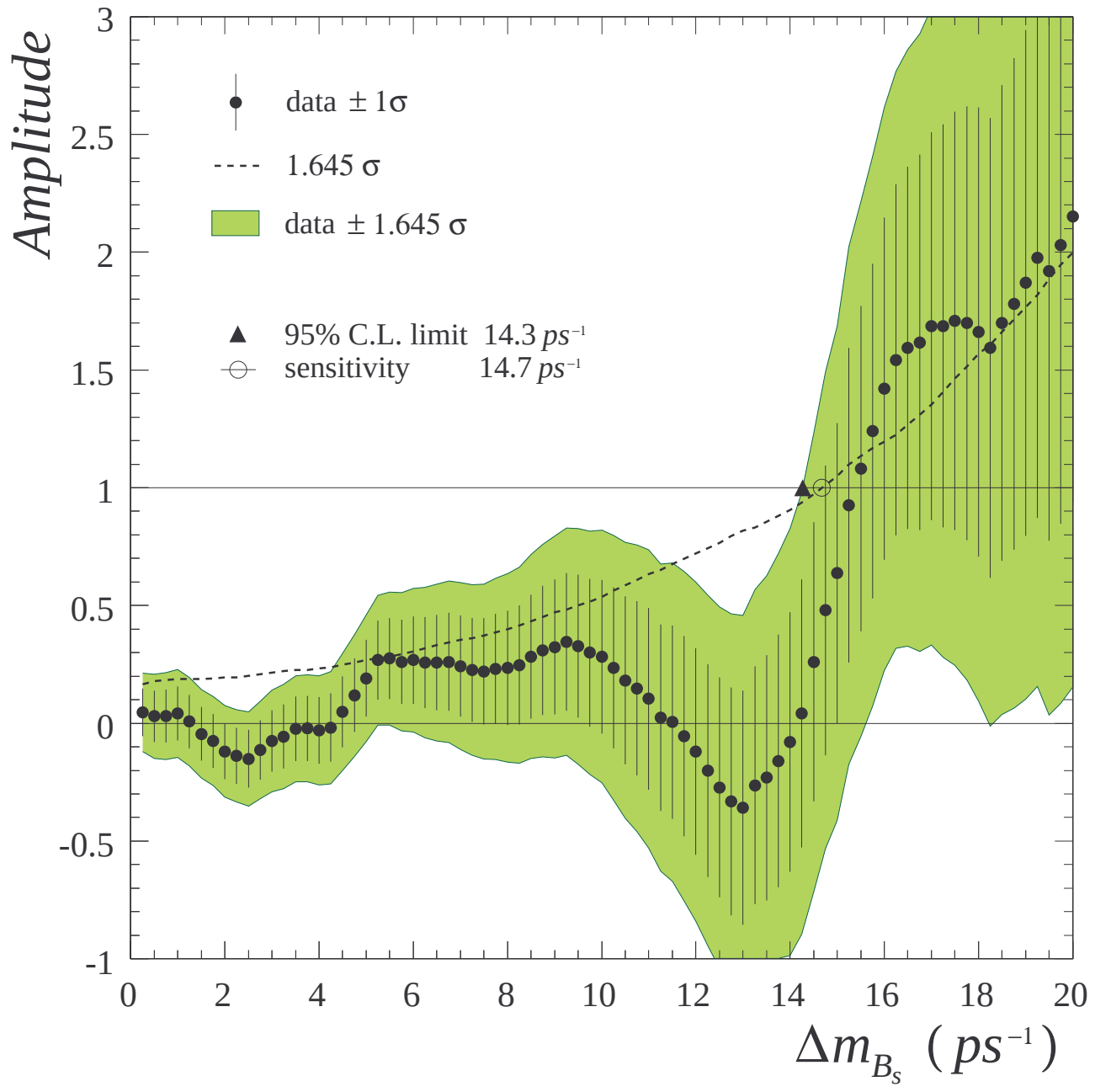

**Figure 5.** – *Spectrum of the $B_s^0-\overline{B}_s^0$ oscillation amplitude, determined by the LEP B Oscillations Working Group as a world average of the results achieved by ALEPH, DELPHI, OPAL, CDF and SLD. The values of the amplitude $\mathcal{A}$ and its error $\sigma_{\mathcal{A}}$ are shown as functions of the oscillation frequency $\Delta m_{B_s}$. The precision of the amplitude measurement decreases as $\Delta m_{B_s}$ increases; the sensitivity $\Delta m_{B_s}=14.7\,ps^{-1}$ is the value at which it is no longer possible to distinguish between $\mathcal{A}=1$ and $\mathcal{A}=0$ because of the increased margin of error. At $\Delta m_{B_s}=14.7\,ps^{-1}$ the dotted line, representing the graph of the function $1.645\sigma_{\mathcal{A}}(\Delta m_{B_s})$, intersects the line $\mathcal{A}=1$, so the values $\mathcal{A}=1$ and $\mathcal{A}=0$ cannot be considered as mutually incompatible at a confidence level higher than $P(x>\bar{x}+1.645\sigma_{\bar{x}})=5\%$. The coloured region represents the confidence interval $\pm1.645\sigma$ (90%); its upper margin is the graph of $\mathcal{A}(\Delta m_{B_s})+1.645\sigma_{\mathcal{A}}(\Delta m_{B_s})$; the value of this function does not exceed 1 until the frequency $\Delta m_{B_s}=14.3\,ps^{-1}$ is reached; therefore, for $\Delta m_{B_s}<14.3\,ps^{-1}$ the probability that $\mathcal{A}$ equals 1 is lower than 5%.*

Using the result in Eq. (2.7-26), the measurement (2.7-28) of the $\Delta m_{B_d}$ oscillation frequency can be translated into the following determination of the product $|V_{tb}V_{td}|$:

$$|V_{tb}V_{td}| = (7.9 \pm 1.4)\cdot 10^{-3} \qquad (2.7\text{-}33)$$

where the values of the parameters occurring in Eq. (2.7-7) have been anticipated from Table XV. This is actually a determination of $|V_{td}|$ itself, since the unitarity condition fixes $|V_{tb}|$ as $1+\mathcal{O}(\lambda^4)$. However, if the result of the direct measurement 2.7-1 is assumed for $|V_{tb}|$, one obtains

$$|V_{td}| = (8.2 \pm 2.0)\cdot 10^{-3}\,. \qquad (2.7\text{-}34)$$

The result 2.7-29 can similarly be used to set a lower limit on the value of $|V_{ts}|$. However, a more tight constraint is provided by the ratio of $\Delta m_{B_s}$ (Eq. 2.7-8) to $\Delta m_{B_d}$ (Eq. 2.7-7):

$$\frac{\Delta m_{B_s}}{\Delta m_{B_d}} = \frac{m_{B_s} f_{B_s}}{m_{B_d} f_{B_d}} \left|\frac{V_{ts}}{V_{td}}\right|^2 . \qquad (2.7\text{-}35)$$

The limit (2.7-29) correspond, with $f_{B_s}/f_{B_d} = 1.149 \pm 0.036$ (obtained by combining Eqs. 2.7-18,19,20) and $\Delta m_{B_d}$ given by Eq. (2.7-28), to

$$|V_{ts}/V_{td}| > 4.5 \quad 95\%\, C.L. \qquad (2.7\text{-}36)$$

## *2.7.4 CP violation in the neutral kaon system; $\varepsilon'_K/\varepsilon_K$*

The CP violation measured by the parameter $\varepsilon_K$ corresponds to an asymmetry in the extent to which the neutral kaons $K_0$ and $\overline{K}_0$ take part in the formation of the $|K_L\rangle$ and $|K_S\rangle$ mass eigenstates (called long-lived and short-lived states since $\tau_{K_L} \approx 600\ \tau_{K_S}$). In fact only the combinations in which $|K_0\rangle$ and $|\overline{K}_0\rangle$ have the same weight,

$$|K_\pm\rangle \propto |K^0\rangle \pm |\overline{K}^0\rangle, \quad CP|K_\pm\rangle = \pm|K_\pm\rangle, \qquad (2.7\text{-}37)$$

are CP eigenstates, while the $\Delta S$= 2 forces which cause the mixing favour the non-homogeneous configurations

$$|K_S\rangle \propto p|K^0\rangle + q|\overline{K}^0\rangle \propto |K_+\rangle + \frac{1-q/p}{1+q/p}|K_-\rangle \qquad (2.7\text{-}38)$$

$$\text{and}\quad |K_L\rangle \propto p|K^0\rangle - q|\overline{K}^0\rangle \propto |K_-\rangle + \frac{1-q/p}{1+q/p}|K_+\rangle . \qquad (2.7\text{-}39)$$

$\varepsilon_K$ is defined in terms of the amplitudes of the processes $K_{L,S} \to \pi^0\pi^0, \pi^+\pi^-$ ($CP|\pi\pi\rangle = +|\pi\pi\rangle$):

$$\eta_{+-} \equiv \frac{A(K_L \to \pi^+\pi^-)}{A(K_S \to \pi^+\pi^-)} \equiv \varepsilon_K + \varepsilon'_K\,, \qquad (2.7\text{-}40)$$

$$\eta_{00} \equiv \frac{A(K_L \to \pi^0\pi^0)}{A(K_S \to \pi^0\pi^0)} \equiv \varepsilon_K - 2\varepsilon'_K\,. \qquad (2.7\text{-}41)$$

Here $\varepsilon'_K$ parametrizes the 'direct' CP violation predicted by the Standard Model, which involves the dynamics of $\Delta S$= 1 decays and thus has not the same effect on the two channels. This asymmetry is accounted for by the different way in which $\eta_{+-}$ and $\eta_{00}$ are dependent on $\varepsilon'_K$. According to a class of models which refer to Wolfenstein's[125] 'superweak' theory, the observable CP asymmetry of neutral kaon decays should arise from the $\Delta S$= 2 sector alone: therefore $\varepsilon'_K$ would vanish, the equality $\eta_{+-} = \eta_{00} = \varepsilon_K$ would apply and the CP violation observed in $K_L \to \pi^0\pi^0, \pi^+\pi^-$ decays should be entirely attributed to a CP-even component already present in the initial state $K_L$ (thus $|q/p| \neq 1$; $q/p$ is not independent of the global phase-convention of the states $|K^0\rangle$ and $|\bar{K}^0\rangle$).

The superweak theory has been unequivocally contradicted by the recent preliminary results of the KteV and NA48 experiments: the new measurements of the ratio $\varepsilon'_K/\varepsilon_K \cong \mathrm{Re}(\varepsilon'_K/\varepsilon_K)$ are incompatible with the value $\varepsilon'_K/\varepsilon_K = 0$ at a virtually 100% confidence level (6.8σ and 2.5σ respectively):

$$\text{KteV:} \qquad \mathrm{Re}\frac{\varepsilon'_K}{\varepsilon_K} = \left(28.0 \pm 3.0_{stat} \pm 2.6_{syst} \pm 1.0_{MC\,stat}\right)\cdot 10^{-4} = (28.0 \pm 4.1)\cdot 10^{-4}$$

$$\text{NA48:} \qquad \mathrm{Re}\frac{\varepsilon'_K}{\varepsilon_K} = \left(18.5 \pm 4.5_{stat} \pm 5.8_{syst}\right)\cdot 10^{-4} = (18.5 \pm 7.3)\cdot 10^{-4} \tag{2.7-42}$$

| Experiment | $\mathrm{Re}(\varepsilon'_K/\varepsilon_K)$ |
|---|---|
| E731[126] | $(7.4 \pm 5.9)\cdot 10^{-4}$ |
| NA31[127] | $(23.0 \pm 6.5)\cdot 10^{-4}$ |
| KteV[128] | $(28.0 \pm 4.1)\cdot 10^{-4}$ |
| NA48[129] | $(18.5 \pm 7.3)\cdot 10^{-4}$ |
| Average | $(21.2 \pm 4.6)\cdot 10^{-4}$<br>$\chi^2/3 = 2.8,\ s = 1.7$ |

**Table XI.** – *Experimental determinations of the parameter* $\mathrm{Re}(\varepsilon'_K/\varepsilon_K)$ *which quantifies the direct CP violation in the neutral kaon system. The scale factor s =1.7 has been applied to the error in the average because of the slight discrepancy between the measurements.*

The margin of certainty reduces slightly if the world average (see Table XI) is assumed as the best value of $\mathrm{Re}(\varepsilon'_K/\varepsilon_K)$:

$$\mathrm{Re}\frac{\varepsilon'_K}{\varepsilon_K} = (21.2 \pm 4.6)\cdot 10^{-4}. \tag{2.7-43}$$

The average is, however, incompatible with the lack of direct CP violation at a 99.9999% (4.6 σ) confidence level.

It has often been argued[130] that a value of $\varepsilon'_K/\varepsilon_K$ greater than $2\cdot 10^{-3}$ cannot be accounted for within the present model of CP violation based on a single complex phase in the CKM matrix. In actual fact, the evidence for direct CP violation is the only straightforward, unambiguous conclusion which can be drawn from the experimental result, while, at present, large theoretical uncertainties affect the Standard Model computations of the magnitude of the CP-violating effect. Most computed values (see Table XII) are consistent with a lower value than the measured one. However, they are strongly dependent on the strange quark mass (roughly $\varepsilon'_K/\varepsilon_K \propto 1/m_s^2$), of which the most recent estimates, based on lattice-QCD calculations, give much lower values than before. The Standard Model prediction is summarized by the following expression:[131]

$$\frac{\varepsilon'_K}{\varepsilon_K} = \mathrm{Im} V_{td} V_{ts}^* \left\{ -1.35 + \left( \frac{150\,MeV}{m_s(m_c)} \right)^2 \left[ 1.1 \cdot S \cdot B_6^{(1/2)} + (1.0 - 0.67 \cdot S) B_8^{(3/2)} \right] \right\}. \quad (2.7\text{-}44)$$

Here $S$ represents the short-distance QCD effects, which have been calculated up to the next-to-leading order:[140,132]

$$6.5 \le S \le 8.5; \quad (2.7\text{-}45)$$

$B_6^{(1/2)}$ and $B_8^{(3/2)}$ are the hadronic matrix elements of the operators $O_6$ and $O_8$ which contribute to the effective Hamiltonian; they are evaluated in the $K^0 \to (\pi\pi)_{I=0}$ ($\Delta I$= 1/2) and $K^0 \to (\pi\pi)_{I=2}$ ($\Delta I$= 3/2) transitions respectively:[133,134,135]

$$0.8 \le B_6^{(1/2)} \le 1.3, \qquad 0.6 \le B_8^{(3/2)} \le 1.0. \quad (2.7\text{-}46)$$

The term containing the CKM matrix elements is equal to

$$\mathrm{Im} V_{td} V_{ts}^* \cong A^2 \lambda^5 \overline{\eta} \cong 1.24 \cdot 10^{-4} \quad (2.7\text{-}47)$$

at the leading order in $\lambda$. The $s$ quark mass has to be evaluated in the $\mu \cong m_c \cong 1.3\,GeV$ scale. If the whole spectrum of the available theoretical predictions for $m_s$ were taken into account (the range $60\,MeV \le m_s \le 170\,MeV$, for $\mu \cong 2\,GeV$, is quoted by the PDG[12]), a rather indefinite estimate of $\varepsilon'_K/\varepsilon_K$ would be obtained. On the other hand, if an 'average' ($\ge 120\,MeV$) value is assumed for $m_s$, Eq. (2.7-44) and the measurement of $\varepsilon'_K/\varepsilon_K$ cannot be made compatible. The lattice QCD results show a much better agreement with the experimental data. The range of the *quenched* results is[136]

| | |
|---|---|
| $-2.1 \cdot 10^{-4} \le \frac{\varepsilon'_K}{\varepsilon_K} \le 13.3 \cdot 10^{-4}$ | [137] |
| $\frac{\varepsilon'_K}{\varepsilon_K} = (4.6 \pm 3.0 \pm 0.4) \cdot 10^{-4}$ | [138] |
| $\frac{\varepsilon'_K}{\varepsilon_K} = \left(17^{+14}_{-10}\right) \cdot 10^{-4}$ | [139] |
| $\frac{\varepsilon'_K}{\varepsilon_K} = \begin{cases} (5.3 \pm 3.8) \cdot 10^{-4} \text{ using } m_s(m_c) = (150 \pm 20)\,MeV \\ (8.5 \pm 5.9) \cdot 10^{-4} \text{ using } m_s(m_c) = (125 \pm 20)\,MeV \end{cases}$ | [140] |
| $\frac{\varepsilon'_K}{\varepsilon_K} = (15.0 \pm 4.8) \cdot 10^{-4}$ | [141] |

**Table XII.** – *Expected values of the parameter $\varepsilon'_K/\varepsilon_K$ according to some Standard Model calculations.*

$$100\ MeV \le m_s(2GeV)_{quenched} \le 150\ MeV; \tag{2.7-48}$$

if the quenching effects are taken into account, there should be a 40% reduction:[142]

$$60\ MeV \le m_s(2GeV) \le 90\ MeV; \tag{2.7-49}$$

the corresponding values in the $\mu = m_c$ scale are

$$70\ MeV \le m_s(m_c) \le 100\ MeV. \tag{2.7-50}$$

The following range of allowed values is obtained from Eqs. (2.7-44,45,46,47):

$$0.7 \cdot 10^{-3} < \mathrm{Re}\frac{\varepsilon'_K}{\varepsilon_K} < 4.3 \cdot 10^{-3}, \tag{2.7-51}$$

which is fully compatible with the average experimental value in Eq. (2.7-43). $\varepsilon'_K/\varepsilon_K$ is directly proportional to the imaginary part of the CKM matrix (Eq. 2.7-47); however, due to the large uncertainty in its expected value, the use of this measurement as a constraint would have a completely negligible effect.

### *2.7.5 The $|\varepsilon_K|$ constraint*

The PDG[12] averages for the CP violation parameters of the kaon system are

$$|\eta_{+-}| = (2.284 \pm 0.018) \cdot 10^{-3}, \tag{2.7-52}$$

$$|\eta_{00}/\eta_{+-}| = 0.9930 \pm 0.020; \tag{2.7-53}$$

their product is

$$|\eta_{00}| = (2.268 \pm 0.018) \cdot 10^{-3}. \tag{2.7-54}$$

Inverting the relations (2.7-40) and (2.7-41) to extract the modulus of $\varepsilon_K$, one gets

$$|\varepsilon_K| = \frac{|\eta_{+-}|}{\left|1 + \frac{\varepsilon'_K}{\varepsilon_K}\right|} \cong \frac{|\eta_{+-}|}{1 + \mathrm{Re}\frac{\varepsilon'_K}{\varepsilon_K}}, \tag{2.7-55}$$

$$|\varepsilon_K| = \frac{|\eta_{00}|}{\left|1 - 2\frac{\varepsilon'_K}{\varepsilon_K}\right|} \cong \frac{|\eta_{00}|}{1 - 2\,\mathrm{Re}\frac{\varepsilon'_K}{\varepsilon_K}} \tag{2.7-56}$$

and using Eq. (2.7-43) together with either (2.7-52) or (2.7-54), from the preceding expressions the measurement

$$|\varepsilon_K| = (2.279 \pm 0.018) \cdot 10^{-3} \tag{2.7-57}$$

is obtained. The contribution of the $c$ quark loop cannot be neglected in the kaon system:[96]

$$\varepsilon_K = e^{i\pi/4} \frac{G_F^2 m_W^2 m_K}{12\sqrt{2}\pi^2 \Delta m_K} f_K^2 B_K \left(\mathrm{Im} M_{12} + 2\xi\,\mathrm{Re} M_{12}\right) \tag{2.7-58}$$

with
$$M_{12} = \eta_{cc} S(x_c)\left(V_{cs}V_{cd}^*\right)^2 + \eta_{tt} S(x_t)\left(V_{ts}V_{td}^*\right)^2 + 2\eta_{ct} S(x_c, x_t) V_{cs}V_{cd}^* V_{ts}V_{td}^*. \tag{2.7-59}$$

$M_{12}$ is proportional to the non-diagonal element of the neutral kaon mass matrix, which represents $K^0 - \bar{K}^0$ mixing: $M_{12}^* \propto \langle \bar{K}^0 | \mathcal{H}_{eff}(\Delta S = 2) | K^0 \rangle$; $\xi$ is the ratio

$$\xi = \frac{\mathrm{Im}\, A[K \to (\pi\pi)_{I=0}]}{\mathrm{Re}\, A[K \to (\pi\pi)_{I=0}]}, \qquad (2.7\text{-}60)$$

$\Delta m_K$ the mass difference between the $K_L$ and $K_S$ autostates; the QCD corrections to the Inami-Lim functions are the factors

$$\eta_{cc} = 1.38 \pm 0.53, \qquad \eta_{tt} = 0.574 \pm 0.004, \qquad \eta_{ct} = 0.47 \pm 0.04, \qquad (2.7\text{-}61)$$

which have been calculated up to the next-to-leading order[95,96]. The decay constant $f_K$ and the bag parameter $B_K$ determine the magnitude of the hadronic matrix element $\langle \overline{K}^0 | [\bar{s}\gamma^\mu (1-\gamma^5) d]^2 | K^0 \rangle \propto m_K f_K^2 B_K$. A precise value of $f_K$ can be deduced from the measurement of the leptonic decay rate of kaons[12]:

$$f_K = (159.8 \pm 1.5) MeV. \qquad (2.7\text{-}62)$$

On the contrary, the parameter $B_K$ cannot be measured. The two most recent lattice QCD computations give perfectly compatible results:

Kilcup *et al.*[143] $\quad B_K(2GeV) = 0.62 \pm 0.02_{stat} \pm 0.02_{syst}$, (2.7-63)

JLQCD[144] $\quad B_K(2GeV) = 0.628 \pm 0.042$. (2.7-64)

JLQCD's datum is the result of an extended simulation, which included a detailed analysis of the systematic effects of discretization and the renormalization scheme conversion. The scale-independent constant $B_K$ entering into the expression 2.7-58 has to be determined by applying a suitable transformation to the value computed at $\mu = 2GeV$. This can be done either with reference to the physical situation in which three light dynamical quarks ($n_f = 3$) exist, or within the quenched approximation ($n_f = 0$) already used to deduce the results 2.7-63 and 2.7-64. Allowing for both possibilities[133], which lead to slightly different results, the relation

$$\frac{B_K}{B_K(2GeV)} = \begin{cases} 1.34 \\ 1.38 \quad (quenched) \end{cases} = 1.36 \pm 0.02 \qquad (2.7\text{-}65)$$

will be used. If the contribution of the dynamic quarks is included (unquenching) and the deviation from the SU(3)-symmetry assumed for light quarks ($m_u = m_d = m_s$) is taken into account, the value of $B_K$ should increase by a further 10% amount:[145]

$$\frac{B_K}{(B_K)_{quenched}} = 1.05 \pm 0.02, \qquad (2.7\text{-}66)$$

$$\frac{B_K}{(B_K)_{SU(3)}} = 1.04 - 1.05. \qquad (2.7\text{-}67)$$

When Eq. (2.7-64) is multiplied by Eq. (2.7-65) and the corrective factor $B_K / (B_K)_{quenched+SU(3)} = 1.10 \pm 0.05$ is applied (with a conservative increase in the margin of uncertainty in order to allow for a possible systematic error in the evaluation of the unquenching correction[145]) the result

$$B_K = 0.94 \pm 0.08 \qquad (2.7\text{-}68)$$

is obtained.

The term proportional to the real part of the box diagram will be neglected in the expression of $\varepsilon_K$, since it is numerically insignificant (it has been estimated[96] as a correction not exceeding 2% compared to the size of the term proportional to the imaginary part). This approximation is justified on account of the considerable theoretical uncertainty in $B_K$, but the whole expression will have to be considered (and the parameter $\xi$ will also have to be evaluated) as soon as the error in the determination of $B_K$ is reduced. For the present, an additional 2% error is attributed to the measurement of $|\varepsilon_K|$.

The following constraint on the CKM matrix elements is provided by the present experimental and theoretical information about CP violation in the $K^0-\overline{K}^0$ system:

$$|\varepsilon_K| = \frac{G_F^2 m_W^2 m_K f_K^2}{12\sqrt{2}\pi^2 \Delta m_K} B_K \left\{\eta_{cc} S(x_c)\,\mathrm{Im}\left[\left(V_{cs}V_{cd}^*\right)^2\right] + \eta_{tt} S(x_t)\,\mathrm{Im}\left[\left(V_{ts}V_{td}^*\right)^2\right] + 2\eta_{ct} S(x_c, x_t)\,\mathrm{Im}\left(V_{cs}V_{cd}^* V_{ts}V_{td}^*\right)\right\} \quad (2.7\text{-}69)$$

with

$$|\varepsilon_K| = (2.28 \pm 0.05)\cdot 10^{-3}. \quad (2.7\text{-}70)$$

## 2.7.6 *$b \to s\gamma$ penguin decays*

The first observation of the $b\to s\gamma$ electromagnetic penguin decays (see Figure 6), performed by CLEO[146], consisted in the identification of the exclusive channel $B \to K^*(892)\gamma$. The measured decay rate of this process cannot be used to extract accurate information about the weak coupling of the quarks involved in the interaction, because of major theoretical uncertainties in the description of the hadronization process. Recently, CLEO and ALEPH have carried out independent measurements of the inclusive branching ratio $\mathcal{Br}(b \to X_s\gamma)$, providing quite a reliable way of constraining the CKM matrix elements $V_{ts}$ and $V_{tb}$. The Standard Model prediction, normalized to the background processes $b \to c\ell\overline{\nu}_\ell$, has been calculated up to the next-to-leading order[147], assuming $|V_{ts}||V_{tb}|/|V_{cb}| = 0.976 \pm 0.010$ and $\mathcal{Br}(b \to X_c\ell\overline{\nu}_\ell) = (10.4 \pm 0.4)\%$ (the values are taken from [67]):

$$\mathcal{Br}(b \to s\gamma) = (3.28 \pm 0.33)\cdot 10^{-4}. \quad (2.7\text{-}71)$$

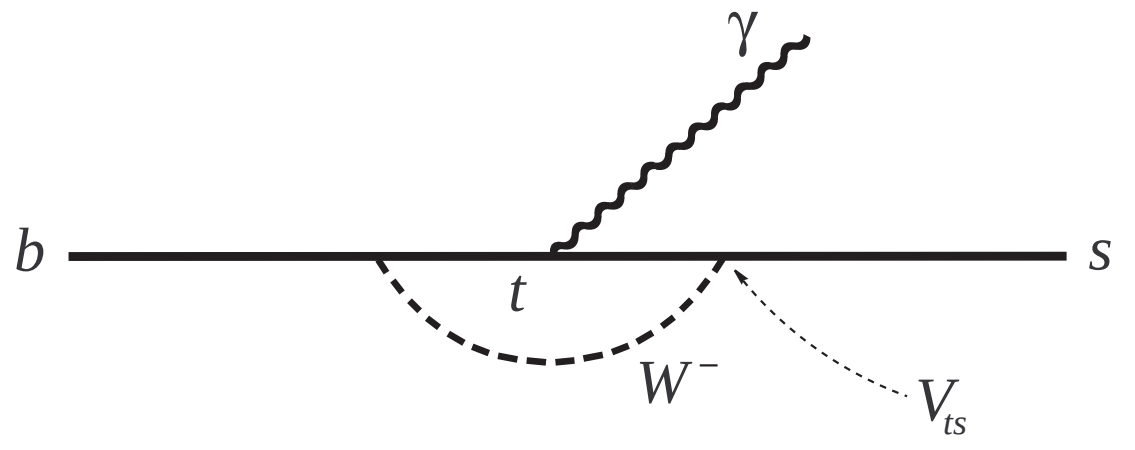

**Figure 6.** – *Example of electromagnetic penguin diagram describing the process $b \to s\gamma$.*

Since

$$\frac{\mathcal{Br}(b \to s\gamma)}{\mathcal{Br}(b \to c\ell\overline{\nu}_\ell)} \propto \frac{|V_{ts}|^2 |V_{tb}|^2}{|V_{cb}|^2}, \quad (2.7\text{-}72)$$

the following expression relates the CKM matrix elements to the measured quantities:

$$\frac{\mathcal{B}r(b \to X_s\gamma)}{(3.28 \pm 0.30)\cdot 10^{-4}} = \frac{\mathcal{B}r(b \to X_c\ell\bar{\nu}_\ell)}{0.104\cdot(0.976)^2}\cdot\frac{|V_{ts}|^2|V_{tb}|^2}{|V_{cb}|^2}; \qquad (2.7\text{-}73)$$

the errors coming from $\mathcal{B}r(b \to X_c\ell\bar{\nu}_\ell)$ and $|V_{ts}||V_{tb}|/|V_{cb}|$ have been subtracted from the total error. CLEO and ALEPH have reported the following results:

$$\text{CLEO}^{148} \quad \mathcal{B}r(B \to X_s\gamma) = (3.15 \pm 0.35_{stat} \pm 0.41_{syst})\cdot 10^{-4}, \qquad (2.7\text{-}74)$$

$$\text{ALEPH}^{149} \quad \mathcal{B}r(b \to X_s\gamma) = (3.11 \pm 0.80_{stat} \pm 0.72_{syst})\cdot 10^{-4}. \qquad (2.7\text{-}75)$$

The different way in which *b* quarks are produced in these experiments (at the $\Upsilon(4S)$ resonance and from *Z* decay) should be taken into account choosing the corresponding measured value for $\mathcal{B}r(b \to X_c\ell\bar{\nu}_\ell)$. However, since the CLEO measurement is the main contribution to the weighted average

$$\mathcal{B}r(b \to X_s\gamma) = (3.14 \pm 0.48)\cdot 10^{-4}, \qquad (2.7\text{-}76)$$

it will be assumed that

$$\mathcal{B}r(b \to X\ell\bar{\nu}_\ell) = \mathcal{B}r(B \to X\ell\bar{\nu}_\ell)_{\Upsilon(4S)} = (10.45 \pm 0.21)\%\ ^{12}, \qquad (2.7\text{-}77)$$

therefore, after subtracting the $b \to X_u\ell\bar{\nu}_\ell$ contribution (Eq. 2.4-2),

$$\mathcal{B}r(b \to X_c\ell\bar{\nu}_\ell) = (10.25 \pm 0.23)\%. \qquad (2.7\text{-}78)$$

By substituting the values 2.7-76 and 2.7-78 into Eq. (2.7-73), the result

$$\frac{|V_{ts}|^2|V_{tb}|^2}{|V_{cb}|^2} = 0.93 \pm 0.14 \pm 0.08 = 0.93 \pm 0.17, \qquad (2.7\text{-}79)$$

is obtained. The first error is the uncertainty in the measurement of the branching ratio; the second error is associated with its theoretical computation. This corresponds, with $|V_{tb}| \approx 1$ and $|V_{cb}|$ given by Eq. (2.5-10), to the following determination of $|V_{ts}|$:

$$|V_{ts}| = (38.1 \pm 3.8)\cdot 10^{-3}. \qquad (2.7\text{-}80)$$

## 2.8 Summary

The best values obtained so far in the present work for the CKM matrix elements are listed in Table XIII, where they are compared to those quoted by the Particle Data Group (updated in January, 1998).

In most cases, a reduction in the experimental uncertainties has been obtained by taking into account the most recent measurements. For example, data from deep inelastic scattering of neutrinos on nucleons and from hadronic $W^{\pm}$ decays have been included in the $|V_{cs}|$ average, leading to a reduction by a factor of 7 in the final uncertainty. A significant improvement has also been achieved in the determination of $|V_{ub}/V_{cb}|$, since new, independent measurements of $|V_{ub}|$ and $|V_{ub}/V_{cb}|$ have made it possible to discriminate between the disagreeing theoretical predictions which were used in the end-point inclusive analyses of non-charmed $B$ decays.

In other cases, the determinations of the matrix elements have not changed considerably; however, all the values obtained are the result of a detailed re-analysis of the available information, in which the relevant experimental parameters have been updated to the most recent world averages.

| | PDG'98 | this analysis |
|---|---|---|
| $\|V_{ud}\|$ | $0.9740 \pm 0.0010$ | $0.9743 \pm 0.0008$ |
| $\|V_{us}\|$ | $0.2196 \pm 0.0023$ | $0.2200 \pm 0.0025$ |
| $\|V_{ub}\|$ | $(3.3 \pm 0.8)\cdot 10^{-3}$ | $(3.6 \pm 0.5)\cdot 10^{-3}$ |
| $\|V_{ub}/V_{cb}\|$ | $0.080 \pm 0.020$ | $0.090 \pm 0.008$ |
| $\|V_{cd}\|$ | $0.224 \pm 0.016$ | $0.225^{+0.013}_{-0.011}$ |
| $\|V_{cs}\|$ | $1.04 \pm 0.16$ | $0.996 \pm 0.024$ |
| $\|V_{cb}\|$ | $(39.5 \pm 1.7)\cdot 10^{-3}$ | $(39.5 \pm 1.7)\cdot 10^{-3}$ |
| $\|V_{tb}\|$ | $0.99 \pm 0.15$ | $0.96^{+0.16}_{-0.12}$ |
| $\|V_{tb}V_{td}\|$ (from $B_d - \bar{B}_d$ oscillation) | $(8.4 \pm 1.8)\cdot 10^{-3}$ | $(7.9 \pm 1.4)\cdot 10^{-3}$ |
| $\|V_{ts}/V_{td}\|$ (from $B_s - \bar{B}_s$ oscillation) | $> 3.7$ 95% *C.L.* | $> 4.5$ 95% *C.L.* |
| $\|V_{ts}V_{tb}/V_{cb}\|$ (from $b \to s\gamma$) | $1.1 \pm 0.43$ | $0.96 \pm 0.09$ |

**Table XIII.** – *Comparison between the values of the CKM matrix elements obtained in the present analysis and those from the Review of Particle Physics (January '98).*

# 3 An up-to-date profile of the unitarity triangle

As far as possible, all the experimental information collected in the previous Section has been treated in such a way that any correlation among the single determinations has been avoided or reduced to a negligible level. As a result, a set of *independent* constraints on the CKM matrix elements has been outlined. In this Section, we refine the determination of the CKM matrix and obtain estimates for other parameters of physical interest by imposing all these constraints simultaneously and requiring explicitly that the three-family unitarity condition of the matrix be satisfied. This further constraint is imposed (together with the removal of the non-physical complex phases) by expressing the matrix elements in terms of a four-variable parametrization (either Wolfenstein's or the canonical parametrization). The number of independent constraints which is needed to make the problem completely determined is then exceeded by a wide margin. Therefore, the precision in the values of the CKM matrix elements can be improved considerably with respect to those obtained from direct measurements. At the same time, the unitarity triangle, which represents one of the unitarity relations on the complex plane, can be determined, and the magnitudes of the CP asymmetries in $B$ decays can be evaluated.

Two different methods are followed independently for the overall determination of the CKM matrix and the unitarity triangle. The first procedure, a $\chi^2$-minimization, is discussed in Sect. 3.1 and the results are reported in Sect. 3.3. A number of additional trials is also performed, in which the principal constraints are removed in turn to show how they affect the results. In Sect. 3.2 the information which is effective in constraining the position of the vertex of the unitarity triangle is represented geometrically on the $(\bar{\rho},\bar{\eta})$ plane using Wolfenstein's parametrization. An account of the second method, based on Bayesian statistics, and the results obtained from its application are presented in Sect. 3.4. Finally, in Sect. 3.5 the impact of the forthcoming experiments at the $B$ factories is evaluated, and the importance they have as essential instruments for the verification of the Standard Model is emphasized.

## *3.1 Description of the fit procedure*

The information used in the present analysis to constrain the CKM matrix is summarized in Table XIV. The matrix elements have been expressed in terms of Wolfenstein's unitary parametrization. Each term has been calculated up to the fourth order in $\lambda$. For this purpose, the expansion of $V_{CKM}$ has been extended further than in Eq. (1.1-7), considering the real and the imaginary parts separately. While it is true that a fourth-order computation is not justified by the present degree of precision of the data, future improvements in the theoretical and experimental uncertainties will increase the sensitivity of the results to the accuracy of these computations. For example, the fourth-order correction is already comparable to the magnitude of the error in the case of $|V_{ud}|$ and it would not even be possible to use the direct measurement of $|V_{tb}|$ as a constraint if the expansion were truncated at the second order.

The maximum-likelihood estimates of the parameters $\lambda$, $A$, $\rho$ and $\eta$ result from the minimization of the function

$$\chi^2(\lambda, A, \rho, \eta) = \sum_i \frac{\left(O_i(\lambda, A, \rho, \eta) - \overline{O}_i\right)^2}{\sigma_{O_i}^2}. \tag{3.1-1}$$

In each term, the unknown member $O_i(\lambda, A, \rho, \eta)$ of a constraint equation is compared with the measured value $\overline{O}_i$ within the error $\sigma_{O_i}$ corresponding to one standard deviation (all the input data of the fit are assumed to be normally distributed).

| Term | Value | Expression | Eq./Ref. |
|---|---|---|---|
| $\lvert V_{ud}\rvert$ | $0.9743 \pm 0.0008$ | $1-\frac{\lambda^2}{2}-\frac{\lambda^4}{8}$ | (2.1-25) |
| $\lvert V_{us}\rvert$ | $0.2200 \pm 0.0025$ | $\lambda$ | (2.2-16) |
| $\lvert V_{ub}\rvert$ | $(3.6 \pm 0.5)\cdot 10^{-3}$ | $A\lambda^3\sqrt{\rho^2+\eta^2}$ | (2.4-4) |
| $\lvert V_{ub}/V_{cb}\rvert$ | $0.090 \pm 0.008$ | $\lambda\sqrt{\rho^2+\eta^2}$ | (2.6-4) |
| $\lvert V_{cd}\rvert$ | $0.225^{+0.013}_{-0.011}$ | $\lambda\left[1-A^2\frac{\lambda^4}{2}(1-2\rho)\right]$ | (2.3-32) |
| $\lvert V_{cs}\rvert$ | $0.996 \pm 0.024$ | $1-\frac{\lambda^2}{2}-\lambda^4\left(\frac{1}{8}+\frac{A^2}{2}\right)$ | (2.3-33) |
| $\lvert V_{cb}\rvert$ | $(39.5 \pm 1.7)\cdot 10^{-3}$ | $A\lambda^2$ | (2.5-10) |
| $\lvert V_{tb}\rvert$ | $0.96^{+0.16}_{-0.12}$ | $1-A^2\frac{\lambda^4}{2}$ | (2.7-1) |
| $\frac{\lvert V_{ts}\rvert^2\lvert V_{tb}\rvert^2}{\lvert V_{cb}\rvert^2}$ | $0.93 \pm 0.17$ | $1-\lambda^2(1-2\rho)-\lambda^4\left[A^2-\eta^2+\rho(1-\rho)\right]$ | (2.7-79) |
| $\Delta m_{B_d}$ | $(0.473 \pm 0.016)ps^{-1}$ | $\Delta m_{B_d}=\frac{G_F^2}{6\pi^2}m_W^2 m_{B_d}\left(\frac{f_{B_d}}{f_{D_s}}\right)^2(f_{D_s})^2 B_B\eta_B S\left(\frac{m_t^2}{m_W^2}\right)A^2\lambda^6\left\{(1-\rho)^2+\eta^2-\lambda^2(\rho^2+\eta^2-\rho)+ +\lambda^4\left[\left(\frac{1}{4}-A^2\right)(\rho^2+\eta^2)+A^2(2\rho-1)\right]\right\}$ | (2.7-7,28) |
| $\Delta m_{B_s}$ | $>14.3ps^{-1}\ 95\%C.L.$ | $\Delta m_{B_s}=\frac{G_F^2}{6\pi^2}m_W^2 m_{B_s}\left(\frac{f_{B_s}}{f_{D_s}}\right)^2(f_{D_s})^2 B_B\eta_B S\left(\frac{m_t^2}{m_W^2}\right)A^2\lambda^4\left\{1-\lambda^2(1-2\rho)-\lambda^4\left[A^2-\eta^2+\rho(1-\rho)\right]\right\}$ | (2.7-8,29) |
| $\lvert\varepsilon_K\rvert$ | $(2.28 \pm 0.05)\cdot 10^{-3}$ | $\lvert\varepsilon_K\rvert=\frac{G_F^2 m_W^2 m_K f_K^2}{6\sqrt{2}\pi^2\Delta m_K}B_K A^2\lambda^6\eta\left\{-\eta_{cc}S\left(\frac{m_c^2}{m_W^2}\right)\left[1-\frac{\lambda^2}{2}-\frac{\lambda^4}{8}\left(4A^2(3-2\rho)+1\right)\right]+ +\eta_{tt}S\left(\frac{m_t^2}{m_W^2}\right)A^2\lambda^4\left[1-\rho-\lambda^2\left(\rho^2+\eta^2-2\rho+\frac{1}{2}\right)+\lambda^4\left[A^2\left(\rho-\frac{1}{2}\right)+\frac{\eta^2}{2}+\frac{1}{8}(4\rho^2-1)\right]\right]+ +\eta_{ct}S\left(\frac{m_c^2}{m_W^2},\frac{m_t^2}{m_W^2}\right)\left[1-\frac{\lambda^2}{2}-\lambda^4\left[A^2\left(\frac{5}{2}-2\rho\right)+\frac{1}{8}\right]\right]\right\}$ | (2.7-69,70) |
| $\sin 2\beta$ | $0.79^{+0.41}_{-0.44}$ | $\frac{2\overline{\eta}(1-\overline{\rho})}{(1-\overline{\rho})^2+\overline{\eta}^2}$ | [13] |

**Table XIV.** – *Constraints on the CKM matrix in Wolfenstein's parametrization. Each single term in the expressions has been calculated up to* $\mathcal{O}(\lambda^4)$ *corrections.*

A number of theoretical or experimental quantities, such as $B_B$, $B_K$, $f_{D_s}$ and $m_t$, on which the expressions of $\Delta m_{B_d}$, $\Delta m_{B_s}$ and $\lvert\varepsilon_K\rvert$ are dependent, are not known precisely enough to be regarded as constants. Therefore, they are handled as variables of the minimization together with $\lambda$, $A$, $\rho$ and $\eta$. Their measurements or theoretical estimates are used as additional constraints, which are imposed by adding further terms to the total $\chi^2$, such as

$$\chi^2_{f_{D_s}} = \frac{\left(f_{D_s} - \overline{f_{D_s}}\right)^2}{\sigma^2_{f_{D_s}}}. \qquad (3.1\text{-}2)$$

The constrained variables of the fit and the parameters which, because of the comparatively small errors (< 1%), are chosen to be fixed at their central values are listed in Table XV.

By allowing some 'known' parameters to vary, a strong coupling between the equations for $\Delta m_{B_d}$ and $\Delta m_{B_s}$ is induced. Therefore, $B_B$, $f_{D_s}$, $f_{B_d}/f_{D_s}$ and $f_{B_s}/f_{D_s}$, instead of $f_{B_d}\sqrt{B_{B_d}}$ and $f_{B_s}\sqrt{B_{B_s}}$, will be used as parameters of the fit when the constraints $\Delta m_{B_d}$ and $\Delta m_{B_s}$ (2.7-7 and 2.7-8) are applied simultaneously: in this way, the correlation arising from the common use of the measurement of $f_{D_s}$ is taken into account. $f_{B_d}/f_{D_s}$ and $f_{B_s}/f_{D_s}$ will be further constrained by means of Eq. (2.7-20) (the determination of $f_{B_s}/f_{B_d}$ is regarded as an independent piece of information):

$$\frac{f_{B_s}/f_{D_s}}{f_{B_d}/f_{D_s}} = 1.15 \pm 0.04 . \qquad (3.1\text{-}3)$$

| Variables | Eq./Ref. | Constants | Ref. |
|---|---|---|---|
| $f_{B_d}/f_{D_s} = 0.76 \pm 0.04$ | (2.7-18) | $\eta_{tt} = 0.574 \pm 0.004$ | [95,96] |
| $f_{B_s}/f_{D_s} = 0.87 \pm 0.04$ | (2.7-19) | $G_F = (1.16639 \pm 0.00001)\cdot 10^{-5} GeV^{-2}$ | [12] |
| $f_{D_s} = (0.254 \pm 0.025) GeV$ | (2.7-17) | $m_W = (80.41 \pm 0.10) GeV$ | [12] |
| $B_B (= B_{B_d} = B_{B_s}) = 1.30 \pm 0.15$ | (2.7-25) | $m_{B_d} = (5.2792 \pm 0.0018) GeV$ | [12] |
| $\eta_B = 0.55 \pm 0.01$ | [95,96] | $m_{B_s} = (5.3692 \pm 0.0020) GeV$ | [12] |
| $\overline{m}_t = (166 \pm 5) GeV$ | [12] | $\Delta m_K = (3.489 \pm 0.009)\cdot 10^{-15} GeV$ | [12] |
| $\overline{m}_c = (1.25 \pm 0.10) GeV$ | [12] | $m_K = (0.497672 \pm 0.000031) GeV$ | [12] |
| $\eta_{cc} = 1.38 \pm 0.53$ | [95,96] | $f_K = (0.1598 \pm 0.0015) GeV$ | [12] |
| $\eta_{ct} = 0.47 \pm 0.04$ | [95,96] | | |
| $B_K = 0.94 \pm 0.08$ | (2.7-68) | | |

**Table XV.** – *Variable and constant parameters of the fit.*

The $\Delta m_{B_s}$ constraint is imposed following the same pattern of the procedure adopted in the search for $B_s^0 - \overline{B}_s^0$ oscillations (see Sect 2.7.3). The variable $\mathcal{A}$ is expected to be distributed normally with mean value equal to 1 in the presence of oscillations. For each value of $\Delta m_{B_s}$, the value assumed by $\mathcal{A}$ is determined using the data represented in Figure 5 and the term

$$\left[\frac{\mathcal{A}(\Delta m_{B_s}) - 1}{\sigma_{\mathcal{A}}(\Delta m_{B_s})}\right]^2 \qquad (3.1\text{-}4)$$

is added to the total $\chi^2$. $\Delta m_{B_s}$ is in turn determined by the current values of the parameters of the minimization using equation (2.7-8). Therefore, the maximum probability ($P \propto e^{-\chi^2/2}$) is attributed to the values of the parameters that yield measurements of $\mathcal{A}$ compatible with 1. Following this method, the information contained in the whole amplitude spectrum become effective, while the single measurement $\mathcal{A}\left(\Delta m_{B_s} = 14.3\, ps^{-1}\right)$ was used to deduce the lower limit (2.7-29).

The complete expression of the total $\chi^2$ is given below. The central values of the experimental or theoretical determinations are indicated with a superscript bar. Gaussian errors corresponding to one standard deviation are assigned to them. The remaining symbols represent the unknown terms of the constraint equations calculated as functions of the parameters.

$$
\begin{aligned}
\chi^2 = & \frac{\left(|V_{ud}| - \overline{|V_{ud}|}\right)^2}{\sigma^2_{|V_{ud}|}} + \frac{\left(|V_{us}| - \overline{|V_{us}|}\right)^2}{\sigma^2_{|V_{us}|}} + \frac{\left(|V_{ub}| - \overline{|V_{ub}|}\right)^2}{\sigma^2_{|V_{ub}|}} + \frac{\left(|V_{ub}|/|V_{cb}| - \overline{|V_{ub}|/|V_{cb}|}\right)^2}{\sigma^2_{|V_{ub}|/|V_{cb}|}} + \\
& + \frac{\left(|V_{cd}| - \overline{|V_{cd}|}\right)^2}{\sigma^2_{|V_{cd}|}} + \frac{\left(|V_{cs}| - \overline{|V_{cs}|}\right)^2}{\sigma^2_{|V_{cs}|}} + \frac{\left(|V_{cb}| - \overline{|V_{cb}|}\right)^2}{\sigma^2_{|V_{cb}|}} + \frac{\left(|V_{tb}| - \overline{|V_{tb}|}\right)^2}{\sigma^2_{|V_{tb}|}} + \\
& + \frac{\left(|V_{ts}|^2|V_{tb}|^2/|V_{cb}|^2 - \overline{|V_{ts}|^2|V_{tb}|^2/|V_{cb}|^2}\right)^2}{\sigma^2_{|V_{ts}|^2|V_{tb}|^2/|V_{cb}|^2}} + \frac{\left(\Delta m_{B_d} - \overline{\Delta m_{B_d}}\right)^2}{\sigma^2_{\Delta m_{B_d}}} + \\
& + \left(\frac{\mathcal{A}(\Delta m_{B_s}) - 1}{\sigma_{\mathcal{A}}(\Delta m_{B_s})}\right)^2 + \frac{\left(|\varepsilon_K| - \overline{|\varepsilon_K|}\right)^2}{\sigma^2_{|\varepsilon_K|}} + \frac{\left(f_{B_d}/f_{D_s} - \overline{f_{B_d}/f_{D_s}}\right)^2}{\sigma^2_{f_{B_d}/f_{D_s}}} + \frac{\left(f_{B_s}/f_{D_s} - \overline{f_{B_s}/f_{D_s}}\right)^2}{\sigma^2_{f_{B_s}/f_{D_s}}} + \\
& + \frac{\left(\dfrac{f_{B_s}/f_{D_s}}{f_{B_d}/f_{D_s}} - \overline{f_{B_s}/f_{B_d}}\right)^2}{\sigma^2_{f_{B_s}/f_{B_d}}} + \frac{\left(f_{D_s} - \overline{f_{D_s}}\right)^2}{\sigma^2_{f_{D_s}}} + \frac{\left(B_B - \overline{B_B}\right)^2}{\sigma^2_{B_B}} + \frac{\left(\eta_B - \overline{\eta_B}\right)^2}{\sigma^2_{\eta_B}} + \frac{\left(m_t - \overline{m_t}\right)^2}{\sigma^2_{m_t}} + \\
& + \frac{\left(m_c - \overline{m_c}\right)^2}{\sigma^2_{m_c}} + \frac{\left(\eta_{cc} - \overline{\eta_{cc}}\right)^2}{\sigma^2_{\eta_{cc}}} + \frac{\left(\eta_{ct} - \overline{\eta_{ct}}\right)^2}{\sigma^2_{\eta_{ct}}} + \frac{\left(B_K - \overline{B_K}\right)^2}{\sigma^2_{B_K}} + \left(\frac{\sin 2\beta - \overline{\sin 2\beta}}{\sigma(\sin 2\beta)}\right)^2
\end{aligned}
\tag{3.1-5}
$$

The asymmetric errors in the measurements of $|V_{cd}|$, $|V_{tb}|$ and $\sin 2\beta$ have been handled as already described in the footnote on page 24.

24 constraints (i.e. the number of $\chi^2$ terms) and 14 parameters ($\lambda$, $A$, $\rho$, $\eta$ and those listed in the first column of Table XV) correspond to 24–14= 10 degrees of freedom in the fit.

The minimization has been performed using the numerical libraries of MINUIT[150]. The errors quoted in the results are the ones, asymmetric, computed by the MINOS subprogram taking into account the non-linearity of the problem.

## 3.2 Constraints on the unitarity triangle in Wolfenstein's parametrization.

One of the advantages in the use of Wolfenstein's parametrization is that the effectiveness of a constraint can be estimated at first sight. In fact, the present status of the uncertainties is such that some constraints are almost ineffective in comparison with the much stronger requirement of unitarity. For example, the $\mathcal{O}(\lambda^5)$ term of

$$|V_{cd}| = \lambda\left[1 - A^2\frac{\lambda^4}{2}(1-2\rho)\right], \tag{3.2-1}$$

i.e. the fourth-order correction in the magnitude of the matrix element, is completely negligible compared to the experimental error. Therefore, the measurement of $|V_{cd}|$ and that of $|V_{us}| = \lambda + \mathcal{O}(\lambda^7)$ impose essentially the same constraint on the parameter $\lambda$, but the former is six times less precise and thus insignificant. Analogously, $|V_{cs}|$, $|V_{tb}|$ and $|V_{ts}|^2|V_{tb}|^2/|V_{cb}|^2$ differ from unity by $\mathcal{O}(\lambda^2)$ terms (see Table XIV), which can be neglected given the comparatively low precision of the measurements. The error in $|V_{ud}|$ is probably just below the level of precision which would make the constraint useful. The expressions of $\Delta m_{B_s}$ and $|V_{ts}|^2|V_{tb}|^2/|V_{cb}|^2$ have the same functional dependence on the parameters, which occurs only at the second order in $\lambda$:

$$\Delta m_{B_s} \propto |V_{ts}|^2|V_{tb}|^2/|V_{cb}|^2 \propto 1 - \lambda^2(1-2\rho) + \mathcal{O}(\lambda^4). \tag{3.2-2}$$

However, the 'measurement' of $\Delta m_{B_s}$ does have a non-negligible effect as a constraint on the vertex of the unitarity triangle. The equations for $\Delta m_{B_s}$ and $\Delta m_{B_d}$ are in fact strongly coupled together by the common factor $f_{D_s}^2 B_B \eta_B S(m_t^2/m_W^2)$ and the additional constraint (3.1-3). The $\Delta m_{B_s}$ constraint can thus be replaced by the equivalent equation

$$\frac{\Delta m_{B_s}}{\Delta m_{B_d}} = \frac{m_{B_s}}{m_{B_d}}\cdot\frac{f_{B_s}}{f_{B_d}}\cdot\left|\frac{V_{ts}}{V_{td}}\right|^2 = \frac{m_{B_s}}{m_{B_d}}\cdot\frac{f_{B_s}}{f_{B_d}}\cdot\frac{1}{\lambda^2}\cdot\left[\frac{1}{(1-\rho)^2+\eta^2} + \mathcal{O}(\lambda^2)\right], \tag{3.2-3}$$

which is significant at the lowest order in $\lambda$.
The way in which the constraints applied to the CKM matrix define an allowed region for the vertex $(\bar\rho,\bar\eta) = (1-\lambda^2/2)(\rho,\eta)$ of the unitarity triangle is shown in Figure 7. The measurement of the ratio $|V_{ub}/V_{cb}|$ defines an annulus centred in the origin on the ($\bar\rho,\bar\eta$) plane:

$$|V_{ub}/V_{cb}| = \lambda\sqrt{\rho^2+\eta^2} = \lambda\sqrt{\bar\rho^2+\bar\eta^2}\Big/\left(1-\frac{\lambda^2}{2}\right). \tag{3.2-4}$$

Since $|V_{ub}| = A\lambda^2|V_{ub}/V_{cb}|$ has the same dependence on $\bar\rho$ and $\bar\eta$, the graph represents the combined measurements of $|V_{ub}/V_{cb}|$, $|V_{ub}|$ and $|V_{cb}|$ (with average $|V_{ub}/V_{cb}|$= $0.090 \pm 0.007$). At the lowest order in $\lambda$, an annular region with centre in ($\bar\rho,\bar\eta$) = (1,0) is favoured by the measurement of $\Delta m_{B_d}$ (see Table XIV):

$$\sqrt{(1-\bar\rho)^2+\bar\eta^2} \cong \sqrt{(1-\rho)^2+\eta^2} \propto \sqrt{\Delta m_{B_d}}\Big/A\lambda^3 . \tag{3.2-5}$$

The dotted arc shown in the figure circumscribes the area defined by the constraint $\Delta m_{B_s} > 14.3\,ps^{-1}$, which has been expressed as a lower bound of $\Delta m_{B_s}/\Delta m_{B_d}$ (3.2-3), with $f_{B_s}/f_{B_d}$ fixed at its central value. The dependence of $|\varepsilon_K|$ on $\bar{\rho}$ and $\bar{\eta}$ is, at the lowest order in $\lambda$,

$$|\varepsilon_K| \propto B_K A^2 \lambda^6 \bar{\eta} \left\{ -\eta_{cc} S(x_c) + \eta_{ct} S(x_c, x_t) + A^2 \lambda^4 (1-\bar{\rho}) \eta_{tt} S(x_t) \right\}, \qquad (3.2\text{-}6)$$

where (see Eqs. 2.7-3, 2.7-4 and the mass values in Table XV)

$$\begin{aligned} S(x_c) &= (2.42 \pm 0.39) \cdot 10^{-4} \\ S(x_t) &= 2.38 \pm 0.11 \\ S(x_c, x_t) &= (2.15 \pm 0.31) \cdot 10^{-3} \end{aligned} \qquad (3.2\text{-}7)$$

The foregoing expressions define the region included between the two lines shaped like hyperbolas. The direct measurement of the angle $\beta$ is represented by a cone with the vertex in (1,0). Clearly, this measurement is not precise enough to constitute an effective constraint.

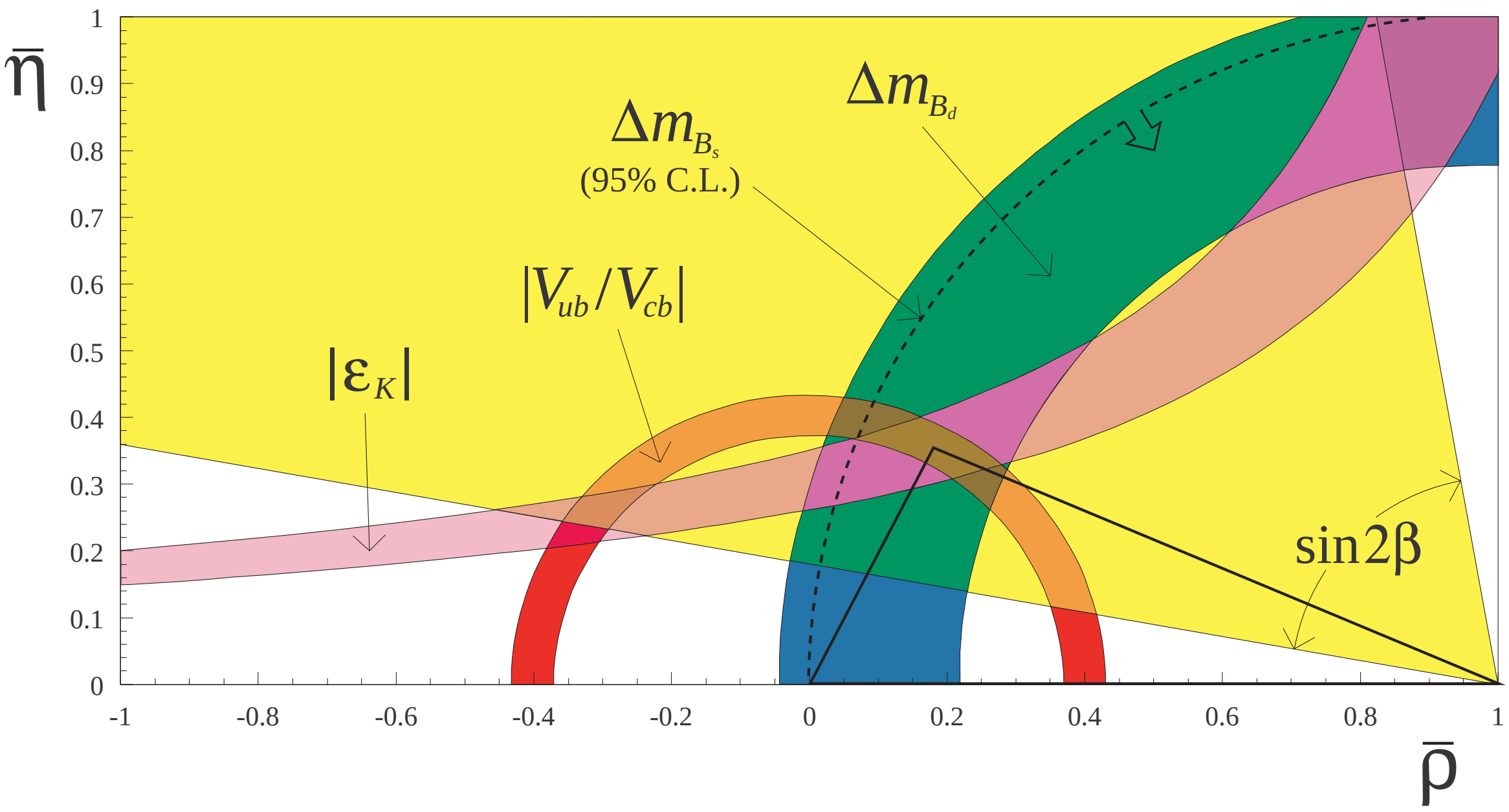

**Figure 7.** – *Graphic representation of the experimental constraints on the vertex of the unitarity triangle. The continuous lines mark the boundary of the regions favoured by the constraints* $|\varepsilon_K|$, $\Delta m_{B_d}$, $|V_{ub}/V_{cb}|$ *and* $\sin 2\beta$ *at the 68% confidence level. They have been calculated by allowing all the parameters to vary inside their respective* $\pm 1\sigma$ *intervals. The limit on* $\Delta m_{B_s}$ *(95% C.L.) is represented by the dotted line. The graph also indicates the most probable shape of the unitarity triangle as determined by the simultaneous application of all constraints.*

## 3.3 *Results*

The complete solution of the fit is given in Table XVI (CKM parameters and matrix elements) and Table XVII (output values of the additional parameters). The most probable shape of the unitarity triangle and the contours of the 68 and 95% confidence regions for the vertex $(\bar\rho,\bar\eta)$ are shown in Figure 8.

The maximum-likelihood estimates of $\bar\rho$, $\bar\eta$, sin2$\alpha$, sin2$\beta$, $\gamma = \delta_{13}$, $\vartheta_{12}$, $\vartheta_{23}$ and $\vartheta_{13}$ have been obtained by repeating the minimization with the use of different parametrizations. Wolfenstein's parameters $\rho$ and $\eta$ can be replaced by the coordinates $\bar\rho,\bar\eta$ of the vertex of the unitarity triangle using the one-to-one relations (1.2-6). A choice between several non-equivalent ways of inverting the relations (1.2-8) has to be made in order to express the parametrization in terms of sin2$\alpha$ and sin2$\beta$. In fact, while $\bar\rho$ and $\bar\eta$ can be placed into one-to-one correspondence with tan$\beta$ and tan$\gamma\equiv$ tan($\pi$–$\alpha$–$\beta$),

$$\begin{cases} \dfrac{\bar\eta}{\bar\rho} = \tan\gamma \\ \dfrac{\bar\eta}{1-\bar\rho} = \tan\beta \end{cases} \Rightarrow \begin{cases} \bar\eta = \dfrac{\tan\beta}{1+\tan\beta/\tan\gamma} \\ \bar\rho = \dfrac{\tan\beta/\tan\gamma}{1+\tan\beta/\tan\gamma} \end{cases} \tag{3.3-1}$$

sin2$\alpha$ and sin2$\beta$ cannot be inverted unless the domains of $\alpha$ and $\beta$ are already known. Assuming that $\alpha$ and $\beta$ belong to the intervals

$$0 < 2\beta < \frac{\pi}{2} \qquad \text{and} \qquad \frac{\pi}{2} < 2\alpha < \frac{3}{2}\pi\,, \tag{3.3-2}$$

which are favoured by the previously obtained results for $\bar\rho$ and $\bar\eta$, the relations

$$\alpha = \frac{\pi}{2} - \frac{1}{2}\arcsin(\sin 2\alpha)\,, \qquad \beta = \frac{1}{2}\arcsin(\sin 2\beta) \tag{3.3-3}$$

hold. This choice is also supported *a posteriori,* for the minimum $\chi^2$ and the output values of all other parameters are the same as those yielded by the primary fit. When the constraint equations are expressed in terms of the canonical parametrization (1-3), the values of the three Euler angles $\vartheta_{12}, \vartheta_{23}, \vartheta_{13}$ and the phase $\delta_{13} = \gamma$ are obtained as a result of the fit. An alternative possibility of estimating the angle $\gamma$ is given by Eqs. (3.3-1) and (3.3-3), from which $\bar\rho$ and $\bar\eta$ can be extracted as functions of sin2$\beta$, $\gamma$ or sin2$\alpha$, $\gamma$. The results are displayed in Figure 8 in terms of three different couples of parameters.

The constraints have in turn been released by subtracting the corresponding terms from the total $\chi^2$. In this way, the influence of each single constraint on the determination of the unitarity triangle has been ascertained. The most significant results are listed in Table XVIII. The way in which the confidence regions for ($\bar\rho,\bar\eta$) vary in the four most interesting cases can be seen in Figure 9.

| | 68% C.L. | 95% C.L. |
|---|---|---|
| $\lambda$ | $0.2219^{+0.0020}_{-0.0021}$ | $0.2179 \div 0.2258$ |
| $A$ | $0.798 \pm 0.029$ | $0.743 \div 0.868$ |
| $\bar{\rho}$ | $0.175^{+0.046}_{-0.034}$ | $0.103 \div 0.288$ |
| $\bar{\eta}$ | $0.354^{+0.031}_{-0.032}$ | $0.275 \div 0.415$ |
| $\sin 2\alpha$ | $-0.11^{+0.20}_{-0.22}$ | $-0.73 \div 0.26$ |
| $\sin 2\beta$ | $0.725^{+0.044}_{-0.046}$ | $0.632 \div 0.809$ |
| $\gamma = \delta_{13}$ | $\left(63.7^{+5.3}_{-7.0}\right)^\circ$ | $45.4^\circ \div 74.4^\circ$ |
| $\vartheta_{12}$ | $(12.82 \pm 0.12)^\circ$ | $12.58^\circ \div 13.05^\circ$ |
| $\vartheta_{23}$ | $\left(2.250^{+0.074}_{-0.071}\right)^\circ$ | $2.12^\circ \div 2.43^\circ$ |
| $\vartheta_{13}$ | $\left(0.202^{+0.014}_{-0.013}\right)^\circ$ | $0.176^\circ \div 0.230^\circ$ |
| $\lvert V_{ij} \rvert$ | $\begin{pmatrix} 0.97508\binom{+45}{-46} & 0.2218\,(20) & 0.00353\binom{+25}{-24} \\ 0.2217\,(20) & 0.97432\,(46) & 0.0393\binom{+12}{-13} \\ 0.00782\binom{+32}{-33} & 0.0386\binom{+13}{-12} & 0.999223\binom{+48}{-51} \end{pmatrix}$ | $\begin{pmatrix} 0.97417 \div 0.97597 & 0.2179 \div 0.2258 & 0.00309 \div 0.00402 \\ 0.2178 \div 0.2257 & 0.97341 \div 0.97522 & 0.0369 \div 0.0424 \\ 0.00701 \div 0.00846 & 0.0363 \div 0.0418 & 0.999118 \div 0.999316 \end{pmatrix}$ |
| | $\chi^2 = 2.6$ | |

**Table XVI.** – *Results of the fit.*

| | 68% C.L. | 95% C.L. |
|---|---|---|
| $f_{B_d} / f_{D_s}$ | $0.760^{+0.29}_{-0.28}$ | $0.705 \div 0.817$ |
| $f_{B_s} / f_{D_s}$ | $0.872 \pm 0.031$ | $0.811 \div 0.933$ |
| $f_{D_s}$ | $257^{+16}_{-15}\ MeV$ | $228 \div 289\ MeV$ |
| $B_B \left(= B_{B_d} = B_{B_s}\right)$ | $1.31 \pm 0.13$ | $1.06 \div 1.58$ |
| $\eta_B$ | $0.550 \pm 0.010$ | $0.531 \div 0.570$ |
| $\overline{m}_t$ | $165.9 \pm 4.9\ GeV$ | $156.3 \div 175.5\ GeV$ |
| $\overline{m}_c$ | $1.244 \pm 0.098\ GeV$ | $1.05 \div 1.44\ GeV$ |
| $\eta_{cc}$ | $1.42 \pm 0.51$ | $0.42 \div 2.42$ |
| $\eta_{ct}$ | $0.468 \pm 0.039$ | $0.391 \div 0.545$ |
| $B_K$ | $0.930^{+0.073}_{-0.072}$ | $0.790 \div 1.074$ |

**Table XVII.** – *Output values of the additional parameters entering into the expressions of* $\lvert \varepsilon_K \rvert$, $\Delta m_{B_d}$ *and* $\Delta m_{B_s}$.

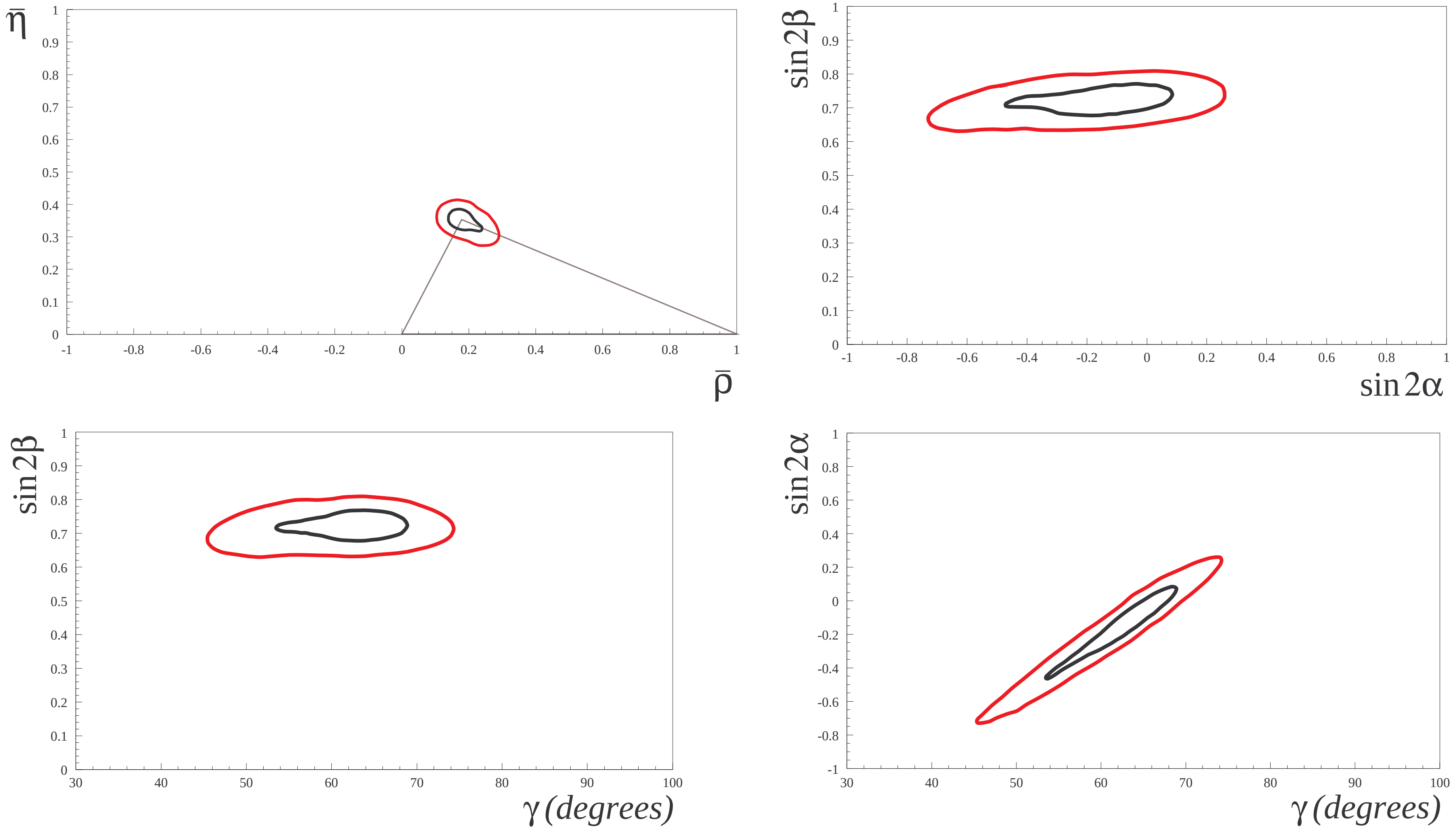

**Figure 8.** – *Contours of the 68 and 95% two-dimensional confidence regions for the vertex* $(\bar{\rho},\bar{\eta})$ *of the unitarity triangle and the observables* $\sin2\alpha$, $\sin2\beta$ , $\gamma$.

| | $\bar{\rho}$ | $\bar{\eta}$ | $\sin 2\alpha$ | $\sin 2\beta$ | $\gamma$ | $\lvert V_{CKM} \rvert$ (68%) |
|---|---|---|---|---|---|---|
| all the constraints | $0.175^{+0.046}_{-0.034}$<br>$0.103 \div 0.288$ | $0.354^{+0.031}_{-0.032}$<br>$0.275 \div 0.415$ | $-0.11^{+0.20}_{-0.22}$<br>$-0.73 \div 0.26$ | $0.725^{+0.044}_{-0.046}$<br>$0.632 \div 0.809$ | $(63.7^{+5.3}_{-7.0})^\circ$<br>$45.4^\circ \div 74.4^\circ$ | $\begin{pmatrix} 0.97508\binom{+45}{-46} & 0.2218(20) & 0.00353\binom{+25}{-24} \\ 0.2217(20) & 0.97432(46) & 0.0393\binom{+12}{-13} \\ 0.00782\binom{+32}{-33} & 0.0386\binom{+13}{-12} & 0.999223\binom{+48}{-51} \end{pmatrix}$ |
| without $\Delta m_{B_s}$ | $0.186^{+0.064}_{-0.090}$<br>$-0.041 \div 0.299$ | $0.350^{+0.042}_{-0.044}$<br>$0.266 \div 0.429$ | $-0.17^{+0.47}_{-0.36}$<br>$-0.83 \div 0.82$ | $0.725^{+0.044}_{-0.048}$<br>$0.624 \div 0.812$ | $(62^{+14}_{-10})^\circ$<br>$43^\circ \div 96^\circ$ | $\begin{pmatrix} 0.97508(46) & 0.2218(20) & 0.00355\binom{+28}{-27} \\ 0.2217(20) & 0.97431\binom{+47}{-46} & 0.0394\binom{+17}{-16} \\ 0.00775\binom{+62}{-49} & 0.0388(17) & 0.999216\binom{+62}{-66} \end{pmatrix}$ |
| without $\Delta m_{B_d}$ and $\Delta m_{B_s}$ | $0.196^{+0.068}_{-0.104}$<br>$-0.386 \div 0.314$ | $0.346 \pm 0.045$<br>$0.261 \div 0.430$ | $-0.22^{+0.53}_{-0.37}$<br>$> -0.82$ | $0.726^{+0.045}_{-0.049}$<br>$0.625 \div 0.811$ | $(60^{+16}_{-10})^\circ$<br>$42^\circ \div 154^\circ$ | $\begin{pmatrix} 0.97508(46) & 0.2218(20) & 0.00357\binom{+29}{-28} \\ 0.2217(20) & 0.97431\binom{+46}{-47} & 0.0395(16) \\ 0.00767\binom{+73}{-53} & 0.0389(17) & 0.999213\binom{+64}{-66} \end{pmatrix}$ |
| without $\lvert \varepsilon_K \rvert$ | $0.172^{+0.046}_{-0.035}$<br>$0.099 \div 0.312$ | $0.358^{+0.032}_{-0.035}$<br>$0.232 \div 0.421$ | $-0.08^{+0.20}_{-0.24}$<br>$-0.91 \div 0.30$ | $0.728^{+0.045}_{-0.048}$<br>$0.629 \div 0.812$ | $(64.3^{+5.3}_{-7.6})^\circ$<br>$38^\circ \div 75^\circ$ | $\begin{pmatrix} 0.97508(46) & 0.2218(20) & 0.00357\binom{+29}{-28} \\ 0.2217(20) & 0.97431\binom{+46}{-47} & 0.0395\binom{+16}{-15} \\ 0.00791(45) & 0.0389\binom{+16}{-15} & 0.999211\binom{+61}{-64} \end{pmatrix}$ |
| without $\lvert V_{ub}/V_{cb} \rvert$ and $\lvert V_{ub} \rvert$ | $0.166^{+0.065}_{-0.048}$<br>$-0.074 \div 0.363$ | $0.340^{+0.081}_{-0.060}$<br>$0.234 \div 0.544$ | $-0.13^{+0.23}_{-0.25}$<br>$-0.74 \div 0.33$ | $0.70^{+0.10}_{-0.11}$<br>$0.495 \div 0.990$ | $(64.0^{+5.6}_{-7.2})^\circ$<br>$44^\circ \div 76^\circ$ | $\begin{pmatrix} 0.97508(46) & 0.2218(20) & 0.00339\binom{+74}{-54} \\ 0.2217(20) & 0.97432\binom{+46}{-47} & 0.0394\binom{+16}{-15} \\ 0.00788\binom{+44}{-43} & 0.0388(15) & 0.999216\binom{+60}{-63} \end{pmatrix}$ |
| only $\lvert V_{ij} \rvert$ and $\lvert V_{ub}/V_{cb} \rvert$ | $-0.427 \div 0.427$<br>$-0.457 \div 0.457$ | $-0.427 \div 0.427$<br>$-0.457 \div 0.457$ | undetermined | $-0.773 \div 0.773$<br>$-0.813 \div 0.813$ | undetermined | $\begin{pmatrix} 0.97508(46) & 0.2218(20) & 0.00357(29) \\ 0.2217\binom{+21}{-20} & 0.97431\binom{+50}{-46} & 0.0395\binom{+17}{-16} \\ 0.00782\binom{+50}{-29} & 0.0389\binom{+21}{-27} & 0.999212\binom{+64}{-67} \end{pmatrix}$ |

**Table XVIII.** – *Results (68 and 95% C.L.) obtained after releasing in turn the main constraints.*

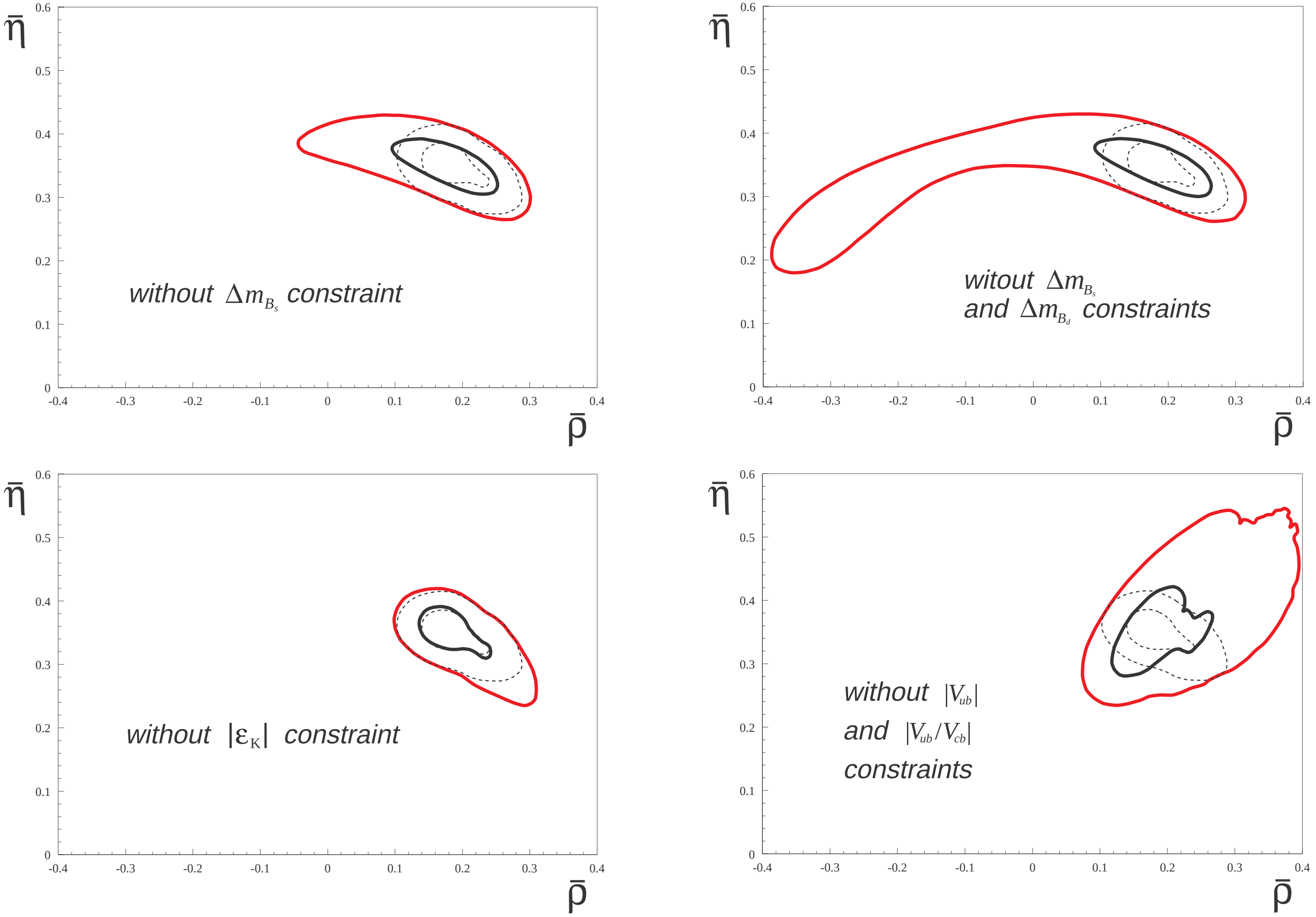

**Figure 9.** – *Regions of maximum probability (68% and 95%) for the vertex $(\bar{\rho},\bar{\eta})$ of the unitarity triangle, as determined by four different sets of constraints. Each graph has been superimposed over the contours obtained in the complete fit (see Figure 8).*

It should be noted that, even when the only constraint which is based on the present experimental evidence for CP violation, namely the measurement of $|\varepsilon_K|$, is released, it is unequivocally predicted that the observables parametrizing the magnitude of CP violation ($\bar{\eta}$, $\gamma$, $\sin 2\beta$) assume non-zero values. (The measurement of $\sin 2\beta$ is not incompatible with the absence of CP-violating effects and, given its precision, has no influence on the results). The current precision in the determination of the phase is almost entirely due to the much more effective $|V_{ub}/V_{cb}|$ and $|V_{ub}|$ constraints: when these are removed, the error in $\sin 2\beta$, for example, is nearly tripled. However, the imaginary part of the CKM matrix becomes totally undetermined after the removal of all constraints the theoretical interpretation of which assumes the dominant contribution of the top quark in the virtual intermediate state of the $\Delta S= 2$ and/or $\Delta B= 2$ processes. On the other hand, it is evident that the remaining constraints (the direct measurements of the moduli $|V_{ij}|$ except $|V_{tb}|$) cannot be used to confine the domain of the vertex of the unitarity triangle to any definite region, except for a spherically symmetrical area around the origin ($|V_{ub}/V_{cb}|$ and $|V_{ub}|$).
As expected, the results have proved to be nearly or completely insensitive to the removal of the constraints $|V_{ud}|$, $|V_{cd}|$, $|V_{cs}|$, $|V_{tb}|$ and $|V_{ts}|^2|V_{tb}|^2/|V_{cb}|^2$.

It is not clear to what extent the minimum $\chi^2$ value can be interpreted as an estimate of the goodness-of-fit. The contribution of non-Gaussian uncertainties should be taken into account. Moreover, the ineffective constraints should reasonably not be numbered among the degrees of freedom of the fit. Whatever its exact statistical interpretation, the small $\chi^2$ value (=2.6 unnormalized) obtained in the fit with all the constraints, is the clear sign of a high degree of consistency in the experimental data and between the data and the model.

The value of $\Delta m_{B_s}$ has been computed using the relation linking $\Delta m_{B_s}$ to $\Delta m_{B_d}$:

$$\Delta m_{B_s} = \Delta m_{B_d} \cdot \frac{m_{B_s}}{m_{B_d}} \cdot \frac{f_{B_s}}{f_{B_d}} \cdot \left| \frac{V_{ts}}{V_{td}} \right|^2 . \tag{3.3-4}$$

The results are

$$\Delta m_{B_s} = 15.4^{+3.0}_{-0.7}\, ps^{-1} \tag{3.3-5}$$

$$\text{with} \quad 14.1 ps^{-1} < \Delta m_{B_s} < 21.5 ps^{-1} \ \ 95\%\ \text{C.L.},$$

when the term depending on the measured amplitude $\mathcal{A}(\Delta m_{B_s})$ is included in the $\chi^2$ and

$$\Delta m_{B_s} = 15.9^{+3.2}_{-3.1}\, ps^{-1} \tag{3.3-6}$$

$$\text{with} \quad 9.9 ps^{-1} < \Delta m_{B_s} < 22.4 ps^{-1} \ \ 95\%\ \text{C.L.},$$

after releasing the direct experimental constraint on $\Delta m_{B_s}$. When the other constraints in turn are removed, the results listed in Table XIX are obtained.

| | 68% C.L. | $\Delta m_{B_s}$ $(ps^{-1})$ | 95% C.L. |
|---|---|---|---|
| complete fit | $15.4^{+3.0}_{-0.7}$ | | $14.1 \div 21.5$ |
| without $\Delta m_{B_s}$ | $15.9^{+3.2}_{-3.1}$ | | $9.9 \div 22.4$ |
| without $\Delta m_{B_d}$ and $\Delta m_{B_s}$ | $15.2^{+4.4}_{-3.5}$ | | $8.9 \div 24.8$ |
| without $\lvert\varepsilon_K\rvert$ | $15.4^{+3.0}_{-0.7}$ | | $14.0 \div 23.5$ |
| without $\lvert V_{ub}/V_{cb}\rvert$ and $\lvert V_{ub}\rvert$ | $15.4^{+3.1}_{-0.7}$ | | $14.0 \div 23.6$ |

**Table XIX.** – $\Delta m_{B_s}$ *predictions.*

The combination of CKM matrix element which determines the magnitude of the direct-CP-violation parameter $\varepsilon'_K/\varepsilon_K$ (Eq. 2.7-44) has also been calculated:

$$\mathrm{Im} V_{td} V_{ts}^* = (1.24 \pm 0.14) \cdot 10^{-4} \qquad (3.3\text{-}7)$$

$$1.0 \cdot 10^{-4} < \mathrm{Im} V_{td} V_{ts}^* < 1.5 \cdot 10^{-4} \quad 95\%\ \text{C.L.}$$

The major theoretical uncertainty in the results of the fit is the one affecting the knowledge of $B_K$ and $B_B$ (the value of $\eta_{cc}$ is even more uncertain, but a more precise determination would be of little importance, since in $\lvert\varepsilon_K\rvert$ the term multiplied by $\eta_{cc}$ is roughly ten times smaller than those containing $\eta_{ct}$ and $\eta_{tt}$). The corresponding constraints (i.e. the terms $\left(B_K - \overline{B_K}\right)^2/\sigma^2_{B_K}$ and $\left(B_B - \overline{B_B}\right)^2/\sigma^2_{B_B}$ of the $\chi^2$) have been removed in order to establish their influence on the results. Moreover, as $B_K$ and $B_B$ are included among the 'free' parameters of the fit, their values are determined, together with those of the CKM parameters, on the basis of the remaining data, which are mostly experimental in origin (see Table XX).

| free parameters | $\bar\rho$ | $\bar\eta$ | $B_B$ | $B_K$ |
|---|---|---|---|---|
| | $0.175^{+0.046}_{-0.034}$<br>$0.103 \div 0.288$ | $0.354^{+0.031}_{-0.032}$<br>$0.275 \div 0.415$ | | |
| $B_B$ | $0.176^{+0.071}_{-0.036}$<br>$0.103 \div 0.303$ | $0.354^{+0.032}_{-0.031}$<br>$0.270 \div 0.415$ | $1.34^{+0.36}_{-0.26}$<br>$0.89 \div 2.19$ | |
| $B_K$ | $0.172^{+0.047}_{-0.035}$<br>$0.099 \div 0.311$ | $0.358^{+0.033}_{-0.035}$<br>$0.236 \div 0.421$ | | $0.89^{+0.17}_{-0.14}$<br>$0.64 \div 1.27$ |
| $B_K$ and $B_K$ | $0.172^{+0.122}_{-0.035}$<br>$0.098 \div 0.426$ | $0.358^{+0.033}_{-0.035}$<br>$0.104 \div 0.421$ | $1.31^{+0.37}_{-0.27}$<br>$0.85 \div 3.90$ | $0.89^{+0.19}_{-0.15}$<br>$> 0.63$ |

**Table XX.** – *Simultaneous determination of the vertex of the unitarity triangle and the bag parameters (68 and 95% C.L.), performed by removing the theoretical constraints on* $B_B$ *and* $B_K$*. The results are perfectly compatible with the (more precise) theoretical determinations* $B_B = 1.30 \pm 0.15$ *and* $B_K = 0.94 \pm 0.08$ *used in the complete fit.*

The purely theoretical determinations of the bag parameters clearly have considerable weight, although this is not evident when the corresponding constraints are released one at a time (the cause may be found in the strong correlation between the constraints $\Delta m_{B_d}$, $\Delta m_{B_s}$ and $|\varepsilon_K|$). In particular, both $B_B$ and the $\bar{\rho}$-coordinate of the vertex (the two parameters are highly correlated) are determined with much less precision when only the 'experimental' constraints are used.
On the other hand, a significant increase in the precision of the results will not be achieved by reducing the theoretical errors in $B_B$ and $B_K$, unless a general improvement in the level of the uncertainties is obtained (particularly in the quark masses, which occur in the factor $S(x_c, x_t)$, and in the decay constants). The values of $B_K$ and $B_B$ are in fact already known at the highest useful level of precision: if the errors in $B_B$ and $B_K$ are simultaneously reduced by a factor of 10 ($B_B = 1.300 \pm 0.015$, $B_K = 0.940 \pm 0.008$), the results

$$\bar{\rho} = 0.176^{+0.044}_{-0.036}, \qquad \bar{\eta} = 0.353 \pm 0.031 \qquad (3.3\text{-}8)$$
$$0.104 < \bar{\rho} < 0.282 \qquad 0.277 < \bar{\eta} < 0.413 \qquad 95\% C.L$$

are obtained, which in practice coincide with the ones reported in Table XVI.

Our results for the CKM parameters and for the vertex of the unitarity triangle are compatible with those obtained in previous analyses[151,152] which made use of similar techniques but different sets of experimental data.

## 3.4 Bayesian determination of the CKM matrix

To have an estimate of the goodness of the values obtained with the $\chi^2$ minimization method and obtain the probability distribution functions (p.d.f.) of the CKM elements and of all the estimated quantities, a complementary method[153], based on Bayesian statistics[154], has been applied to the data set of Table XIV. This method has the considerable advantage that it keeps track of all the assumptions made from beginning to end, whereas other methods usually do not. For example, the standard $\chi^2$-minimization method starts from the assumption that all the input values are Gaussian-distributed and provide estimates that are assumed to be Gaussian-distributed. This is reliable only at a first approximation, since the asymmetric errors and the shapes of the maximum probability regions clearly indicate that at least some quantities depart from Gaussianity.
The Bayesian approach for the statistic estimations starts from the definition of a set of parameters ($\bar{x}$) with a prior probability $P(\bar{x}, i)$ and a set of experimental data $e$ with a probability distribution function $P(e, \bar{x} \mid i)$, which depends on the $\bar{x}$ parameters. If $\bar{x}$ is a set of continuous variables, their final ('posterior') probability distribution function is given by

$$P(\bar{x}, i \mid e) = \frac{P(e, \bar{x} \mid i) \times P(\bar{x}, i)}{\int P(e, \bar{x} \mid i) \times P(\bar{x}, i) d\bar{x}}. \qquad (3.4\text{-}1)$$

Hence, if an initial probability distribution for certain variables ($\bar{x}$) is used and new experimental data are expressed by $P(e,\bar{x}\,|\,i)$, from Eq. (3.4-1) we get a new distribution of probability for the initial variables (the posterior p.d.f.). From the shape of this distribution, the best estimations or the allowed ranges of the starting variables can be extracted. The prior p.d.f. $P(\bar{x},i)$ and the allowed range for $\bar{x}$ are generally a matter of assumption. As long as the $P(e,\bar{x}\,|\,i)$ function contains more information than the prior p.d.f., these assumptions do not play any role in the final results[155].

Eq. (3.4-1) can be used to estimate the p.d.f. of the CKM elements. To allow the comparison with the $\chi^2$ minimization method, we consider exactly the same experimental data set (Table XIV). Therefore, in *e* we consider the quantities $|V_{ud}|$, $|V_{us}|$, $|V_{cd}|$, $|V_{cs}|$, $|V_{cb}|$, $|V_{ub}/V_{cb}|$, $|V_{ub}|$, $|V_{ts}|^2|V_{tb}|^2/|V_{cb}|^2$ and $\sin 2\beta$, which depend on the four CKM parameters only, and the quantities $|\varepsilon_K|$, $\Delta m_{B_d}$, $\mathcal{A}(\Delta m_{B_s})$ which depend also on other experimental or theoretical parameters. These parameter dependencies fix the set of the $\bar{x}$ quantities. Concerning the CKM parameters, we take the four angles that define completely a unitarized three-family CKM matrix (Eq. 1.1-3): $\Theta=(\vartheta_{12},\vartheta_{13},\vartheta_{23},\delta_{13})$. The $\Theta$ parameter space is given by all the allowed space, that is $\Theta\in[0,\pi/2[^3\times[0,2\pi[$. For the prior distribution $P(\Theta,i)$ of $\Theta$, since we have no information to use, we assume a uniform distribution in the allowed space. As is customary in the Bayesian method, this general and weak assumption has no (or weak) effect on the final result[154]. As far as $|\varepsilon_K|$, $\Delta m_{B_d}$, and $\Delta m_{B_s}$ are concerned, we use the following expressions:

$$|\varepsilon_K|=\frac{G_F^2 f_K^2 m_K m_W^2}{12\sqrt{2}\pi^2\Delta m_K}B_K\cdot\mathrm{Im}\,M_{12}=\Phi_K\cdot\mathrm{Im}\,M_{12} \tag{3.4-2}$$

$$\Delta m_{B_d}=\frac{G_F^2 m_{B_d} m_W^2 f_{B_d}^2 B_B\eta_B}{6\pi^2}S(x_t)\cdot\left|V_{td}^*V_{tb}\right|^2=\Phi_B\cdot S(x_t)\cdot\left|V_{td}^*V_{tb}\right|^2 \tag{3.4-3}$$

$$\Delta m_{B_s}=\frac{G_F^2 m_{B_s} m_W^2 f_{B_s}^2 B_B\eta_B}{6\pi^2}S(x_t)\cdot\left|V_{ts}^*V_{tb}\right|^2=\Phi_B\cdot\Phi_{ds}\cdot S(x_t)\cdot\left|V_{ts}^*V_{tb}\right|^2 \tag{3.4-4}$$

where $M_{12}$ has the same expression as in Eq. (2.7-57), while $\Phi_K$, $\Phi_B$ and $\Phi_{ds}$ are given by

$$\Phi_K=\frac{G_F^2 f_K^2 m_K m_W^2}{12\sqrt{2}\pi^2\Delta m_K}B_K,\qquad \Phi_B=\frac{G_F^2 m_{B_d} m_W^2 f_{B_d}^2 B_B\eta_B}{6\pi^2},\qquad \Phi_{ds}=\frac{m_{B_s} f_{B_s}^2}{m_{B_d} f_{B_d}^2}. \tag{3.4-5}$$

The factors $\Phi_i$, which have been isolated in these expressions, will be considered as three other initial parameters in the set $\bar{x}$, which is finally composed by $(\Theta,\Phi)=(\vartheta_{12},\vartheta_{13},\vartheta_{23},\delta_{13},\Phi_K,\Phi_B,\Phi_{ds})$. The distinction between $\Phi_B$ and $\Phi_{ds}$ in $\Delta m_{B_s}$ takes correctly into account that a strong correlation of the theoretical errors between $\Delta m_{B_d}$ and $\Delta m_{B_s}$ exists. For each of them, we assume a Gaussian prior p.d.f. defined on the positive side only with the mean values and standard deviations given in table XVI.

| parameter | value |
|---|---|
| $\Phi_K$ | $(18.0 \pm 1.6)\cdot 10^3$ |
| $\Phi_B$ | $(3.10 \pm 0.69)\cdot 10^3\ GeV$ |
| $\Phi_{ds}$ | $1.34 \pm 0.09$ |

**Table XXI.** – *The* $\Phi$ *factors, part of the* $\bar{x}$ *parameter set, which enter into the expressions of* $|\varepsilon_K|$, $\Delta m_{B_d}$ *and* $\Delta m_{B_s}$. *Values and errors listed have been used to define the prior p.d.f..*

For the expression of the probability distribution function $P(e,\bar{x}\,|\,i)$ which enters in Eq. (3.4-1), it is possible to take advantage of the fact that all the measurements in the set $e$ are independent. Therefore, we use:

$$P(e,\bar{x}\,|\,i) = \prod_i P_i(\mu_{X_i},\bar{x}\,|\,i) \qquad (3.4\text{-}6)$$

where the product runs over all the measurements considered and $P_i$ are their probabilities. These are given by the Gaussian distribution functions, associated to the $\mu_X$ measured values, times an interval value $\Delta X$ :

$$P_i(\mu_X,\bar{x}\,|\,i) = g(X(\bar{x}),\mu_X,\sigma_X)\Delta X \qquad (3.4\text{-}7)$$

with

$$g(X(\bar{x}),\mu_X,\sigma_X) = \frac{1}{\sqrt{2\pi}\sigma_X} e^{-\frac{(X(\bar{x})-\mu_X)^2}{2\sigma_X^2}} \qquad (3.4\text{-}8)$$

The $P_i(\mu_X,\bar{x}\,|\,i)$ written in this way represents the probability to measure $X$ in an interval $[\mu_X,\mu_X+\Delta X]$ when the mean value and the standard deviation of the Gaussian distribution are given by $X(\bar{x})$ and $\sigma_X$. The actual choice of the interval width $\Delta X$ is not relevant since, when $P(e,\bar{x}\,|\,i)$ is built and substituted into Eq. (3.4-1), all the $\Delta X_i$ cancel out in the ratio. Apart from these $\Delta X_i$ terms, $P(e,\bar{x}\,|\,i)$ has the structure of a likelihood.

For the posterior probability function $P(\bar{x},i\,|\,e)$, we can calculate the probability distribution of any variable $z$ which is a function of $\bar{x}$, $z = f(\bar{x})$, by means of the following relation:

$$P(z,i\,|\,e) = \frac{d}{dz}\int \theta(z - f(\bar{x}))P(\bar{x},i\,|\,e)\,d\bar{x} = \frac{d}{dz}\int_{z>f(\bar{x})} P(\bar{x},i\,|\,e)\,d\bar{x} \qquad (3.4\text{-}9)$$

where the integral must be evaluated in all the allowed space and $\theta(x)$ is the usual Heavyside function, that is 1 for $x > 0$ and 0 for $x < 0$.

In order to obtain the p.d.f. for any quantity $z$ which we are interested in using Eq. (3.4-9), we applied extensively the Monte Carlo method for the integral evaluation, generating a sample of $\approx 10^9$ points in the $\bar{x}$ space with a probability distribution given by the $P(\bar{x},i)$.

The p.d.f. for the most important quantities, which are evaluated simultaneously, are shown in Figures 10-13. As can be seen in Figure 10, the three angles $\vartheta_{ij}$ can vary in a rather small range and have an approximate Gaussian distribution, whereas the phase $\delta_{13}$ can vary in a much wider range and it is clearly not Gaussian-distributed. The p.d.fs. for $\sin 2\beta$, $\sin 2\alpha$, $\Delta m_{B_s}$ and the vertex of the unitarity triangle are shown in Figure 11 and Figure 13. They have a marked non-Gaussian behaviour, which is particularly evident in the shape of $\sin 2\alpha$. The p.d.f. for the moduli of the CKM elements are shown in Figure 12. As it can be seen, the elements $|V_{cb}|$, $|V_{td}|$, $|V_{ts}|$ and $|V_{tb}|$ are clearly not Gaussian-distributed.

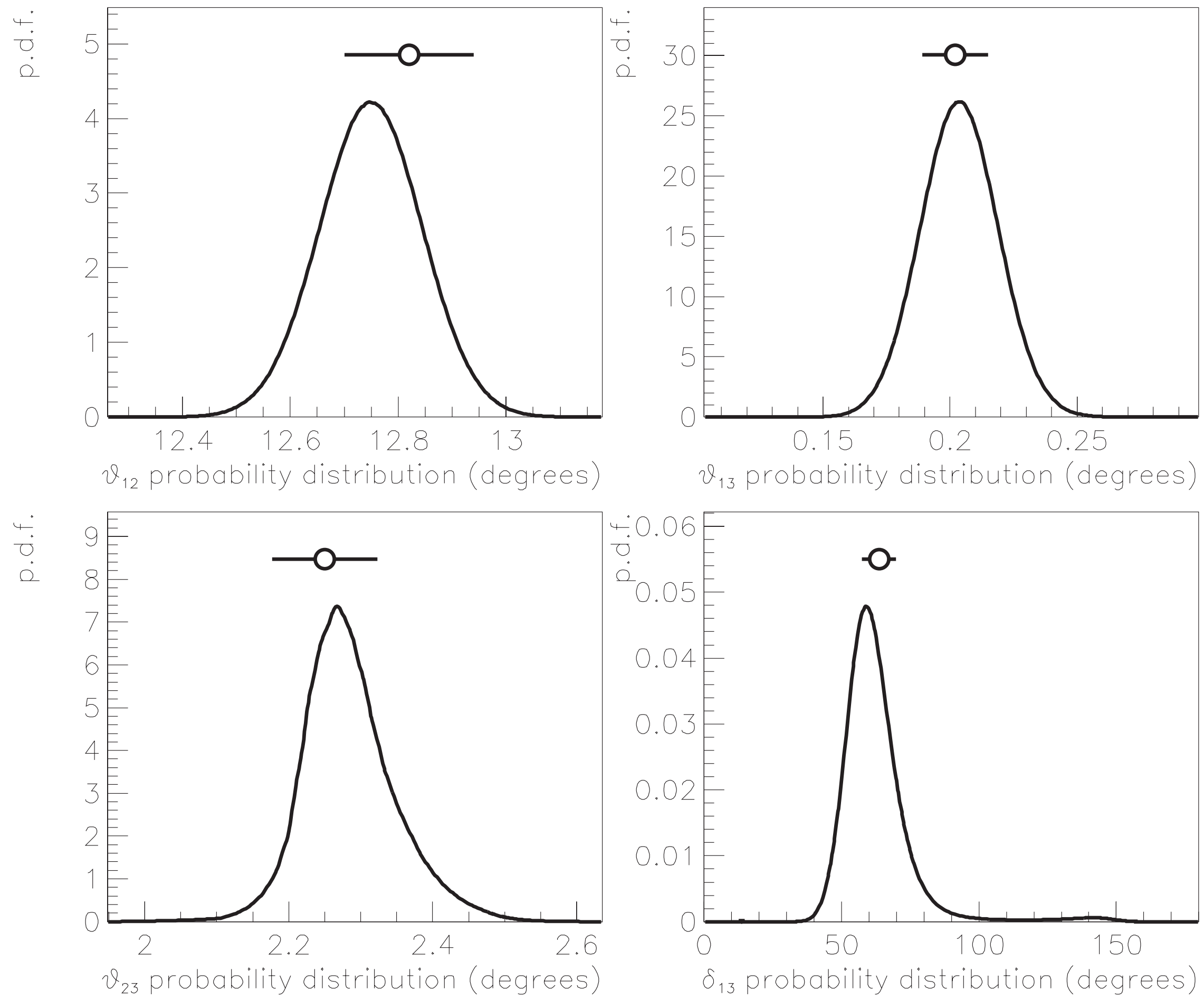

**Figure 10.** – *Posterior probability distribution functions for $\vartheta_{ij}$ and $\delta_{13}$. Points with error bars refer to the solution obtained with the $\chi^2$-minimization method.*

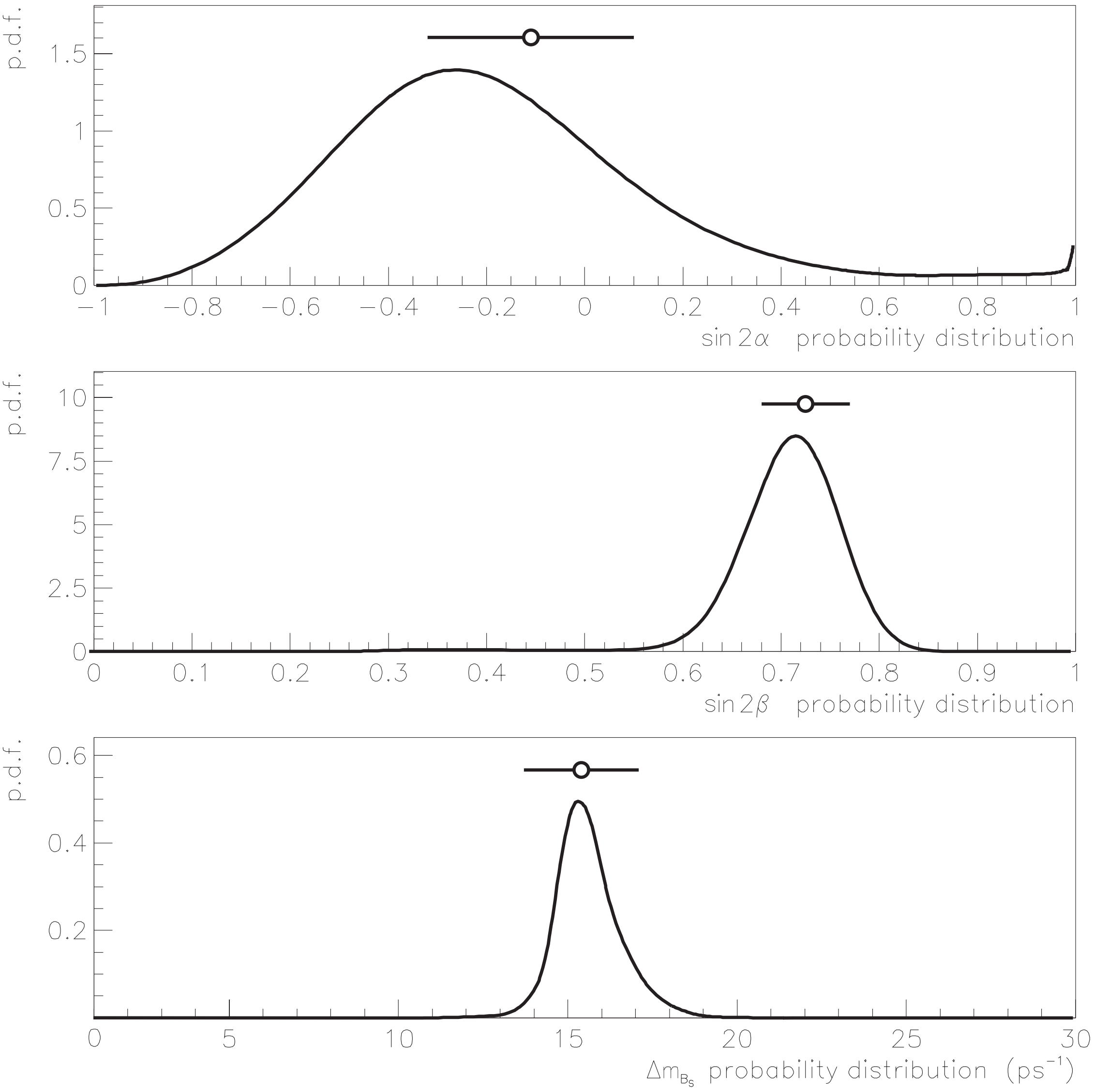

**Figure 11.** – *Posterior probability distribution functions for* $\sin 2\alpha$, $\sin 2\beta$ *and* $\Delta m_{B_s}$ .

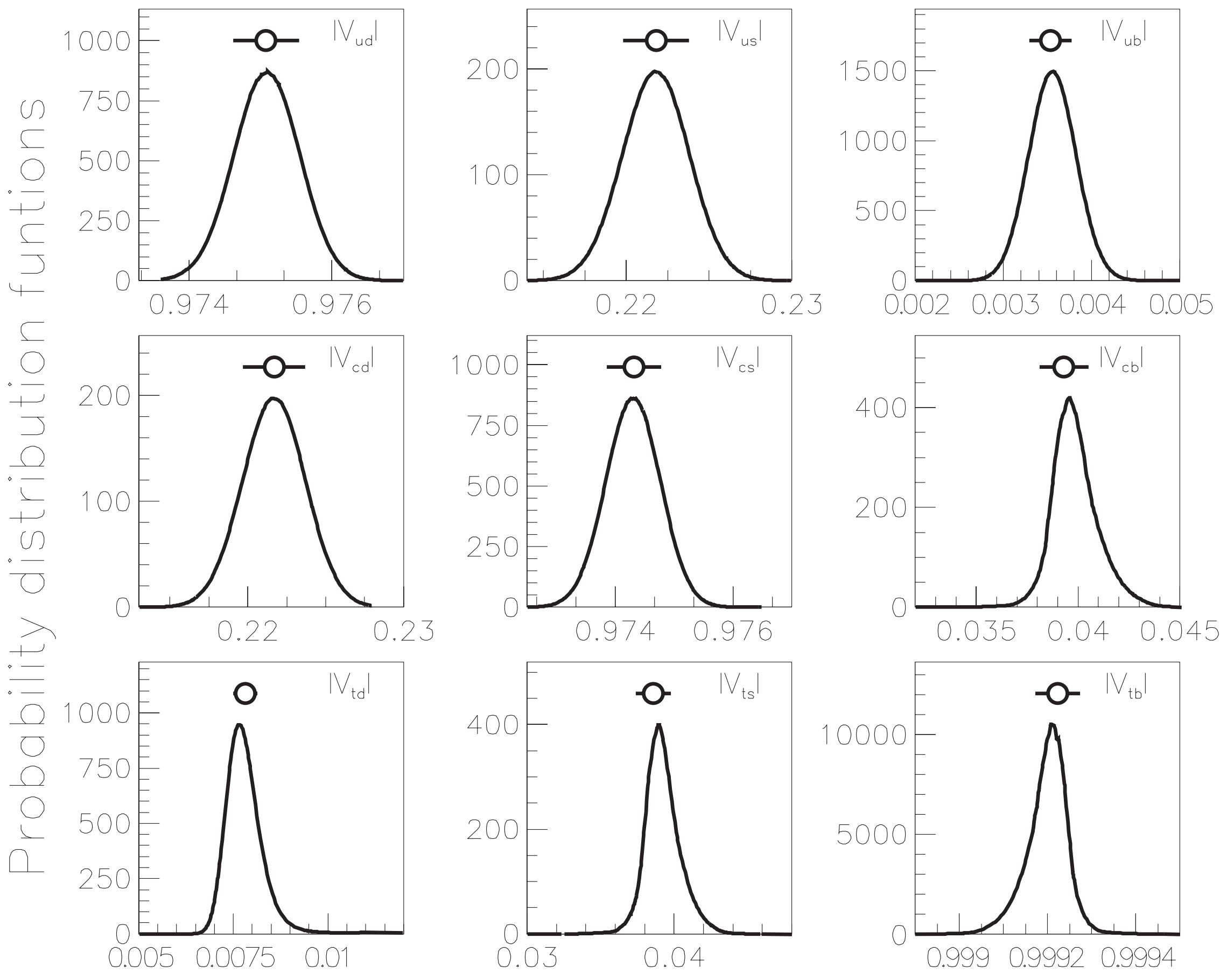

**Figure 12.** – *Posterior probability distribution functions for* $\left|V_{ij}\right|$.

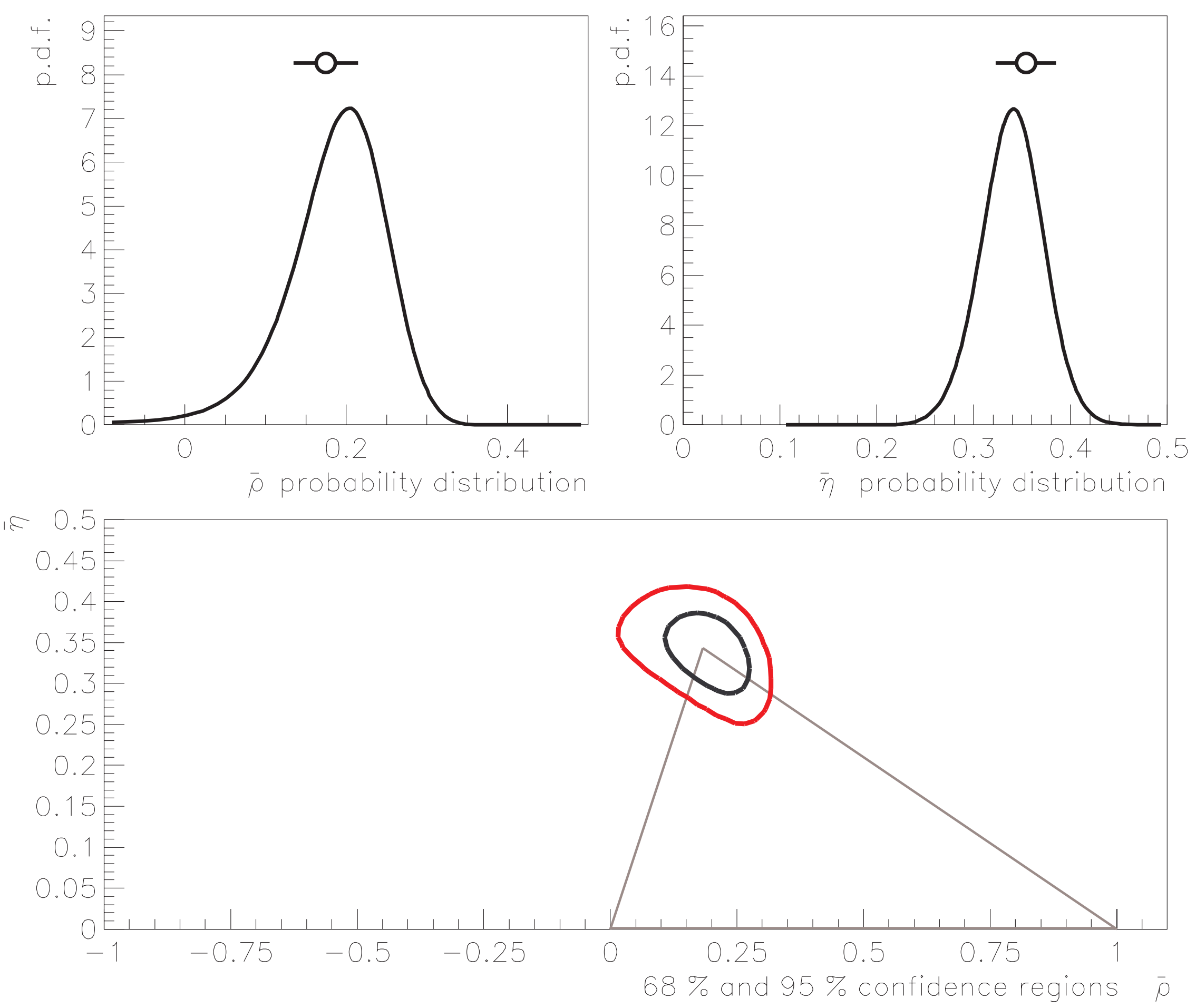

**Figure 13.** – *Posterior probability distribution functions for* $\overline{\rho}$ *and* $\overline{\eta}$ *, 68% and 95% probability contours for the vertex of the unitarity triangle.*

In Table XXII, we present the complete result of this method for comparison with the one obtained with the $\chi^2$-minimization method (Table XVI). As Bayesian estimators of the p.d.f. parameters, we take the *mean* and *root mean square* of the relative p.d.f.. The comparison with the results given in Table XVI shows a good agreement on all data within one sigma. As expected, the biggest differences can be observed on the variables having the clearest non-Gaussian behaviour, for which the Bayesian estimation of the root mean square gives always a bigger value. As regards the 95% limits, since these are mainly determined by the properties of the p.d.f. tails (which are assumed to be Gaussian in the $\chi^2$ method) the Bayesian estimation always favour wider intervals. The 95% regions quoted are central-interval estimates, computed as in the previous analysis of Ref. 153.

| parameter | mean ± r.m.s. | 95% C.L. |
|---|---|---|
| $\lambda$ | $0.2218 \pm 0.0020$ | $0.2178 \div 0.2257$ |
| $\overline{\rho}$ | $0.183 \pm 0.063$ | $0.045 \div 0.293$ |
| $\overline{\eta}$ | $0.343 \pm 0.032$ | $0.283 \div 0.412$ |
| $\gamma = \delta_{13}$ | $(64 \pm 13)^{\circ}$ | $47^{\circ} \div 96^{\circ}$ |
| $\vartheta_{12}$ | $(12.81 \pm 0.12)^{\circ}$ | $12.6^{\circ} \div 13.0^{\circ}$ |
| $\vartheta_{13}$ | $(2.27 \pm 0.06)^{\circ}$ | $2.16^{\circ} \div 2.42^{\circ}$ |
| $\vartheta_{23}$ | $(0.203 \pm 0.015)^{\circ}$ | $0.175^{\circ} \div 0.234^{\circ}$ |
| $\sin 2\alpha$ | $-0.15 \pm 0.33$ | $-0.67 \div 0.74$ |
| $\sin 2\beta$ | $0.71 \pm 0.06$ | $0.60 \div 0.80$ |
| $\lvert V_{ud} \rvert$ | $0.9751 \pm 0.0005$ | $0.9742 \div 0.9760$ |
| $\lvert V_{us} \rvert$ | $0.2218 \pm 0.0020$ | $0.2178 \div 0.2258$ |
| $\lvert V_{ub} \rvert$ | $0.0036 \pm 0.0003$ | $0.0030 \div 0.0041$ |
| $\lvert V_{cd} \rvert$ | $0.2217 \pm 0.0020$ | $0.2177 \div 0.2256$ |
| $\lvert V_{cs} \rvert$ | $0.9743 \pm 0.0003$ | $0.9734 \div 0.9752$ |
| $\lvert V_{cb} \rvert$ | $0.0398 \pm 0.0011$ | $0.0379 \div 0.0424$ |
| $\lvert V_{td} \rvert$ | $0.0078 \pm 0.0006$ | $0.0071 \div 0.0093$ |
| $\lvert V_{ts} \rvert$ | $0.0392 \pm 0.0012$ | $0.0371 \div 0.0420$ |
| $\lvert V_{tb} \rvert$ | $0.99921 \pm 0.00004$ | $0.99910 \div 0.99928$ |
| $\lvert V_{ub}/V_{cb} \rvert$ | $0.0894 \pm 0.0066$ | $0.0765 \div 0.1025$ |
| $\Delta m_{B_s}$ | $(15.5 \pm 0.9)\ ps^{-1}$ | $13.8 \div 17.5\ ps^{-1}$ |

**Table XXII.** – *Results of the Bayesian estimation method. The results can be directly compared with those obtained with the $\chi^2$-minimization method (see Table XVI), with which they are fully compatible.*

## 3.5 New prospects

According to the experimental results concerning kaon decays, a single complex phase in the CKM matrix seems a suitable explanation for CP violation. On the other hand, insofar as the baryon number of the universe has to be considered as a dynamically generated quantity rather than an initial condition, the Standard Model fails to account for the present abundance of matter by several orders of magnitude[156]. The measurement of the CP-violating phenomena which are expected to occur in heavier quark sectors will provide an independent way of testing the present theoretical knowledge. The study of *B* mesons, which are ten times heavier than kaons and thus undergo a wide range of (mostly rare) decays, seems to be an almost ideal field of research. Large CP asymmetries are predicted to occur in the decays of *B* mesons and the experiments will have many possibilities of revealing the presence of new physics. The measurement of CP violation in the $b$ quark sector will be the subject of an intense activity in the so-called $B$ factories after 2000.

A quite clear strategy can be adopted in the measurement of sin2$\beta$. The already mentioned CP asymmetry (1-17) which is expected in the rates of $B_d^0/\overline{B}_d^0 \to J/\psi\, K_S$ decays will be measured either as an oscillating function of time, whose amplitude is equal to sin2$\beta$, or by integration over a definite interval of time.

The extraction of sin2$\alpha$ from the measurement of the asymmetry

$$a_{\pi^+\pi^-}^{CP} = \frac{\Gamma\left(\overline{B}_d^0 \to \pi^+\pi^-\right) - \Gamma\left(B_d^0 \to \pi^+\pi^-\right)}{\Gamma\left(\overline{B}_d^0 \to \pi^+\pi^-\right) + \Gamma\left(B_d^0 \to \pi^+\pi^-\right)} \tag{2.4-1}$$

is less straightforward, since ‘tree’ and ‘penguin’ amplitudes interfere with different weak and strong phases in $B_d^0/\overline{B}_d^0 \to \pi^+\pi^-$ decays, which therefore violate CP symmetry directly. Additional measurements of the isospin-related processes $B_d^0 \to \pi^+\pi^-$, $B^+ \to \pi^+\pi^0$, $B_d^0 \to \pi^0\pi^0$ and their CP-conjugates will be required in order to estimate the relative phase of the amplitudes $\mathcal{A}\left(B_d^0 \to \pi^+\pi^-\right)$ and $\mathcal{A}\left(\overline{B}_d^0 \to \pi^+\pi^-\right)$.

An even greater uncertainty affects the prospects for the determination of the angle $\gamma$. A measurement of $\gamma$ can in principle be obtained from the observation of the time-dependent CP asymmetries involved in $B_s^0 - \overline{B}_s^0$ oscillations, such as those between the rates of $B_s^0(t) \to K_s\rho^0$ and $\overline{B}_s^0(t) \to K_s\rho^0$, or $B_s^0(t) \to D_s^- K^+$ and $\overline{B}_s^0(t) \to D_s^+ K^-$. The major experimental difficulty is due to the rapidity of the oscillations, which have not yet been resolved. The most promising method in the short run is probably the one consisting in the measurement of the direct CP asymmetry between the decays $B_d^0 \to K^+\pi^-$ and $\overline{B}_d^0 \to K^-\pi^+$; the extraction of $\gamma$ requires some additional branching ratio measurements in order to eliminate the hadronic uncertainties which are introduced by a phase difference between tree and penguin amplitudes.

A distinction can be made between the experiments at the hadronic machines (HERA-B, CDF, LHCb) and those exploiting the $B$ production induced by electromagnetic interactions between colliding $e^+$ and $e^-$ (BaBar and Belle). While the experiments in the first class can detect the production of $b\overline{b}$ couples with a high cross-section, but with a low signal-to-background ratio, a low cross-section $\sigma_{B\overline{B}}$ is expected at the electromagnetic $B$ factories, which, on the other hand, will be operating in a frame of quite a few background events.

The measurement of the $B_s^0 - \overline{B}_s^0$ oscillation frequency $\Delta m_{B_s}$ is within reach of the hadronic $B$ factories only. The chances of success of each experiment are restricted by the maximum time resolution it can achieve, which imposes an upper limit on the measurable value of $\Delta m_{B_s}$.

HERA-B will study the reactions $pN \to b\overline{b} \to B_d^0, B^\pm, B_s^0, \Lambda_b$ in fixed-target mode ($\sqrt{s} \approx 42GeV$). The expected cross-section is high ($\sigma_{b\overline{b}} \approx 12nb$), but, because of a very slow signal-to-background ratio ($\sigma_{b\overline{b}}/\sigma_{tot} \approx 10^{-6}$), an extremely sensitive trigger is needed.

At BaBar and Belle, the $B$ mesons will be produced by electromagnetic interaction in collider mode ($e^+e^- \to \Upsilon(4S) \to B_d^0, B^\pm$). The favourable signal-to-background ratio ($\sigma_{B\bar{B}}/\sigma_{tot} \approx 0.28$) is counterbalanced by a very low production cross-section ($\sigma_{B\bar{B}} \approx 1.1nb$), so that a quite high luminosity is required: integrated luminosities of respectively 30 and 100 $fb^{-1}$ can be reached after one year by the two experiments. For both BaBar and Belle the $B_s$ physics is inhibited.

By the end of 2000, CDF Run II will become operative. The reactions $p\bar{p} \to b\bar{b} \to B_d^0, B^\pm, B_s^0, \Lambda_b$ will be studied in collider, with $\sqrt{s} \approx 2.0TeV$. The expected cross-section is $\sigma_{b\bar{b}} \approx 50\mu b$, but a large amount of background events will have to be rejected ($\sigma_{b\bar{b}}/\sigma_{tot} \approx 10^{-3}$).

Finally, LHCb is expected to start in 2005. $b$-flavoured hadrons will be produced in $pp$ collisions at $\sqrt{s} \approx 14TeV$. With such a high energy in the centre of mass, the $b\bar{b}$ cross section will reach the value $\sigma_{b\bar{b}} \approx 500\mu b$. At the same time, the background sensitivity will be confined to $\sigma_{b\bar{b}}/\sigma_{tot} \approx 1/160$.

The predictions obtained in the present analysis,

$$\sin 2\alpha = -0.11^{+0.20}_{-0.22}, \quad \sin 2\beta = 0.725^{+0.044}_{-0.046}, \quad \gamma = \left(63.7^{+5.3}_{-7.0}\right)^\circ, \quad \Delta m_{B_s} = 15.4^{+3.0}_{-0.7}\, ps^{-1}, \qquad (2.4\text{-}2)$$

already provide an estimate of the limits beyond which the new measurements will have to be definitely interpreted as signals of new physics. For example, the prediction for $\sin 2\beta$ is precise enough to define a wide range of values which the Standard Model would be unable to account for, even within the ‘physical’ range [0,1].

The most general supersymmetric (SUSY) extensions of the Standard Model would be able to change the expected shape of the unitarity triangle (the existence of only three generations of particles is not called into question) in quite an unpredictable way, due to the contribution of new, presently unconstrained phases. Some restrictive assumptions have to be made in order to give a quantitative account of the possible modifications of the present scenario, as allowed by the currently available experimental data. In the Minimal Supersymmetric Standard Model (MSSM) and its several variants (see Ref. 152 for a recent review), the flavour–changing processes are controlled by essentially the same mixing matrix ($V_{CKM}$) as in the Standard Model. The direct measurements of the CKM matrix elements are not affected by these theories, while the contribution of the supersymmetric particles to the effective FCNC processes ($B$-meson and $K$-meson mixing) can be parametrized by adding a ‘supersymmetric’ term to the top-quark loop function in the expressions of $\Delta m_{B_d}$, $\Delta m_{B_s}$ and $|\varepsilon_K|$. Equations (1.7-7), (1.7-8) and (1.7-65) have thus to be modified by making the substitutions

$$\begin{aligned} \Delta m_{B_d}: &\qquad S(x_t) \to \left(1 + \Delta_{SUSY}^{B_d}\right) \cdot S(x_t) \\ \Delta m_{B_s}: &\qquad S(x_t) \to \left(1 + \Delta_{SUSY}^{B_s}\right) \cdot S(x_t) \qquad (2.4\text{-}3) \\ |\varepsilon_K|: &\qquad S(x_t) \to \left(1 + \Delta_{SUSY}^{K}\right) \cdot S(x_t) \end{aligned}$$

The quantities $\Delta^{B_d}_{SUSY}$, $\Delta^{B_s}_{SUSY}$ and $\Delta^{K}_{SUSY}$ are positive definite functions of the masses of the supersymmetric particles involved (top squark, chargino and charged Higgs) and of other SUSY parameters. To an excellent approximation, they are equal to each other[152]:

$$\Delta^{B_d}_{SUSY} = \Delta^{B_s}_{SUSY} = \Delta^{K}_{SUSY} = \Delta_{SUSY} \,. \qquad (2.4\text{-}4)$$

Therefore, as far as the determination of the vertex of the unitarity triangle is concerned, only one additional degree of freedom ($\Delta_{SUSY}$) has to be considered, so that a suitable level of predictivity can be reached in the computation of the relevant parameters. The experimental constraints on the masses of the supersymmetric particles, on the electric dipole moments of electron and neutron and on the branching ratio of $B \to X_s\gamma$ decays can be used to set an upper bound on $\Delta_{SUSY}$: the values allowed in different MSSM models range[157] from $\Delta_{SUSY} = 0$ (Standard Model) to $\Delta_{SUSY} = 0.6$.

| | $\Delta_{SUSY}$ | $\bar{\rho}$ | $\bar{\eta}$ | $\sin 2\alpha$ | $\sin 2\beta$ | $\gamma$ | $\Delta m_{B_s}$ $(ps^{-1})$ |
|---|---|---|---|---|---|---|---|
| FREE | $-0.03^{+0.39}_{-0.19}$<br>$< 1.9$ 95% | $0.174^{+0.087}_{-0.036}$<br>$0.099 \div 0.385$ | $0.355^{+0.032}_{-0.033}$<br>$0.171 \div 0.417$ | $-0.10^{+0.23}_{-0.56}$<br>$-1 \div 0.34$ | $0.726^{+0.045}_{-0.048}$<br>$0.629 \div 0.811$ | $\left(64.0^{+5.4}_{-15.1}\right)^\circ$<br>$24.8^\circ \div 75.0^\circ$ | $15.4^{+6.8}_{-0.7}$<br>$14.0 \div 32.3$ |
| FIXED | 0.0 (SM) | $0.175^{+0.046}_{-0.034}$<br>$0.103 \div 0.288$ | $0.354^{+0.031}_{-0.032}$<br>$0.275 \div 0.415$ | $-0.11^{+0.20}_{-0.22}$<br>$-0.73 \div 0.26$ | $0.725^{+0.044}_{-0.046}$<br>$0.632 \div 0.809$ | $\left(63.7^{+5.3}_{-7.0}\right)^\circ$<br>$45.4^\circ \div 74.4^\circ$ | $15.4^{+3.0}_{-0.7}$<br>$14.1 \div 21.5$ |
| | 0.2 | $0.187^{+0.103}_{-0.038}$<br>$0.113 \div 0.333$ | $0.345^{+0.033}_{-0.071}$<br>$0.236 \div 0.409$ | $-0.19^{+0.24}_{-0.59}$<br>$-0.98 \div 0.26$ | $0.720^{+0.047}_{-0.049}$<br>$0.620 \div 0.809$ | $\left(61.5^{+5.9}_{-17.2}\right)^\circ$<br>$37.4^\circ \div 72.8^\circ$ | $15.7^{+6.2}_{-0.8}$<br>$14.2 \div 25.3$ |
| | 0.4 | $0.242^{+0.078}_{-0.031}$<br>$0.140 \div 0.356$ | $0.311^{+0.034}_{-0.049}$<br>$0.213 \div 0.387$ | $-0.52^{+0.21}_{-0.40}$<br>$-1 \div 0.102$ | $0.702^{+0.053}_{-0.061}$<br>$0.569 \div 0.799$ | $\left(52.0^{+5.1}_{-13.3}\right)^\circ$<br>$33.0^\circ \div 68.6^\circ$ | $18.5^{+5.8}_{-1.4}$<br>$14.7 \div 27.7$ |
| | 0.6 | $0.299^{+0.040}_{-0.052}$<br>$0.192 \div 0.373$ | $0.267^{+0.042}_{-0.039}$<br>$0.197 \div 0.358$ | $-0.82^{+0.25}_{-0.17}$<br>$-1 \div -0.216$ | $0.665^{+0.063}_{-0.068}$<br>$0.530 \div 0.783$ | $\left(41.7^{+9.1}_{-6.6}\right)^\circ$<br>$29.8^\circ \div 60.2^\circ$ | $22.6^{+3.5}_{-3.6}$<br>$15.9 \div 29.7$ |

**Table XXIII.** – *MSSM predictions.*

The data displayed in the first row of Table XXIII are the result of a new fit[1] performed using the modified formulas for $\Delta m_{B_d}$, $\Delta m_{B_s}$, $|\varepsilon_K|$ and treating $\Delta_{SUSY}$ as an additional free parameter. Although values of $\Delta_{SUSY}$ greater than 0.36 are disfavoured at the 84% C.L., SUSY effects as large as 190% of the Standard Model expectation are possible according to the 95% result. The prediction for the observable $\sin 2\beta$ is completely insensitive to the new physics at the present level of precision of the data, while the lowest values of $\sin 2\alpha$ and $\gamma$ and the highest ones of $\Delta m_{B_s}$ allowed at the 95% C.L. ($\gamma < 45°$, $\sin 2\alpha < -0.73$, $\Delta m_{B_s} > 22 ps^{-1}$) fall outside the corresponding ranges of the Standard Model predictions. A lower value of the imaginary part $\bar{\eta}$ and, in general, *less* marked CP-violating effects are predicted in the MSSM than in the Standard Model itself.

Figure 14 shows the 95% contours obtained after fixing the parameter at four different values belonging to the allowed range ($\Delta_{SUSY}$ = 0.0, 0.2, 0.4, 0.6). The variation of the minimum $\chi^2$ value with respect to the Standard Model ($\Delta_{SUSY} = 0$) result ($\chi_0^2$) is also indicated. The numerical results are given in Table XXIII. Even though the Standard Model scenario is slightly favoured by the present experimental constraints, the two extreme regions $\Delta_{SUSY} = 0.0$ and $\Delta_{SUSY} = 0.6$ of the $(\overline{\rho}, \overline{\eta})$ plane are not mutually exclusive. It will not be possible to test the reliability of the Standard Model with respect of its minimal extensions until more precise (and independent) experimental data are made available by the experiments at the $B$ factories. Figure 15 shows how the improved determination of the unitarity triangle may be able to exhibit significant divergences between the models already in the first years of running of the experiments.

[1] The constraints $\sin 2\beta$ and $\mathcal{Br}(B \to X_s\gamma)$ ($|V_{ts}|^2 |V_{tb}|^2 / |V_{cb}|^2$) have been removed.

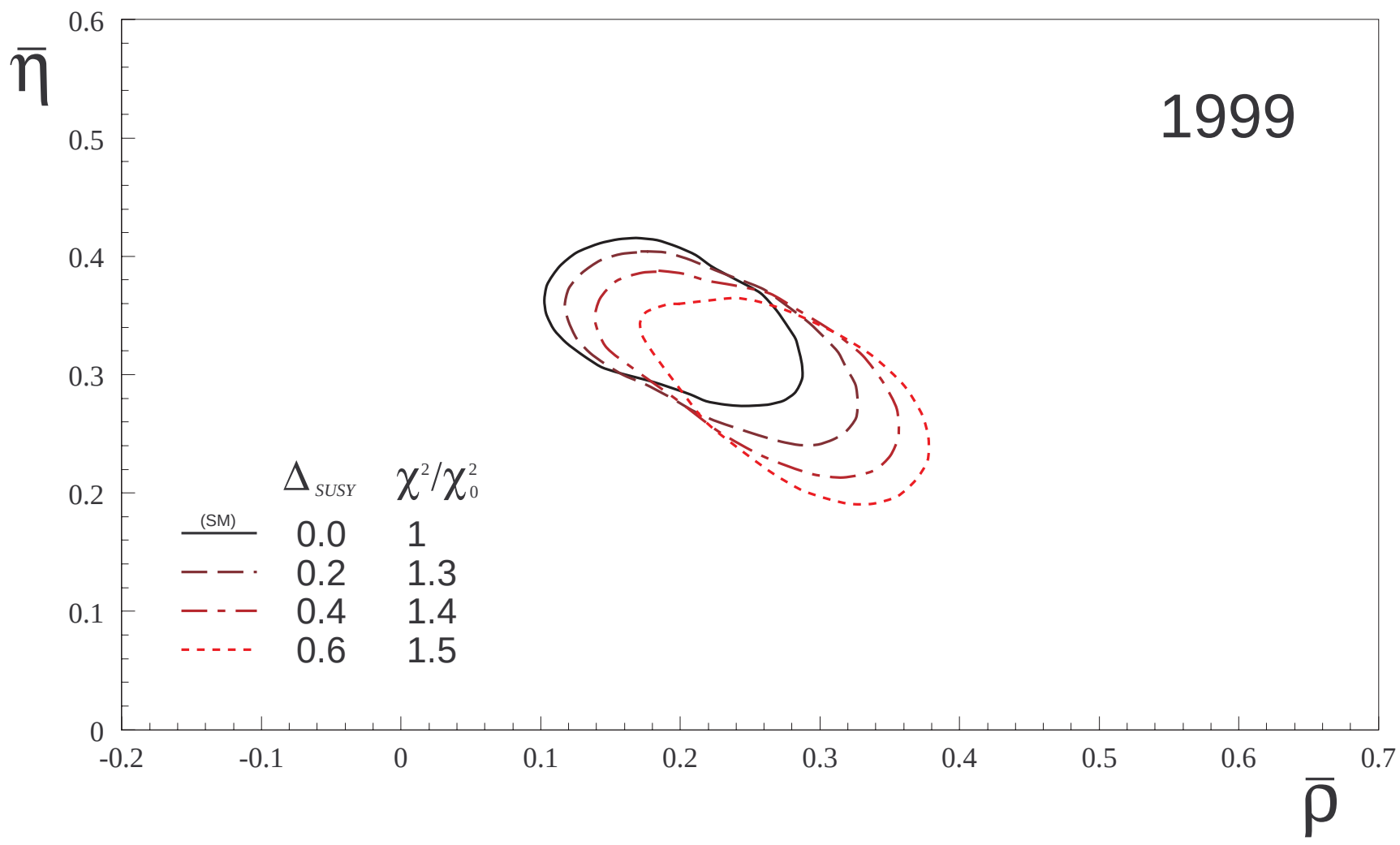

**Figure 14.** – *95% C.L. allowed regions for the vertex of the unitarity triangle in four different SUSY scenarios.*

The approximate error magnitudes expected in each experiment for the measurements of $\sin 2\beta$ and $\sin 2\alpha$ are reported in Table XXIV as functions of the year. The error values in $\sin 2\alpha$ are the results of very unequal estimations, in which different weights have been attributed to the unknown contribution of the penguin amplitude. The measurement of $\gamma$ has not been taken into account, since most of the experiments refrain from making predictions on the error. The largest measurable values of $\Delta m_{B_s}$ are also indicated. Using these data, the projections represented in Figure 15 have been calculated. They describe the probable way in which the results of the experiments running at the $B$ factories will progressively increase the accuracy in the determination of the vertex of the unitarity triangle, according to different Minimal Supersymmetric models (including the Standard Model itself). It has been assumed that the new measurements are compatible with the values predicted inside each respective model (Table XXIII) and that the first measurement of the $B_s^0 - \overline{B}_s^0$ oscillation frequency is provided by LHCb in 2006 with the error $\sigma(\Delta m_{B_s}) = 0.009\, ps^{-1}$. The results expected from single experiments have been combined. These hypothetical measurements have been used as the additional constraints of a new fit procedure, as a result of which the contours shown in Figure 15 have been calculated.

| | | $\Delta m_{B_s}$ | 2000 | 2002 | 2004 | 2006 |
|---|---|---|---|---|---|---|
| | | | $\sigma(\sin 2\beta)/\sigma(\sin 2\alpha)$ | | | |
| HERA-B | $pN \to b\bar{b} \to B_d^0, B^\pm, B_s^0, \Lambda_b$ | $\leq 22 ps^{-1}$ | 0.17 / 0.35 | 0.12 / 0.25 | 0.09 / 0.18 | 0.07 / 0.14 |
| BaBar | $e^+e^- \to \Upsilon(4S) \to B_d^0, B^\pm$ | – | 0.08 / 0.31 | 0.05 / 0.18 | 0.04 / 0.14 | 0.03 / 0.12 |
| Belle | $e^+e^- \to \Upsilon(4S) \to B_d^0, B^\pm$ | – | 0.10 / 0.15 | 0.06 / 0.09 | 0.04 / 0.07 | 0.04 / 0.06 |
| CDF | $p\bar{p} \to b\bar{b} \to B_d^0, B^\pm, B_s^0, \Lambda_b$ | $\leq 35 ps^{-1}$ | – | 0.07 / 0.09 | 0.05 / 0.06 | 0.04 / 0.05 |
| LHCb | $pp \to b\bar{b} \to B_d^0, B^\pm, B_s^0, \Lambda_b$ | $\leq 50 ps^{-1}$ | – | – | – | 0.023 / 0.06 |
| *Average* | | | **0.06 / 0.13** | **0.03 / 0.06** | **0.02 / 0.04** | **0.01 / 0.03** |

**Table XXIV.** – *Expected uncertainties in the measurements of* $\sin 2\beta$ *and* $\sin 2\alpha$ *and largest measurable values of* $\Delta m_{B_s}$ *in the experiments at the B factories.*

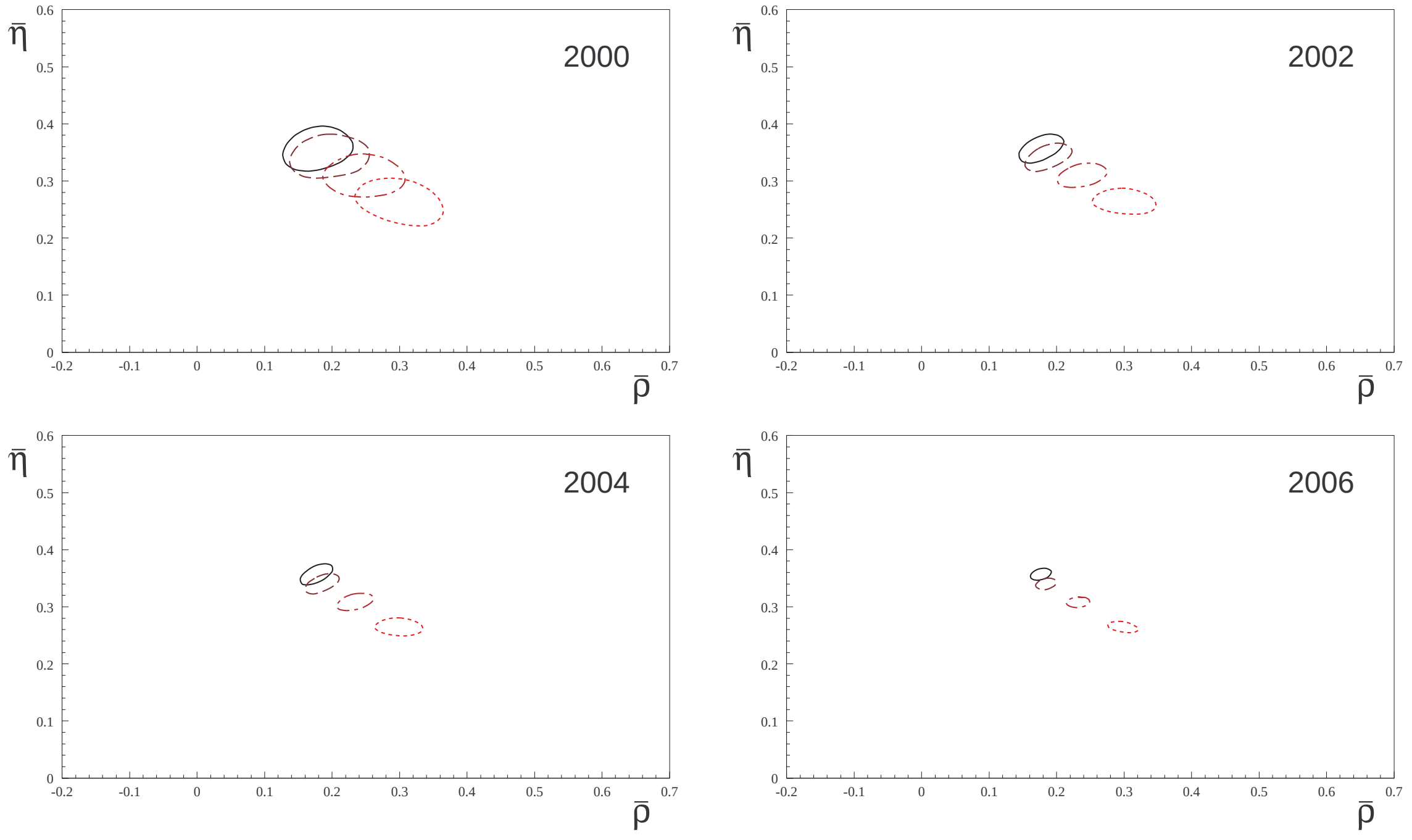

**Figure 15.** – *Expected evolution in the determination of the vertex of the unitarity triangle resulting from forthcoming measurements of* $\sin 2\alpha$, $\sin 2\beta$ *and (in 2006)* $\Delta m_{B_s}$ *at the B factories. Each graph represents the 95% C.L. regions corresponding respectively to* $\Delta_{SUSY} = 0$ *(Standard Model scenario, the upper left plot in all four cases) and* $\Delta_{SUSY} = 0.2, 0.4, 0.6.$ *The present (1999) scenarios were shown in Figure 14.*

# Conclusions

An intensive program for the study of CP violation in weak decays involving the *b*-quark is about to be carried on by dedicated experiments. Only indirect estimates can be given at present for the magnitude of these phenomena; moreover, they are obtained in the framework of a critical sector of the Standard Model, the one describing the flavour-changing processes and parametrized by the CKM matrix, the verification of which offers the most favourable prospects for the discovery of new physics. With the aim of setting the status of the present knowledge about the CKM matrix before the start of the new experimental venture, a detailed and comprehensive review of the theoretical and experimental information until now available in this field has been presented.

A great amount of precise information, obtained in some cases by exploiting new experimental sources, has been made available by the exceptionally thriving activity of the experiments in the very last years, leading in some cases to a noticeable improvement in the direct determination of the CKM matrix elements. For example, the precise measurement of $W^{\pm}$ hadronic decays can be translated into a determination of the element $|V_{cs}|$ being a factor of 6 more precise than the former best value derived from the analysis of *D* semileptonic decays. Moreover, thanks to a new measurement of the size of the nucleon strange-quark sea, it has been possible to extract a value for $|V_{cs}|$ from the neutrino-nucleon scattering data. Especially valuable information comes from the recent progress in the observation of *B* decays. More stringent constraints on $|V_{td}|$ and (in the form of a lower limit) $|V_{ts}|$ are now provided by the study of $\overline{B}_d^0 - \overline{B}_d^0$ and $\overline{B}_s^0 - \overline{B}_s^0$ oscillations, while the first determination of $|V_{ts}|$ has been made possible by new measurements and accurate theoretical calculations of the $b \to s\gamma$ decay rate. Furthermore, new or recently improved measurements of *B* semileptonic decays have led to better determinations of $|V_{ub}|$, $|V_{cb}|$ and $|V_{ub}/V_{cb}|$.

Both the latest measurements and earlier (but not always unproblematic) results have been critically reviewed in the present work, calling attention to the least clearly defined experimental situations and pointing out the main theoretical uncertainties and questions. Among the results obtained, the theoretical uncertainty on the earliest experimental determinations of $|V_{ub}/V_{cb}|$ has been reduced by exploiting the results of the recent measurements of $|V_{ub}|$, $|V_{cb}|$ and $|V_{ub}/V_{cb}|$ to discriminate between the disagreeing predictions of different models. The resulting information provides at present the most effective constraint on the values of the imaginary part of the CKM matrix and of the CP-violating parameters.

A set of independent constraints has then been defined on the basis of the collected information. The unitarity condition of the 3×3 CKM matrix has been imposed as a further constraint by expressing each quantity in terms of either Wolfenstein's or the canonical parametrization. With the explicit assumption of unitarity, an over-determined problem is obtained, the solution of which gives much more precise values for the matrix elements; at the same time, other quantities of physical interest, such as the angles of the unitarity triangle and the $B_s$-oscillation parameter, can be determined.

Two independent procedures have been followed, consisting respectively in a $\chi^2$-minimization and in a more refined approach based on Bayesian statistics. The results obtained from the application of the two methods are in good agreement and, in some cases, constitute a great improvement with respect to the direct experimental determinations of the CKM elements. For example, the errors in $|V_{cd}|$ and $|V_{cs}|$ are reduced respectively by factors of 6 and 8, while the precision on $|V_{tb}|$ is increased by 4 orders of magnitude. The probability distribution functions of the CKM elements and of all other relevant parameters are obtained as a result of the Bayesian determination. Highly asymmetric distributions are found for some quantities, such as the angles of the unitarity triangle, indicating that a complete description of the final knowledge of these parameters cannot be given if a simple confidence interval is quoted, as in the results of the $\chi^2$ minimization. On the other hand, the $\chi^2$ method has the advantage of providing a measure of the compatibility between the data and the model. The low $\chi^2$ value at the minimum denotes a high degree of consistency between the experimental data and the description of the CP violating phenomena considered inside the Standard Model.

The following predictions concerning the physics of *B*-meson decays have been obtained:

$$\sin 2\alpha = -0.11^{+0.20}_{-0.22}, \quad \sin 2\beta = 0.725^{+0.044}_{-0.046}, \quad \gamma = \left(63.7^{+5.3}_{-7.0}\right)^{\circ}, \; \Delta m_{B_s} = 15.4^{+3.0}_{-0.7}\, ps^{-1}.$$

They indicate with absolute certainty the presence of CP violation in *B* decays according to the Standard Model. Moreover, this conclusion is independent of the experimental evidence for the parallel phenomenon already observed in the kaon system, since the removal of the $|\varepsilon_K|$ constraint does not alter significantly the results.

The above-mentioned results constitute the reference values to which the direct CP-violation and $B_s^0$-mixing measurements will have to be compared in the search for new physics. On the other hand, their constraining effect on at least one class of *minimal* extensions of the Standard Model is still rather weak. In fact, the predictions of the Standard Model and those of its Minimal Supersymmetric extensions are at present nearly indistinguishable. However, as it has been shown in the last Section of this work, the direct measurements of CP violation expected in the first years of running of the *B* factories should already be able to bring the knowledge of the unitarity triangle to a substantially higher level of precision, offering the chance of detecting even the smallest non-Standard-Model effects.

# Acknowledgments

We would like to thank F. Buccella, N. Cabibbo, E.A. Paschos, P. Santorelli and H. Schröder for stimulating discussions and many helpful suggestions.

# References

[1] N. Cabibbo, *Phys. Rev. Lett.* **10** (1963) 351
[2] M. Kobayashi and T. Maskawa, *Prog. Theor. Phys.* **49** (1973) 652
[3] P. Faccioli, Thesis (in Italian), Bologna University (1999)
[4] M. Bargiotti *et al.*, *Nuovo Cimento* **112** A (1999), in press
[5] S.L. Glashow, J. Iliopoulos and L. Maiani, *Phys. Rev.* D **2** (1970) 1285
[6] M.L. Perl *et al.* (MARK I Collab.), *Phys. Rev. Lett.* **35** (1975) 1489; M.L. Perl *et al.* (MARK I Collab.), *Phys. Lett.* B **63** (1976) 466; M.L. Perl *et al.* (MARK I Collab.), *Phys. Lett.* B **70** (1977) 487
[7] M. Bernardini *et al.*, *"A proposal to search for leptonic quarks and heavy leptons produced by ADONE"*, I.N.F.N./AE-67/3 (1967); V. Alles Borelli *et al.*, *Lett. Nuovo Cimento* **4** (1970) 1156; M. Bernardini *et al.*, *Nuovo Cimento* A **17** (1973) 383; V. Alles Borelli *et al.*, *Phys. Lett.* B **59** (1975) 201
[8] S.W. Herb *et al.* (Fermilab P.C. Collab.), *Phys. Rev. Lett.* **39** (1977) 252; W.R. Innes *et al.* (Fermilab P.C. Collab.), *Phys. Rev. Lett.* **39** (1977) 1240
[9] L.-L. Chau and W.-Y. Keung, *Phys. Rev. Lett.* 53 (1984) 1802
[10] L. Maiani, *Phys. Lett.* **B** 62 (1976) 183
[11] L. Wolfenstein, *Phys. Rev. Lett.* **51** (1983) 1945
[12] C. Caso *et al.* (Particle Data Group), *Eur. Phys. J.* C **3** (1998) 1
[13] CDF Collab., *"A measurement of* $\sin 2\beta$ *from* $B \to J/\psi K_S^0$ *with the CDF detector"*, CDF/PUB/BOTTOM/CDF/4855
[14] A. Bertin and A. Vitale, *Riv. Nuovo Cimento* **7** (1984) No. 7
[15] F.J.M. Farley *et al.*, *Proceedings of the International Conference on High Energy Physics, Geneva* (CERN, Geneva, 1962), p.415; A. Zichichi, *Suppl. Nuovo Cimento* **1** (1964) 11
[16] G. Bardin *et al.*, *Phys. Lett.* B **137** (1984) 135
[17] K.L. Giovanetti *et al.*, *Phys. Rev.* D **29** (1984) 343
[18] I. S. Towner and J.C. Hardy, *"The current status of* $|V_{ud}|$*"*, preprint nucl-th/9809087
[19] $\Delta_R$: Ref. 44 and A. Sirlin, in *"Precision Tests of the Standard Electroweak Model"*, ed. P. Langacker (World-Scientific, Singapore, 1994). $\delta_C$: I. S. Towner, J.C. Hardy and M. Harvey, *Nucl. Phys.* A **284** (1977) 269; W.E. Ormand and B.A. Brown, *Phys. Rev.* C **52** (1995) 2455. $\delta_R$: A. Sirlin, *Phys. Rev.* **164** (1967) 1767; A. Sirlin and R. Zucchini, *Phys. Rev. Lett.* **57** (1986) 1994; W. Jaus and G. Rasche, *Phys. Rev.* D **35** (1987) 3420; A. Sirlin, *Phys. Rev.* D **35** (1987) 3423; I. S. Towner, *Nucl. Phys.* A **540** (1992) 478; I. S. Towner, *Phys. Lett.* B **333** (1994) 13
[20] D.H. Wilkinson, *Nucl. Phys.* A **511** (1990) 301; *Nucl. Instrum. Meth.* A **335** (1993) 172, 182, 201; *Z. Phys.* A **348** (1994) 129
[21] K. Saito and A.W. Thomas, *Phys. Lett.* B **363** (1995) 157
[22] D.H. Wilkinson, *Nucl. Phys.* A **377** (1982) 474; I. S. Towner and J.C. Hardy, in *"Symmetries and Fundamental Interactions in Nuclei"*, eds. E.M. Henley and W.C. Haxton (World-Scientific, Singapore, 1995), pp. 183-249
[23] C.J. Christensen *et al.*, *Phys. Rev.* D **5** (1972) 1628
[24] P.E. Spivak, *Sov. Phys. JETP* **67** (1988) 1735

[25] J. Last *et al.*, *Phys. Rev. Lett.* **60** (1988) 995

[26] R. Kossakowski *et al.*, *Nucl. Phys.* A **503** (1989) 473

[27] J. Byrne *et al.*, *Europhys. Lett.* **33** (1996) 187

[28] Yu. Yu. Kosvintsev *et al.*, *JETP Lett.* **44** (1986) 571

[29] W. Paul *et al.*, *Z. Phys.* A **45** (1989) 25

[30] W. Mampe *et al.*, *Phys. Rev. Lett.* **63** (1989) 593

[31] V.V. Nesvizhevskii *et al.*, *Sov. Phys. JETP* **75** (1992) 405

[32] W. Mampe *et al.*, *JETP Lett.* **57** (1993) 82

[33] V.E. Krohn *et al.*, *Phys. Lett.* B **55** (1975) 175

[34] B.G. Erozolimskii *et al.*, *Sov. J. Nucl. Phys.* **30** (1979) 356

[35] P. Bopp *et al.*, *Phys. Rev. Lett.* **56** (1986) 919

[36] B. Yerozolimsky *et al.*, *Phys. Lett.* B **412** (1997) 240

[37] P. Liaud *et al.*, *Nucl. Phys.* A **612** (1997) 53

[38] H. Abele *et al.*, *Phys. Lett.* B **407** (1997) 212

[39] C. Stratowa *et al.*, *Phys. Rev.* D **18** (1978) 3970

[40] A. Bertin and A. Vitale, *Riv. Nuovo Cimento* **7** (1984) No. 8

[41] W.K. McFarlane *et al.*, *Phys. Rev.* D **32** (1985) 547

[42] V.P. Koptev *et al.*, *JETP Lett.* **61** (1995) 877; T. Numao *et al.*, *Phys. Rev.* D **52** (1995) 4855

[43] H. Leutwyler and M. Roos, *Z. Phys.* C **25** (1984) 91

[44] W.J. Marciano and A. Sirlin, *Phys. Rev. Lett.* **56** (1986) 22

[45] H.H. Williams *et al.*, *Phys. Rev. Lett.* **33** (1974) 240; T. Becherrawy, *Phys. Rev.* D **1** (1970) 1452

[46] W. Jaus, *Phys. Rev.* D **44** (1991) 2851

[47] E.A. Paschos and U. Türke, *Phys. Rep.* **178** (1989) 145

[48] J.F. Donoghue, B.R. Holstein and S.W. Klimt, *Phys. Rev.* D **35** (1987) 934

[49] M. Bourquin *et al.* (BGHORS Collab.), *Z. Phys.* C **21** (1983) 27;
Particle Data Group, *Rev. Mod. Phys.* **56** (1984) 1

[50] R. Flores-Mendieta, A. García and G. Sánchez-Colón, *Phys. Rev.* D **54** (1996) 6855

[51] J.A. Thompson et al., *Phys. Rev.* D **21** (1980) 25

[52] J. Anderson and M.A. Luty, *Phys. Rev.* D **47** (1993) 4975

[53] H. Abramowicz *et al.* (CDHS), *Z. Phys.* C **15** (1982) 19

[54] A.O. Bazarko *et al.* (CCFR Collab.), *Z. Phys.* C **65** (1995) 189

[55] P. Vilain *et al.* (CHARM II Collab.), *"Leading-order QCD analysis of neutrino-induced dimuon events"*, CERN-EP/98-128, August 1998

[56] T. Bolton, *"Determining the CKM parameter $V_{cd}$ from $\nu N$ charm production"*, hep-ex/9708014

[57] N. Ushida *et al.* (Fermilab E531 Collab.), *Phys. Lett.* B **206** (1988) 375

[58] Ref. 67 and Y. Kubota *et al.* (CLEO Collaboration), *Phys. Rev.* D **54** (1996) 2994

[59] S. J. Brodsky and B.-Q. Ma, *Phys. Lett.* B **381** (1996) 317

[60] F. Buccella, O. Pisanti and L. Rosa, *"The strange quark problem in the framework of statistical distributions"*, in preparation

[61] U.K. Yang *et al.* (CCFR), *"Measurement of the longitudinal structure function and $|V_{cs}|$ in the CCFR experiment"*, preprint hep-ex/9806023

[62] L. Montanet *et al.* (Particle Data Group), *Phys. Rev.* D **50** (1994) 1173

[63] P. Abreu *et al.* (DELPHI Collab.), *Phys. Lett.* B **439** (1998) 209
[64] R. Barate *et al.* (ALEPH Collab.), *Phys. Lett.* B **453** (1999) 107; ALEPH Collaboration, *"Measurement of W-pair production in $e^+e^-$ collisions at 189 GeV", International Europhysics Conference on High Energy Physics, Tampere 1999 (HEP'99),* ALEPH 99-064 CONF 99-038, July 1999; ALEPH Collaboration, *"A direct measurement of $|V_{cs}|$ in hadronic W decays using a charm tag",* CERN-EP-99-124, September 1999
[65] The L3 Collaboration, *"Preliminary results on charm production in W decays", XXIX International Conference on High Energy Physics, Vancouver 1998 (ICHEP'98*), L3 Note 2232, June 1998; *"Preliminary results on the measurement of W-pair cross sections in $e^+e^-$ interactions at $\sqrt{s} = 189$ GeV and W-decay branching fractions", HEP'99,* L3 Note 2376, March 1999
[66] The OPAL Collaboration, *"$W^+W^-$ production in $e^+e^-$ collisions at 189 GeV",* OPAL Physics Note PN378 March 1999; *"A measurement of charm production in $e^+e^- \to W^+W^-$ at LEP2", HEP'99,* OPAL Physics Note PN402, July 1999
[67] R.M. Barnett *et al.* (Particle Data Group), *Phys. Rev.* D **54** (1996) 1
[68] R. Fulton *et al.* (CLEO Collab.), *Phys. Rev. Lett.* **64** (1990) 16
[69] H. Albrecht *et al.* (ARGUS Collab.), *Phys. Lett.* B **234** (1990) 409
[70] H. Albrecht *et al.* (ARGUS Collab.), *Phys. Lett.* B **255** (1991) 297
[71] I.I. Bigi, M. Schifman and N.G. Uraltsev, *Annu. Rev. Nucl. Part. Sci.* **47** (1997) 591; N. Uraltsev, *"Theoretical uncertainties in $\Gamma_{sl}(b \to u)$",* hep-ph/9905520 June 1999; I.I. Bigi, *"Memo on extracting $|V_{cb}|$ and $|V_{ub}/V_{cb}|$ from semileptonic B decays",* hep-ph/9907270 July 1999
[72] D. Abbaneo et al. (The LEP $V_{ub}$ Working Group), *"Determination of the LEP average* BR$(b \to X_u \ell \nu)$ *and derivation of $|V_{ub}|$",* LEPVUB-99/01 June 1999
[73] J.P. Alexander *et al.* (CLEO Collab.), *Phys. Rev. Lett.* **77** (1996) 5000
[74] C.P. Jessop *et al.* (CLEO Collaboration), *"Measurement of $\mathcal{B}(B \to \rho \ell \nu)$, $|V_{ub}|$ and the Form Factor Slope in $B \to \rho \ell \nu$ Decay", ICHEP'98* #855, CLEO CONF 98-18; B.H. Behrens *et al.* (CLEO Collab.), *"Measurement of $B \to \rho \ell \nu$ decay and $|V_{ub}|$",* preprint hep-ex/9905056
[75] M. Acciarri *et al.* (L3 Collab.), *Phys. Lett.* B **436** (1998) 174
[76] R. Barate *et al.* (ALEPH Collab.), *Eur. Phys. J.* C **6** (1999) 555
[77] N. Isgur and M. Wise, *Phys. Lett.* B **232** (1989) 113, *ibid.* **237** (1990) 527; H. Georgi, *Phys. Lett.* B **238** (1990) 395
[78] I. Caprini and M. Neubert, *Phys. Lett.* B **380** (1996) 376
[79] M.E. Luke, *Phys. Lett.* B **252** (1990) 447
[80] H. Albrecht *et al.* (ARGUS Collab.), *Z. Phys.* C **57** (1993) 533, *Phys. Rept.* **276** (1996) 223
[81] M. Athanas *et al.* (CLEO Collab.), *"Measurement of the $\overline{B} \to D \ell \overline{\nu}$ partial width and form factor parameters",* hep-ex/9705019; J. Bartelt *et al.* (CLEO Collab.), *"Measurement of the $B \to D \ell \nu$ branching fractions and form factor",* hep-ex/9811042
[82] B. Barish *et al.* (CLEO Collab.), *Phys. Rev.* D **51** (1995) 1014

[83] K. Ackerstaff *et al.* (OPAL Collab.), *Phys. Lett.* B **395** (1997) 128
[84] D. Buskulic *et al.* (ALEPH Collab.), *Phys. Lett.* B **395** (1997) 373
[85] P. Abreu *et al.* (DELPHI Collab.), *Z. Phys.* C **71** (1996) 539; DELPHI Collaboration, *"New precise measurement of $V_{cb}$", HEP'99* #4_518, DELPHI 99-107 CONF 294, June 1999
[86] J.G. Körner and G.A. Schuler, *Z. Phys.* C **38** (1988) 511
[87] M. Wirbel, B. Stech and M. Bauer, *Z. Phys.* C **29** (1985) 637
[88] G. Altarelli, N. Cabibbo, G. Corbò, L. Maiani and G. Martinelli, *Nucl. Phys.* B **208** (1982) 365
[89] N. Isgur, D. Scora, B. Grinstein and M.B. Wise, *Phys. Rev.* D **39** (1989) 799
[90] J. Bartelt *et al.* (CLEO Collab.), *Phys. Rev. Lett.* **71** (1993) 4111
[91] M. Battaglia *et al.* (DELPHI Collab.), *"Measurement of $\mathcal{Br}(b \to X_u \ell \nu)$ and determination of $|V_{ub}|/|V_{cb}|$ with DELPHI at LEP", HEP'99* #4_521, DELPHI 99-110 CONF 297, June 1999
[92]The CDF Collaboration, *"First direct measurement of the ratio of branching fractions $B(t \to Wb)/B(t \to Wq)$ and of the CKM element $|V_{tb}|$ in $p\bar{p}$ collisions at $\sqrt{s} = 1.8 TeV$ "*, CDF/PUB/TOP/CDFR/5028 (1999)
[93] A.P. Heinson, *"Measuring the CKM matrix element $V_{tb}$ at D0 and CDF"*, D0-Conf-97-26 UCR/D0-97-12, FERMILAB-Conf-97/238-E, July 97
[94] T. Inami, C.S. Lim, *Prog. Theor. Phys.* **65** (1981) 297
[95] G. Buchalla, A.J. Buras and M.K. Harlander, *Nucl. Phys.* B **337** (1990) 313; W.A. Kaufman, H. Steger and Y.P. Yao, *Mod. Phys. Lett.* A **3** (1989) 1479; J.M. Flynn, *Mod. Phys. Lett.* A **5** (1990) 877; A. Datta, J. Frölich and E.A. Paschos, *Z. Phys.* C **46** (1990) 63; A.J. Buras, M. Jasmin and P.H. Weisz, *Nucl. Phys.* B **347** (1990) 491; S. Herrlich and U. Nierste, *Nucl. Phys.* B **419** (1994) 292; A.J. Buras, MPI-PHT/95-88, TUM-T31-97/95 (1995)
[96] G. Buchalla, A.J. Buras, M.E. Lautenbacher, *Rev. Mod. Phys.* **68** (1996) 1125
[97] P. Paganini, F. Parodi, P. Rondeau and A. Stocchi, *Physica Scripta* **58** (1998) 556
[98] S. Aoki *et al.* (WA75 Collab.), *Prog. Theor. Phys.* **89** (1993) 131
[99] J.Z. Bai *et al.* (BES Collab.), *Phys. Rev. Lett.* **74** (1995) 4599
[100] K. Kodama *et al.* (Fermilab E653 Collab.), *Phys. Lett.* B **382** (1996) 299
[101] M. Acciarri *et al.* (L3 Collab.), *Phys. Lett.* B **396** (1997) 327
[102] F. Parodi, P. Rondeau, A. Stocchi (DELPHI Collab.) *"Measurement of the branching fraction $D_s^+ \to \tau^+ \nu_\tau$ ", HEP'97 Conference, Jerusalem,* #455, DELPHI 97-105 CONF 87
[103] M. Chada *et al.* (CLEO Collab.), *Phys. Rev.* D **58** (1998) 032002
[104] ALEPH Collaboration, *"Leptonic decays of the $D_s$ meson", ICHEP'98* #937
[105] F Buccella and P. Santorelli, *Nuovo Cimento* A **107** (1994) No. 5
[106] C. Bernard *et al.* (MILC Collab.), *Phys. Rev. Lett.* **81** (1998) 4812
[107] C. Bernard *et al.* (MILC Collab.), *"Heavy-light decay constants: conclusion from the Wilson action"*, hep-lat/9809109
[108] S Aoki *et al.* (JLQCD Collab.), *Phys. Rev. Lett.* **80** (1998) 5711

[109] K. I. Ishikawa et al. (JLQCD Collab.), *"Scaling behaviour of* $f_B$ *with NRQCD"*, hep-lat/9809152

[110] L. Lellouch and C.J. David Lin (UKQCD Collab.), *"Neutral B meson mixing and heavy-light decay constants from quenched lattice QCD"*, hep-lat/9809018

[111] A. X. El-Khadra *et al.*, *Phys. Rev.* D **58** (1998) 014506

[112] J. Hein, *"Pseudoscalar decay constant in heavy-light systems"*, hep-lat/9809051

[113] N. Yamada *et al.*, *"Preliminary study of* $B_B$ *parameter using Lattice NRQCD"*, hep-lat/9809156

[114] C.R. Allton *et al.*, *Phys. Lett.* B **405** (1997) 133; L. Conti, *Nucl. Phys.* B (*Proc. Suppl.*) **63A-C** (1998) 359

[115] V. Giménez, G. Martinelli, *Phys. Lett.* B **398** (1997) 135

[116] A. Ali Khan *et al.*, *Phys. Lett.* B **427** (1998) 132

[117] D. Becirevic *et al.*, *"Non-perturbatively improved heavy-light mesons: masses and decay constants"*, hep-lat/9811003

[118] V.M. Braun, *"QCD sum rules for heavy flavors"*, Invited talk at the 8th International Symposium on Heavy Flavour Physics (Heavy Flavours 8), Southampton, July 1999

[119] A. Soni, *Nucl. Phys.* B (*Proc. Suppl.*) **47** (1996) 43

[120] V. Giménez and J. Reyes, *"Calculation of the continuum-lattice HQET matching for the complete basis of four-fermion operators: reanalysis of the* $B^0-\overline{B}^0$ *mixing"*, hep-lat/9806023

[121] J. Christensen, T. Draper and C. McNeile, *Phys. Rev.* D **56** (1997) 6993

[122] C. Bernard, T. Blum and A. Soni, *"SU(3) flavour breaking in hadronic matrix elements for* $B-\overline{B}$ *oscillations"*, hep-lat/9801039

[123] The LEP B Oscillations Working Group, *"Combined results on* $B^0$ *oscillations: results for Lepton-Photon 1999"*, *XIX International Symposium on Lepton and Photon Interactions at High Energies, Stanford University, August 1999*, http://www.cern.ch/LEPBOSC/

[124] H.G. Moser, A. Roussarie, *Nucl. Instrum. Meth.* A **384** (1997) 491

[125] L. Wolfenstein, *Phys. Rev. Lett.* **13** (1964) 380

[126] L.K. Gibbons *et al.* (E731), *Phys. Rev. Lett.* **70** (1993) 70

[127] G.D. Barr *et al.* (NA31), *Phys. Lett.* B **317** (1993) 233

[128] A. Alavi-Harati *et al.*, *Phys. Rev. Lett.* **83** (1999) 22

[129] V. Fanti *et al.*, *"A new measurement of direct CP violation in two pion decays of the neutral kaon"*, submitted to Physics Letters B, hep-ex/9909022

[130] I. Bigi, *"CP Violation – A Probe of Nature's Grand Design"*, UND-HEP-97-BIG 09

[131] Y. Keum, U. Nierste and A.I. Sanda, *"A short look at* $\varepsilon'/\varepsilon$ *"*, hep-ph/9903230

[132] A.J. Buras, M. Jamin and M.E. Lautenbacher, *Nucl. Phys.* B **370** (1992) 69, *Nucl. Phys.* B **400** (1993) 37, *Nucl. Phys.* B **400** (1993) 75, *Nucl. Phys.* B **408** (1993) 209; M. Ciuchini, E. Franco, G, Martinelli and L. Reina, *Phys. Lett.* B **301** (1993) 263, *Nucl. Phys.* B **415** (1994) 403

[133] R. Gupta, *"Quark masses, B-parameters, and CP violation parameters* $\varepsilon$ *and* $\varepsilon'/\varepsilon$ *"*, hep-ph/9801412

[134] L. Conti *et al.*, *Phys. Lett.* B **421** (1998) 273

[135] W.A. Bardeen, A.J. Buras and J.-M. Gérard, *Phys. Lett.* B **180** (1986) 133, *Nucl. Phys.* B **293** (1987) 787, *Phys. Lett.* B **192** (1987) 138; T. Hambye *et al.*, *Phys. Rev.* D **58** (1998) 014017
[136] S.R. Sharpe, *"Progress in lattice gauge theory"*, hep-lat/9811006
[137] A.J. Buras, M, Jamin, M.E. Lautenbacher, *Phys. Lett.* B **389** (1996) 749
[138] M. Ciuchini, *Nucl. Phys.* B *(Proc. Suppl.)* **59** (1997) 149
[139] S. Bertolini *et al.*, *Nucl. Phys.* B **514** (1998) 93
[140] A.J. Buras, *"Weak hamiltonian, CP violation and rare decays"*, TUM-HEP-316-98, hep-ph/9806471
[141] E.A. Paschos, *"Rescattering effects for $\varepsilon'/\varepsilon$ "*, in preparation
[142] R. Burkhalter, *"Recent results from the CP-PACS Collaboration"*, hep-lat/9810043; S. Aoki *et al.*, *"Full QCD light hadron spectrum from the CP-PACS"*, hep-lat/9809120
[143] G. Kilcup, R. Gupta, S.R. Sharpe, *Phys. Rev.* D **57** (1998) 1654
[144] S. Aoki *et al.* (JLQCD Collab.), *Phys. Rev. Lett.* **80** (1998) 5271
[145] S.R. Sharpe, *Nucl. Phys.* B (*Proc. Suppl.*) **53** (1997) 181
[146] R. Ammar *et al.* (CLEO Collab.), *Phys. Rev. Lett.* **71** (1993) 674
[147] K. Chetyrkin, M. Misiak, M. Münz, *Phys. Lett.* B **400** (1997) 206
[148] M.S. Alam *et al.* (CLEO Collab.), *Phys. Rev. Lett.* **74** (1995) 2885; S. Glenn *et al.* (CLEO Collab.), *"Improved measurement of $\mathcal{B}(b \to s\gamma)$"*, *ICHEP'98* #1011, CLEO CONF 98-17
[149] R. Barate *et al.* (ALEPH Collab.), *Phys. Lett.* B **429** (1998) 169
[150] F. James, *MINUIT – Function Minimization and Error Analysis*, CERN Program Library Long Writeup D506 (1994)
[151] F. Parodi, P. Rondeau, A. Stocchi, *Nuovo Cimento* **112** A (1999) No. 8
[152] A. Ali and D. London, *"Profiles of the unitarity triangle and CP-violating phases in the Standard Model and supersymmetric theories"*, hep-ph/9903535
[153] D.E. Jaffe and S. Youssef, *Computer Physics Communications* **11** (1997) 206
[154] W.T. Eadie et al., *Statistical Methods in Experimental Physics*, North-Holland Publ. Comp., Amsterdam, 1998
[155] G. D'Agostini, *"Bayesian reasoning in high-energy physics: principles and applications"*, CERN 99-03 July 1999
[156] P. Huet and E. Sather, *Phys. Rev.* D **51** (1995) 379
[157] S. Baek and P. Ko, *Phys. Rev. Lett.* **83** (1999) 488